\pdfoutput=1
\documentclass[cernpreprint, atlasdraft=false, , UKenglish, texmf, orcidlogo]{atlasdoc}
\usepackage[subcaption]{atlaspackage}

\usepackage{multirow}
\usepackage{makecell}
\usepackage{duckuments}
\usepackage{longtable}
\usepackage{booktabs}
\usepackage[export]{adjustbox}
\usepackage{float}
\usepackage{tabularx}
\usepackage{dirtytalk}
\usepackage{enumitem}
\usepackage{atlasbiblatex}

\usepackage[jetetmiss=true, process=true]{atlasphysics}
\graphicspath{{logos/}{figures/}}

\usepackage{ANA-EXOT-2024-12-PAPER-defs}

\AtlasTitle{Search for resonant leptoquark production via lepton--jet signatures in $pp$ collisions at $\sqrt{s} = 13$~TeV and $\sqrt{s} = 13.6$~TeV with the ATLAS detector}

\AtlasAbstract{%
This paper presents a search for physics beyond the Standard Model targeting a heavy resonance visible in the invariant mass of the lepton--jet system.
The analysis focuses on final states with a high-energy lepton and jet, and is optimised for the resonant production of leptoquarks—a novel production mode mediated by the lepton content of the proton originating from quantum fluctuations.
Four distinct and orthogonal final states are considered: $e$+light jet, $\mu$+light jet, $e$+$b$-jet, and $\mu$+$b$-jet, constituting the first search at the Large Hadron Collider for resonantly produced leptoquarks with couplings to electrons and muons.
Events with an additional same-flavour lepton, as expected from higher-order diagrams in the signal process, are also included in each channel.
The search uses proton--proton collision data from the full Run~2, corresponding to an integrated luminosity of 140~fb$^{-1}$ at a centre-of-mass energy of $\sqrt{s} = 13$~TeV, and from a part of Run~3 (2022--2023), corresponding to 55~fb$^{-1}$ at $\sqrt{s} = 13.6$~TeV.
No significant excess over Standard Model predictions is observed.
The results are interpreted as exclusion limits on scalar leptoquark ($\tilde{S}_1$) production, substantially improving upon previous ATLAS constraints from leptoquark pair production for large coupling values.
The excluded $\tilde{S}_1$ mass ranges depend on the coupling strength, reaching up to 3.4~TeV for quark--lepton couplings $y_{de} = 1.0$, and up to 4.3~TeV, 3.1~TeV, and 2.8~TeV for $y_{s\mu}$, $y_{be}$, and $y_{b\mu}$ couplings set to 3.5, respectively.
}

\AtlasRefCode{EXOT-2024-12}

\PreprintIdNumber{CERN-EP-2025-142}

\AtlasJournal{JHEP}
\AtlasJournalRef{JHEP 12 (2025) 180}
\AtlasDOI{10.1007/JHEP12(2025)180}

\hypersetup{pdftitle={ATLAS document},pdfauthor={The ATLAS Collaboration}}

\begin{document}

\maketitle

\tableofcontents

\section{Introduction}

%

While the discovery of the Higgs boson~\cite{HIGG-2012-27,CMS-HIG-12-028} and subsequent precision measurements of its properties~\cite{HIGG-2021-23,CMS-HIG-22-001} remain landmark achievements, an equally fundamental scientific motivation for the Large Hadron Collider (LHC)~\cite{Evans:2008zzb} is the search for physics beyond the Standard Model (BSM). Numerous BSM theories predict the existence of new massive states that could be produced resonantly in proton--proton ($pp$) collisions at the \TeV scale, provided their masses lie within the experimentally accessible range of the LHC\@. When produced through resonant mechanisms, these states would manifest as localised excesses in the invariant mass spectra of their decay products. Such distinctive peaks superimposed on the smoothly falling background predicted by the Standard Model (SM) would constitute an unambiguous signature of new physics. Consequently, resonance searches constitute a cornerstone of the research programs for both the ATLAS and CMS experiments.

Leptoquarks (LQs) represent a prominent category of hypothetical heavy states predicted by various grand unified theories featuring extended gauge groups~\cite{PhysRevD.10.275,BUCHMULLER1987442,Dorsner:2016wpm}.
These colour-charged particles carry both baryon and lepton numbers, naturally reflecting the underlying symmetry between the two sectors as predicted by unifying theories and suggested by observed patterns in nature.
Characterised by fractional electric charges and postulated in either scalar or vector forms, LQs decay into distinctive lepton--quark pairs whose flavour composition depends on the LQs' coupling parameters.

Extensive investigations of LQ signatures were conducted using the datasets collected during the LHC Run 1 and Run 2 data-taking periods.
Previous LHC searches have primarily focused on two production mechanisms: pair production (PP) mediated by quantum chromodynamics (QCD) and coupling-dependent single production (SP) in association with a lepton, as illustrated in Figure~\ref{fig:intro:LQ_prod_modes} (a) and (b), respectively.
While PP cross-sections depend predominantly on the strong coupling constant and
remain largely independent of LQ coupling parameters, production via SP becomes more and more relevant with increasing coupling values due to its direct dependence on these parameters.
Additional LQ sensitivity also arises from non-resonant contributions to dilepton production via $t$-channel processes (Drell--Yan, DY) seen in Figure~\ref{fig:intro:LQ_prod_modes}~(c).
A summary of the search strategies for pair-, single- and DY production of LQs can be found in Refs.~\cite{Diaz_2017,Schmaltz_2019}.

This paper introduces a novel probe of $s$-channel resonant LQ production at hadron colliders~\cite{Ohnemus:1994xf} shown in Figure~\ref{fig:intro:LQ_prod_modes} (d) and (e), enabled by recent advancements in understanding the proton's lepton content that arises from quantum fluctuations. State-of-the-art next-to-leading-order (NLO) calculations of lepton parton distribution functions (PDFs)~\cite{Buonocore_2020} show that, despite the relative scarcity of leptonic constituents of the proton compared to quarks and gluons, this previously unexplored production channel offers competitive sensitivity relative to established search strategies~\cite{LQ_Pheno_2020_scalar}.
The resonant production mechanism yields a relatively clean final state comprising a lepton--jet system whose invariant mass distribution could reveal a characteristic resonance peak, similar to single LQ production. This distinctive signature motivates the development of dedicated search algorithms within the ATLAS experimental programme.

The analysis employs a \Stilde LQ with absolute electric charge $\frac{4}{3}e$ as a benchmark signal. This LQ is a singlet under the $SU(2)$ SM gauge group, with purely right-handed Yukawa couplings to down-type quarks and leptons.
Therefore, it is characterised by a simple, exclusive decay topology into a down-type quark and a charged lepton, with no additional competing decay channels into neutral leptons.
Four distinct coupling scenarios are investigated where the LQ exclusively interacts through single
non-zero coupling parameters: $y_{de}$, $y_{s\mu}$, $y_{be}$ and $y_{b\mu}$,
corresponding to first/second-generation lepton ($e,\mu$) couplings with $d/s$- or $b$-quark partners.
Alternative models featuring couplings to $u/c$-quark flavours show detector-level signatures kinematically
indistinguishable with the studied scenarios, permitting phenomenological re-interpretation
through coupling parameter rotation in the quark flavour space.
The ATLAS Collaboration has performed a pair-production search focusing on scalar LQ decays
into quark--lepton ($q\ell$) final states with $\ell = e,\mu$, establishing mass exclusion limits up
to 1.8~\TeV (electron channel) and 1.7~\TeV (muon channel)
assuming a branching ratio into a charged lepton and a quark of 100\%,
with minimal dependence on the quark flavour~\cite{EXOT-2019-13}.
These results also served as input for a statistical combination of LQ searches~\cite{EXOT-2020-27} to enhance sensitivity across different decay modes.
Searches for LQ pair-production from the CMS Collaboration constrain scalar LQ masses up to 1.4~\TeV and 1.5~\TeV for couplings to electrons and muons, respectively~\cite{CMS-EXO-17-009,CMS-EXO-17-003}. For couplings to muons and $b$-quarks, these constraints extend up to 1.8~\TeV~\cite{CMS-EXO-21-019}.
A recent search by the CMS Collaboration for $t$-channel LQ exchange strengthens the constraints on large couplings involving up- and down-type quarks and electrons or muons, probing LQ masses up to 5~\TeV~\cite{CMS-EXO-22-013}.
The CMS Collaboration also conducted a search for resonant LQs with couplings to a $\tau$-lepton and a $u/d/s/b$
quark using data collected in 2016--2018~\cite{CMS-EXO-22-018}.

The present analysis introduces the first dedicated exploration of resonant LQ production mechanisms coupling to first- and second-generation leptons ($e/\mu$). %
It is conducted using $\sqrt{s} = 13~\TeV$ and $\sqrt{s} = 13.6~\TeV$ $pp$ collision data collected with the ATLAS experiment from 2015 to 2018 in LHC Run~2 (140~\ifb) and from 2022 to 2023 in Run~3 (55~\ifb), respectively.

\begin{figure}[tb]
\centering
\subfloat[]{
\includegraphics[width=0.3\textwidth,valign=m]{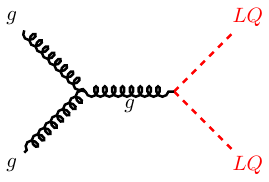}
}\hfil
\subfloat[]{
\includegraphics[width=0.3\textwidth,valign=m]{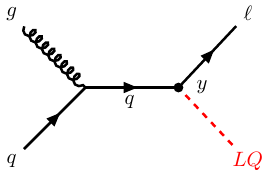}
}\hfil
\subfloat[]{
\includegraphics[width=0.2\textwidth,valign=m]{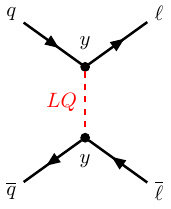}
} \\
\subfloat[]{
\includegraphics[width=0.3\textwidth,valign=c]{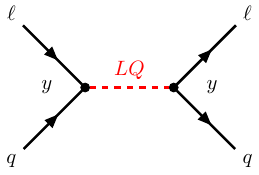}
}
\hfil
\subfloat[]{
\includegraphics[width=0.3\textwidth,valign=c]{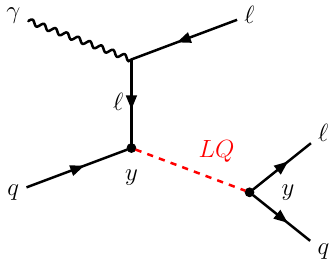}
}
\caption{Example Feynman diagrams for the (a) pair, (b) single, (c) Drell--Yan, as well as (d) leading-order and (e) next-to-leading-order resonant LQ production modes.
The symbol $y$ marks interactions mediated by a LQ Yukawa coupling to quarks and leptons.
All shown LQ production modes except pair production are considered for the interpretation of the analysis results.}
\label{fig:intro:LQ_prod_modes}
\end{figure}

The analysis incorporates four mutually exclusive detection channels: \eljet, \mujet, \elbjet, and \mubjet, where light-jet refers to jets that are not tagged as containing a $b$-hadron decay.
To ensure methodological consistency, a similar analysis strategy is implemented across all channels
while preserving channel-specific optimisations. Each channel features dedicated 1-lepton (1L) and 2-lepton (2L) signal regions (SRs), where the 2L topology is specifically designed to account for NLO contributions~\cite{Greljo:2020tgv,Resonant_LQ_NLO_POWHEG} as shown in
Figure~\ref{fig:intro:LQ_prod_modes} (e).
These diagrams are mediated via the photon constituent of the proton which is less suppressed than the lepton content and therefore their contributions are approximately on equal footing with the ones from the $s$-channel LQ production.
Signal extraction employs a shape-based discrimination strategy through a multi-bin template likelihood
analysis of the lepton--jet invariant mass (\mlj) spectrum. This approach combines channel-specific
background modelling, kinematic selection thresholds, and independent treatments of systematic uncertainties,
enabling simultaneous constraints on potential signals across the full \mlj phase space covered in this analysis.
Backgrounds with prompt, genuine leptons are constrained using dedicated control regions (CRs), complemented with data-driven techniques to estimate contributions from misidentified (\textit{\say{fake}}) or non-prompt (FNP) leptons.
An individual analysis selection is developed for each channel and is applied separately on the Run-2 and Run-3 datasets, enabling cross-validation between the two data-taking periods.
For the interpretation of results, the observed \mlj spectra from Run~2 and Run~3 are fitted simultaneously to maximise the sensitivity to the benchmark signal.

The rest of this paper is organised as follows. A brief description of the ATLAS detector is given in Section~\ref{sec:detector}, and the data and simulation samples used are discussed in Section~\ref{sec:datasets}. Overviews of the reconstruction of physics objects and the event selection are presented in Sections~\ref{sec:objects} and ~\ref{sec:selection}, respectively. The background estimation strategy is described in Section~\ref{sec:bkg_estimation}. The systematic uncertainties related to this search are described in Sections~\ref{sec:systematics}. The results of the search are given in Section~\ref{sec:results}. Finally, Section~\ref{sec:conclusion} presents the conclusions.


%

%
\section{ATLAS detector}

\label{sec:detector}

The ATLAS detector~\cite{PERF-2007-01,GENR-2019-02} at the LHC covers nearly the entire solid angle around the collision point.\footnote{\AtlasCoordFootnote}
It consists of an inner tracking detector surrounded by a thin superconducting solenoid, electromagnetic and hadronic calorimeters,
and a muon spectrometer incorporating three large superconducting air-core toroidal magnets.

The inner-detector system (ID) is immersed in a \qty{2}{\tesla} axial magnetic field
and provides charged-particle tracking in the range \(|\eta| < 2.5\).
The high-granularity silicon pixel detector covers the vertex region and typically provides four measurements per track,
the first hit generally being in the insertable B-layer (IBL).
It is followed by the SemiConductor Tracker (SCT), which usually provides eight measurements per track.
These silicon detectors are complemented by the transition radiation tracker (TRT),
which enables radially extended track reconstruction up to \(|\eta| = 2.0\).
The TRT also provides electron identification information
based on the fraction of hits (typically 30 in total) above a higher energy-deposit threshold corresponding to transition radiation.

The calorimeter system covers the pseudorapidity range \(|\eta| < 4.9\).
Within the region \(|\eta|< 3.2\), electromagnetic calorimetry is provided by barrel and
endcap high-granularity lead/liquid-argon (LAr) calorimeters,
with an additional thin LAr presampler covering \(|\eta| < 1.8\)
to correct for energy loss in material upstream of the calorimeters.
Hadronic calorimetry is provided by the steel/scintillator-tile calorimeter,
segmented into three barrel structures within \(|\eta| < 1.7\), and two copper/LAr hadronic endcap calorimeters.
The solid angle coverage is completed with forward copper/LAr and tungsten/LAr calorimeter modules
optimised for electromagnetic and hadronic energy measurements, respectively.

The muon spectrometer (MS) comprises separate trigger and
high-precision tracking chambers measuring the deflection of muons in a magnetic field generated by the superconducting air-core toroidal magnets.
The field integral of the toroids ranges between \num{2.0} and \qty{6.0}{\tesla\metre}
across most of the detector.
Three layers of precision chambers, each consisting of layers of monitored drift tubes, cover the region \(|\eta| < 2.7\).
These were complemented in the innermost layer of the endcap region by cathode-strip chambers in Run~2,
which were replaced in Run~3 by layers of small-strip thin-gap chambers and Micromegas chambers, both providing precision tracking in the region \(1.3 < |\eta| < 2.7\).
The muon trigger system covers the range \(|\eta| < 2.4\) with resistive-plate chambers in the barrel region, thin-gap chambers in the endcap regions,
and, in Run~3, the aforementioned small-strip thin-gap chambers and Micromegas chambers in the innermost layer of the endcap.

The luminosity is measured mainly by the LUCID--2 detector that records Cherenkov light produced in the quartz windows of photomultipliers located close to the beampipe.

Events were selected by the first-level trigger system implemented in custom hardware,
followed by selections made by algorithms implemented in software in the high-level trigger~\cite{TRIG-2016-01,TRIG-2022-01}.
The first-level trigger accepted events from the \qty{40}{\MHz} bunch crossings at a rate close to \qty{100}{\kHz},
which the high-level trigger further reduced in order to record complete events to disk at about \qty{1.25}{\kHz} and \qty{3}{\kHz} in Run~2 and Run~3, respectively.

The Run-3 detector configuration benefits from several upgrades compared with that of Run~2 to maintain high detector performance at the higher pile-up levels of Run~3.
The improvements include a new innermost layer of the muon spectrometer in the endcap region, which provides higher redundancy and a large reduction in fake muon triggers.
The trigger system also benefits from new digital electronics readout of the LAr calorimeters with significantly increased granularity.
Other updates and further details are provided in Ref.~\cite{GENR-2019-02}.

A software suite~\cite{SOFT-2022-02} is used in data simulation, in the reconstruction
and analysis of real and simulated data, in detector operations, and in the trigger and data acquisition
systems of the experiment.


%

%
\section{Data and simulated event samples}

\label{sec:datasets}

The data used in this analysis were collected using a set of single-electron and single-muon triggers~\cite{TRIG-2018-05,TRIG-2018-01}.
The transverse momentum thresholds of the online leptons vary across data-taking periods and depend on whether isolation requirements are applied at the trigger level.
For electrons (muons), the \pt thresholds range from 24--26 (20--24)~\GeV for triggers with isolation, and increase up to 120--140 (50)~\GeV for those without isolation requirements.
Application of data-quality requirements~\cite{DAPR-2018-01} results in data samples corresponding to integrated luminosities of 140~\ifb and 55~\ifb for Run~2 and Run~3, respectively.

While background contributions from misidentified or non-prompt leptons are mainly estimated by using data-driven techniques, Monte Carlo (MC) simulations are used to estimate the event yields and systematic uncertainties for both the signal processes and SM backgrounds featuring prompt-lepton production.
These simulated event samples include the effect of multiple $pp$ interactions in the same or neighbouring bunch crossings (\pileup), as well as the effect on the detector response due to interactions from bunch crossings before or after the one containing the hard interaction.
All MC events were then re-weighted to match the pile-up distribution observed in the data.
To simulate the detector response, background and signal MC samples were processed through the ATLAS simulation framework~\cite{SOFT-2010-01} in \GEANT~\cite{Agostinelli:2002hh}.

Simulated signal samples of resonant LQ production are used to optimise the event selection and interpret the results.
Events are generated with an implementation of this process in \POWHEGBOXRES at NLO~\cite{Resonant_LQ_NLO_POWHEG} with the LUXlep-\NNPDF[3.1nlo] PDF set\footnote{This PDF set includes both leptons and photons in the proton content.}~\cite{Buonocore_2020} and are interfaced with \Herwig[7.2.3] for parton shower, hadronisation, and underlying event using the H7.2-Default set of tuned parameters~\cite{Bellm_2020}.
This implementation models both the LQ production and subsequent decay.
Decays of bottom and charm hadrons are performed by \EVTGEN[2.1.1]~\cite{Lange:2001uf}.

The 2L selections of the analysis are sensitive to the photon-induced diagrams of resonant LQ production but also to other LQ production modes that feature a second lepton in the final state such as DY and single production.
Therefore, dedicated combined DY+SP LQ samples are generated with a \POWHEGBOX implementation at NLO~\cite{LQ_DYSP_Pheno_2022} and used additionally in the interpretation of the results.
These samples use the same PDF set as the resonant LQ samples and are similarly interfaced to \Herwig[7.2.3] for the parton shower.
These samples also take the interference from DY LQ production with SM DY $Z/\gamma^{*}\to\ell\ell$ into account. As the latter is estimated from dedicated samples, the LQ signal samples only include the interference but not the SM contribution.
The DY+SP samples include a $\mll > 100~\GeV$ requirement at matrix-element level to increase the acceptance of the generated events for the signal selections in this analysis.

The \Stilde LQ is used as benchmark and four different types of minimal LQ scenarios are considered. Only one LQ coupling is considered at a time while all other couplings are set to zero. The couplings considered are $y_{de}$, $y_{s\mu}$, $y_{be}$ and $y_{b\mu}$ that result in an \eljet, \mujet, \elbjet and \mubjet signature, respectively.
For the scenarios with $y_{de}$ and $y_{s\mu}$ couplings, samples with LQ masses between 1 and 5~\TeV are generated.
Models with couplings to $b$-quarks have lower production cross-sections due to the smaller $b$-quark PDF, hence the generated samples cover LQ masses between 1 and 3.5~\TeV.
The signal samples cover the LQ coupling range from 0.1--1.0 for $y_{de}$ and from 0.5--3.5 for the other scenarios.
The ranges of the signal parameters are motivated by the expected sensitivity of the search, taking into account existing constraints from collider and low-energy (for $y_{de}$) experiments, and restricting to the perturbative regime of the theory~\cite{Dorsner:2016wpm}. %
The signal production cross-sections and uncertainties are taken from the \POWHEGBOXRES implementations and evaluated according to prescriptions from Ref.~\cite{Ball:2022oua}. %
As an example, the production cross-section for a 2~\TeV LQ at $\sqrt{s} = 13~\TeV$ for $y_{de}$ couplings between 0.1 and 1.0 range from $0.024~\fb$ to $2.5~\fb$. At $\sqrt{s} = 13.6~\TeV$ these cross-sections increase by approximately 16\%.

A variety of MC generators is utilised to model the SM backgrounds involving the production of prompt leptons.
The event generator configurations for SM processes are mostly identical between the Run~2 and Run~3 MC samples, except that more recent generator versions are used for some of the latter.
Table~\ref{tab:mc_backgr} provides a comprehensive overview of the SM background samples utilised in the analysis.
Additional information regarding ATLAS simulations of \Wjets, \Zjets, \ttbar, single-top ($Wt$, $t$-channel, $s$-channel), and diboson processes is available in Refs.~\cite{ATL-PHYS-PUB-2017-006,ATL-PHYS-PUB-2018-009,ATL-PHYS-PUB-2016-004,ATL-PHYS-PUB-2017-005,}.
The decays of bottom and charm hadrons are performed by \EVTGEN versions 1.2.0, 1.6.0 and 2.1.1 ~\cite{Lange:2001uf}, except for the
backgrounds modelled using \SHERPA, for which the decays are performed internally.

In addition to the hypothesized resonant LQ production, taking into account the lepton PDF in the proton also predicts a SM \ljSM process at the LHC\@.
Such a process has not been observed yet but features the same lepton--jet signature as the targeted LQ model.
To take this contribution into account, an \ljSM sample is generated at leading-order using \MADGRAPH[3.3.4]~\cite{Alwall:2014hca} and interfaced with \Herwig[7.2.3] and the LUXlep-\NNPDF[3.1nlo] PDF set. To enrich the sample in events with large \mlj, the transverse momentum (\pt) of the lepton is required to be greater than 50~\GeV and the generation is performed in bins of lepton \pt.

\begin{table}[tb]
\scriptsize
\centering
\caption{Simulated signal and background event samples with the corresponding matrix element and parton shower (PS) generators, cross-section order in $\alphas$ used to normalise the event yield, set of tuned parameters (tune) for the underlying-event and the generator PDF sets used. For diboson samples,  $V\in\{W,Z\}$. \protect\say{Default} refers to the default tune of the \SHERPA generator. When different, the settings used for the simulation of samples compared with Run-3 data are mentioned in parentheses.
Abbreviations used are defined as: leading-order (LO), next-to-leading-order (NLO), next-to-next-to-leading-order (NNLO), next-to-leading-logarithmic (NLL), next-to-next-to-leading-logarithmic (NNLL). The PDF set employed in the PS is generally the same as in the generator except for the \ttbar, single-top and the \AtlasMC{MG5\_aMC@NLO} $t\bar{t}{+}Z$ and $t\bar{t}{+}W$ samples where the \NNPDF[2.3lo]~\cite{Ball:2012cx} set is used. }
\label{tab:mc_backgr}
\adjustbox{max width=\textwidth}{
\renewcommand{\arraystretch}{1.5}
\begin{tabular}{llllll}
\toprule
Physics process          			& Generator 	& Parton shower 	&  Normalisation 		& Tune 			& PDF (generator) \\
\midrule
Resonant LQ signal & \POWHEGBOXRES~\cite{Resonant_LQ_NLO_POWHEG} &  \Herwig[7.2.3]~\cite{Bellm_2020} & NLO & H7.2-Default~\cite{Bellm_2020} & LUXlep-\NNPDF[3.1nlo]~\cite{Buonocore_2020} \\
DY+SP LQ signal & \POWHEGBOX[v2]~\cite{LQ_DYSP_Pheno_2022} &  \Herwig[7.2.3] & NLO & H7.2-Default & LUXlep-\NNPDF[3.1nlo] \\
\midrule
\multirow{2}{*}{$Z/\gamma^*(\rightarrow \ell\ell)$+jets} 	& \SHERPA[2.2.11]~\cite{Bothmann:2019yzt,Gleisberg:2008fv} & \SHERPA[2.2.11]~\cite{Schumann:2007mg}  & \multirow{2}{*}{NNLO~\cite{ATL-PHYS-PUB-2023-026}} & \multirow{2}{*}{Default~\cite{ATL-PHYS-PUB-2017-005}}	& \multirow{2}{*}{\NNPDF[3.0nnlo]~\cite{Ball:2014uwa}} \\
& (\SHERPA[2.2.14])                       &  (\SHERPA[2.2.14])		                & 			                                         &                                                         	& \\
$Z/\gamma^*(\rightarrow \tau\tau)$+jets 	& \SHERPA[2.2.14]			& \SHERPA[2.2.14]		& NNLO~\cite{ATL-PHYS-PUB-2023-026}			& Default	& \NNPDF[3.0nnlo] \\
\multirow{2}{*}{$W(\rightarrow \ell\nu,\tau\nu)$+jets} 	    & \SHERPA[2.2.11]~\cite{Bothmann:2019yzt} & \SHERPA[2.2.11]    & \multirow{2}{*}{NNLO~\cite{ATL-PHYS-PUB-2023-026}} & \multirow{2}{*}{Default}	& \multirow{2}{*}{\NNPDF[3.0nnlo]} \\
& (\SHERPA[2.2.14])                       &  (\SHERPA[2.2.14]) &                                                    &                         	& \\
\ttbar 				& \POWHEGBOX[v2]~\cite{Frixione:2007nw,Nason:2004rx,Frixione:2007vw,Alioli:2010xd}       		& \PYTHIA[8.230]~\cite{Sjostrand:2014zea}  		& NNLO+NNLL~\cite{Beneke:2011mq,Cacciari:2011hy,Baernreuther:2012ws,Czakon:2012zr,Czakon:2012pz,Czakon:2013goa,Czakon:2011xx} 		& A14~\cite{ATL-PHYS-PUB-2014-021} & \NNPDF[3.0nlo] \\
Single-top 				& \POWHEGBOX[v2]~\cite{Re:2010bp,Nason:2004rx,Frixione:2007vw,Alioli:2010xd} 	& \PYTHIA[8.230]	&NLO+NNLL~\cite{Kidonakis:2010ux}	& A14 			& \NNPDF[3.0nlo] \\
\multirow{2}{*}{Diboson $VV$} &  \SHERPA[2.2.11],\,2.2.12   & \SHERPA[2.2.11],\,2.2.12   & \multirow{2}{*}{LO--NLO~\cite{Hoeche:2011fd,Hoeche:2012yf,Catani:2001cc,Hoeche:2009rj}}  &  \multirow{2}{*}{Default} & \multirow{2}{*}{\NNPDF[3.0nnlo]} \\
&  (\SHERPA[2.2.14],\,2.2.16) & (\SHERPA[2.2.14],\,2.2.16) &                    &  &  \\
$t\bar{t}{+}Z$ 			& \AtlasMC[2.3.3]{MG5\_aMC@NLO}~\cite{Alwall:2014hca,Frederix:2018nkq}			& \PYTHIA[8.210]~\cite{Sjostrand:2014zea}						& NLO~\cite{Alwall:2014hca} 			& A14 & \NNPDF[3.0nlo] \\
\multirow{2}{*}{$t\bar{t}{+}W$} & \AtlasMC[2.3.3]{MG5\_aMC@NLO} & \PYTHIA[8.210]   & \multirow{2}{*}{NLO} & A14       & \multirow{2}{*}{\NNPDF[3.0nlo]} \\
& (\SHERPA[2.2.14])		        & (\SHERPA[2.2.14]) &                      & (Default) &  \\
\multirow{2}{*}{Multijet} & \PYTHIA[8.230] &  \PYTHIA[8.230] & \multirow{2}{*}{LO~\cite{ATL-PHYS-PUB-2019-017}} & \multirow{2}{*}{A14} & \multirow{2}{*}{\NNPDF[2.3lo]} \\
& (\PYTHIA[8.308]) &  (\PYTHIA[8.308]) &                                                  &                      & \\
\ljSM                & \MADGRAPH[3.3.4]~\cite{Alwall:2014hca} &  \Herwig[7.2.3] & LO & H7.2-Default & LUXlep-\NNPDF[3.1nlo] \\
\bottomrule
\end{tabular}
}
\end{table}


%

%
\section{Object reconstruction}

\label{sec:objects}

Events are required to contain a primary vertex built from at least two associated tracks with $\pT > 0.5~\GeV$.
The primary vertex with the highest sum of squared transverse momenta $\sum \pt^2$ of its associated tracks~\cite{ATL-PHYS-PUB-2015-026} is identified as the hard-scatter vertex of interest in each event.
A set of basic data-quality requirements is applied to ensure a fully operating detector and to suppress contributions from detector noise or non-collision backgrounds~\cite{ATLAS-CONF-2015-029}.

Two categories of analysis objects are defined and utilised to define the search regions. Leptons and jets are first “preselected” using loose selection criteria; those that satisfy additional, tighter requirements are designated as “signal” objects.
For electrons, also an intermediate category—falling between the preselected and signal definitions—is introduced which is employed in the data-driven estimate of the FNP electron background, as detailed in Section~\ref{sec:bkg_estimation}.

Preselected electrons are reconstructed using ID tracks matched to energy clusters in the electromagnetic calorimeter. These satisfy $\pt > 10~\GeV${} and $|\eta| < 2.47$ with a \wp{Loose} operation point of likelihood-based identification criteria and the requirement of a hit in the innermost pixel layer~\cite{PERF-2017-01,EGAM-2021-01}.
Electrons reconstructed in the calorimeter transition region, $1.37 < |\eta| < 1.52$, are not considered.
The longitudinal impact parameter $z_0$ of preselected electron tracks is required to satisfy $|z_0\sin\theta| < 0.5$~mm.
Signal electrons must also satisfy $\pt > 25~\GeV$, the \wp{Tight} likelihood-based identification criteria and have a transverse impact parameter $d_0$ with uncertainty $\sigma(d_0)$ satisfying $\left|d_0/\sigma(d_0)\right| < 5$.
To further reject FNP electrons, the \wp{HighPtCaloOnly} isolation discriminant~\cite{EGAM-2021-01} is employed that is calculated from energy deposits in the calorimeter cells in a cone around the electron candidate.

Preselected muons are reconstructed by combining tracks from the ID and the muon spectrometer subsystems.
These are required to have $\pt > 10~\GeV$, $|\eta| < 2.5$, satisfy the \wp{High-\pt} identification criteria~\cite{MUON-2018-03} and $|z_0\sin\theta| < 0.5$~mm.
Signal muons must have $\pt > 25~\GeV$, impact parameter significance $\left|d_0/\sigma(d_0)\right| < 3$ and must satisfy an isolation requirement with a similar performance to the \wp{PflowTight} criterion described in Ref.~\cite{MUON-2018-03}.

Hadronic jets are reconstructed using the anti-$k_t$ algorithm~\cite{Cacciari:2008gp} as implemented in \texttt{FastJet}~\cite{Fastjet} with a jet radius parameter of $R=0.4$.
The inputs to this algorithm are particle-flow objects~\cite{PERF-2015-09} that combine measurements from the ATLAS inner detector and calorimeters~\cite{PERF-2014-07}.
The jet energy scale and resolution are calibrated using simulations, with in situ corrections obtained from data~\cite{JETM-2018-05}.
Preselected jets are required to satisfy $\pt > 20~\GeV$ and $|\eta|<4.5$.
Signal jets are required to have $|\eta|<2.5$ and must additionally satisfy a pile-up jet rejection criterion based on a neural-network variant of the jet vertex tagger~\cite{PERF-2014-03} if they have $\pt<60~\GeV$.
Signal jets that satisfy the 85\% efficiency working point of the \wp{GN2} algorithm~\cite{FTAG-2023-05} are considered to likely contain $b$-hadrons and are referred to as $b$-tagged jets.

The missing transverse momentum \met is calculated as the magnitude of the negative vector sum of the transverse momenta of all identified hard physics objects (preselected leptons and jets) calibrated to their respective energy scales, with a contribution from an additional soft term~\cite{JETM-2020-03}.
This soft term is constructed from ID tracks matched to the hard-scatter vertex but not associated with any of the hard reconstructed objects.

Since the object reconstruction algorithms are applied independently, lepton and jet candidates may share contributions from the same detector signals.
To resolve such ambiguities, an overlap removal procedure is applied to the preselected leptons and signal jets in the following order.
First, any electron sharing an ID track with a muon is removed. Next, jets are removed if they are within $\Delta R < 0.2$ from a remaining electron.
After this, electrons are in turn rejected if they are within $\Delta R < 0.4$ of any remaining jet.
Subsequently, jets with any ghost-associated~\cite{Cacciari_2008} muon or within $\Delta R < 0.2$ are removed if the jet has fewer than three associated tracks with $\pT > 500 ~\MeV$.
Finally, any muon within $\Delta R < 0.4$ of a jet is removed.
Only objects that satisfy this overlap removal procedure are retained for the event selection.


%

%
\section{Event selection}

\label{sec:selection}

A set of analysis variables is derived from the physics objects that satisfy the identification criteria and are used in the event selections.
These are briefly summarised below, where the leading lepton (jet) refers to the lepton (jet) with the largest \pt in the event:

\begin{description}[style=multiline,leftmargin=2.5cm]
\item[\mlj:] Invariant mass of the system built from the leading lepton and leading jet in the event to reconstruct the LQ mass.
\item[\LepOverMlj:] Ratio of the $\pt$ of the leading lepton and the invariant mass of the lepton--jet system $\mlj$.
\item[\dRlj:] Angular separation $\dRlj = \sqrt{\Delta \phi(\ell,j)^2+\Delta\eta(\ell,j)^2}$ between the leading lepton and leading jet .
\item[\dPhilv:] Azimuthal angle between the leading lepton and $\met$.
\item[\mll:] Invariant mass of the lepton pair in the 2L selections.
\item[\metsig:] Object-based \met significance~\cite{JETM-2020-03} that provides a measure for the likelihood of the reconstructed \ptmiss to originate from real invisible particles instead of from detector effects.
\end{description}

All four channels share a common preselection designed to enrich the selected events in the kinematic phase space of interest and to serve as basis to define signal, control, and validation regions.
Each event must contain at least one signal lepton (electron or muon) and at least one jet, both with $\pt > 130~\GeV$. The invariant mass of the lepton--jet system, \mlj, is required to exceed 700~\GeV.
Although the leptons and jets from the LQ decays of interest typically have transverse momenta of several hundreds of \GeV, the \pt requirements at the preselection level are relaxed to retain a sufficiently large sample of events  for reliable background estimation.

The 1L selection requires exactly one signal lepton, while the 2L selection requires exactly two same-flavour signal leptons ($ee$ or $\mu\mu$).
Events with additional preselected leptons beyond those satisfying the signal lepton criteria are vetoed.
For the 2L selections, an additional requirement of $\mll > 70$~\GeV is imposed, as the signal of interest with a resonantly produced LQ does not contribute much at low $\mll$.

To ensure that events are selected in the plateau region of the trigger efficiency, a reconstructed lepton is required to be matched to the trigger-level lepton and to have a sufficiently high \pt above the corresponding online threshold, as described in Section~\ref{sec:datasets}.
For muons, the \pt requirement of the event preselection alone guarantees that the single-muon triggers operate within their efficiency plateau.
For electrons, the required \pt reaches up to 141~\GeV in events selected by single-electron triggers that employ only a loose identification and no isolation criteria at the trigger level.

Events satisfying this preselection are then assigned to one of four analysis channels based on the lepton flavour and the number of $b$-tagged jets \nb.
The electron-based (muon-based) channels are restricted to events with either one or two electrons (muons).
The \eljet and \mujet channels require the absence of any $b$-tagged jet ($b$-veto) while the \elbjet and \mubjet channels require the leading jet to be $b$-tagged.

Building on the preselection, additional requirements are applied—individually optimised for each channel and for the 1L and 2L selections—to enhance the separation between signal and background events.
The resulting signal regions are referred to as \SROneL and \SRTwoL, respectively.

\begin{table}[tb]
\centering
\caption{Definitions of the preselection together with the 1L and 2L SRs for the \eljet, \mujet, \elbjet and \mubjet channels. The lepton requirements indicate an additional veto of preselected leptons, e.g. $\nel = 1$ requires the presence of exactly one signal electron but no other preselected lepton in the event.
}
\label{tab:selection:SRs}
\begingroup
\renewcommand{\arraystretch}{1.25}
\adjustbox{max width=\textwidth}{
\begin{tabular}{c cc cc cc cc}
\toprule
\multicolumn{9}{c}{Preselection} \\
\cmidrule(lr){1-9}
\multicolumn{9}{c}{$\mlj > 700~\GeV \quad\quad \pTlepone > 130~\GeV \quad\quad \pt^{\jetone} > 130~\GeV \quad\quad \mll > 70~\GeV~(\text{if } N_{\ell}=2)$ } \\
\midrule
\midrule
\multirow{2}{*}{Region}  & \multicolumn{2}{c}{\eljet} & \multicolumn{2}{c}{\mujet} & \multicolumn{2}{c}{\elbjet} & \multicolumn{2}{c}{\mubjet}\\
\cmidrule(lr){2-3} \cmidrule(lr){4-5} \cmidrule(lr){6-7}  \cmidrule(lr){8-9}
& \SROneLej & \SRTwoLej & \SROneLmj & \SRTwoLmj & \SROneLeb & \SRTwoLeb & \SROneLmb & \SRTwoLmb \\
\midrule
\mlj[\GeV] & \multicolumn{2}{c}{$\geq 950$} & \multicolumn{2}{c}{$\geq 900$} & \multicolumn{2}{c}{$\geq 900$} & \multicolumn{2}{c}{$\geq 900$} \\
\nel  & 1 & 2 & \multicolumn{2}{c}{0} & 1 & 2 & \multicolumn{2}{c}{0} \\
\nmu  & \multicolumn{2}{c}{0} & 1 & 2 & \multicolumn{2}{c}{0} & 1 & 2 \\
\nb   & \multicolumn{2}{c}{0} & \multicolumn{2}{c}{0} & \multicolumn{2}{c}{1} & \multicolumn{2}{c}{1} \\
\metsig & \multicolumn{2}{c}{$<3.5$} & \multicolumn{2}{c}{$<3.5$} & $<3.0$ & $<5.0$ & $<3.0$ & $<5.0$ \\
\dRlj & $<3.7$ & -- & [2.9, 3.6] & -- & \multicolumn{2}{c}{--} & \multicolumn{2}{c}{[2.4, 4.2]} \\
\LepOverMlj & $>0.4$ & $>0.3$ & $>0.4$ & $>0.3$ & $>0.3$ & $>0.25$ & $>0.3$ & $>0.2$ \\
\mll[\GeV] & -- & $>160$  & -- & $>160$ & -- & $>150$ & -- & $>120$ \\
\bottomrule
\end{tabular}
}
\endgroup
\end{table}

A summary of the SR definitions for each channel is provided in Table~\ref{tab:selection:SRs}.
Since the LQ signal does not produce genuine missing transverse momentum, an upper requirement on \metsig is applied to efficiently suppress \Wjets and top-related backgrounds while retaining most of the signal events.
The lepton and jet originating from a LQ decay are typically produced back-to-back in the detector, motivating a selection on their angular separation \dRlj.
In addition, the LQ decay products are expected to share the parent particle's energy approximately equally, resulting in similar lepton and jet momenta.
Therefore, each channel imposes a lower bound on \LepOverMlj to reject SM events with a large \pt imbalance between the leading lepton and jet.
Finally, in the \SRTwoL selections, a lower requirement on \mll is applied to suppress \Zjets events.

To maximise sensitivity across a broad range of potential LQ masses, the SRs are further binned in \mlj, targeting LQ masses of approximately 1~\TeV and above.
The binning choice is derived by the experimental resolution in \mlj, defined by the relative difference between the invariant mass of the lepton--jet system at reconstruction- and at particle-level.
This resolution is evaluated using LQ signal samples and found to be largely independent of the LQ coupling, with only a mild dependence on the LQ mass.
Figure~\ref{fig:selection:mlj_resolution} shows the relative difference between the reconstructed and the particle-level \mlj for example LQ signals after application of the preselection requirements.
Since the absolute \mlj resolution worsens with increasing LQ mass, the \mlj bin widths are gradually broadened across the spectrum.
Binning continues until the SM background expectation falls below approximately one event, with the final bin in both \SROneL and \SRTwoL being inclusive and capturing all overflow events.
No substantial difference between the \mlj resolution between Run~2 and Run~3 is observed; therefore, the same binning strategy is applied to both data-taking periods.

\begin{figure}[tb]
\centering
\subfloat[]{
\includegraphics[width=0.45\textwidth]{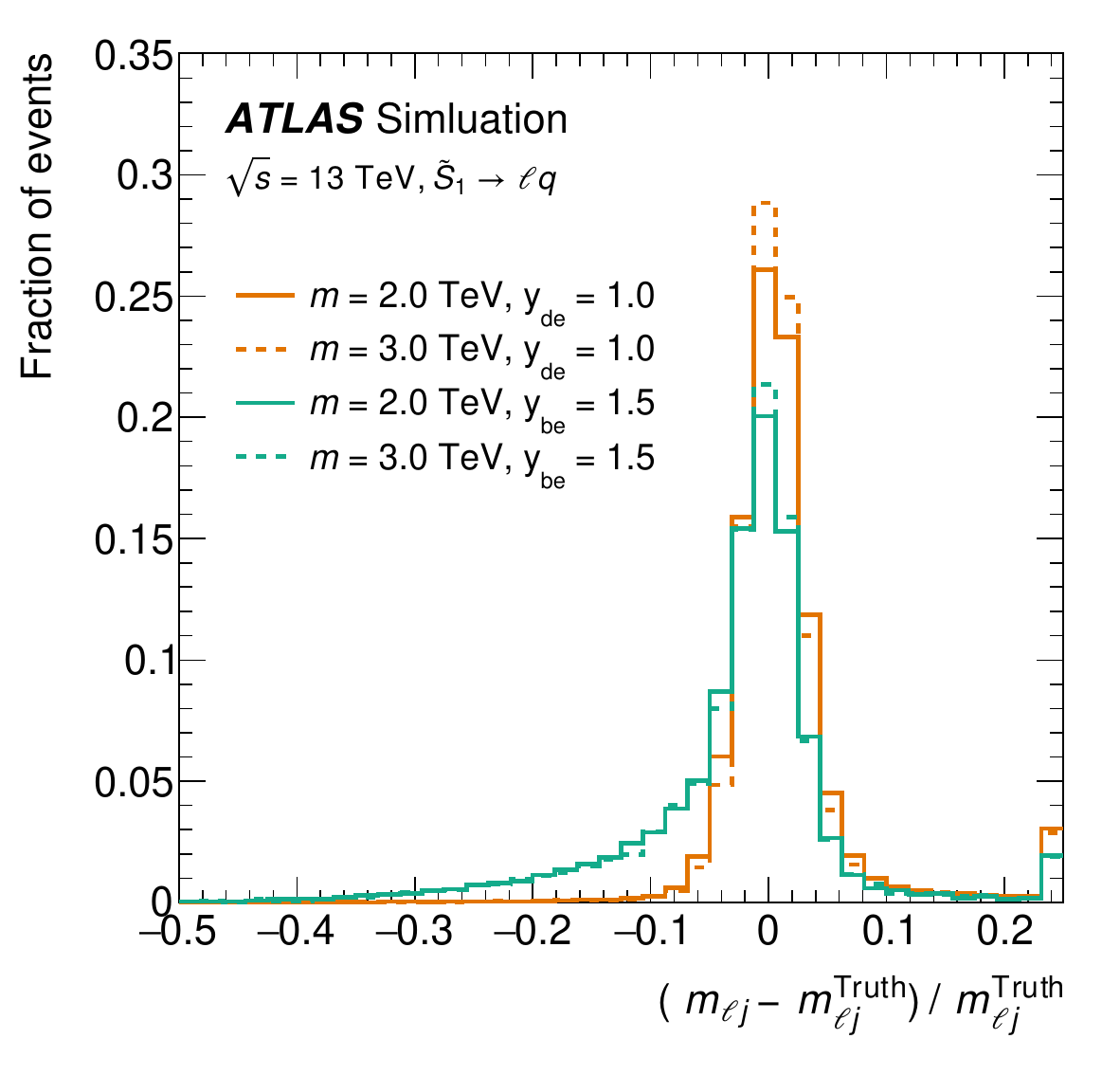}
}
\subfloat[]{
\includegraphics[width=0.45\textwidth]{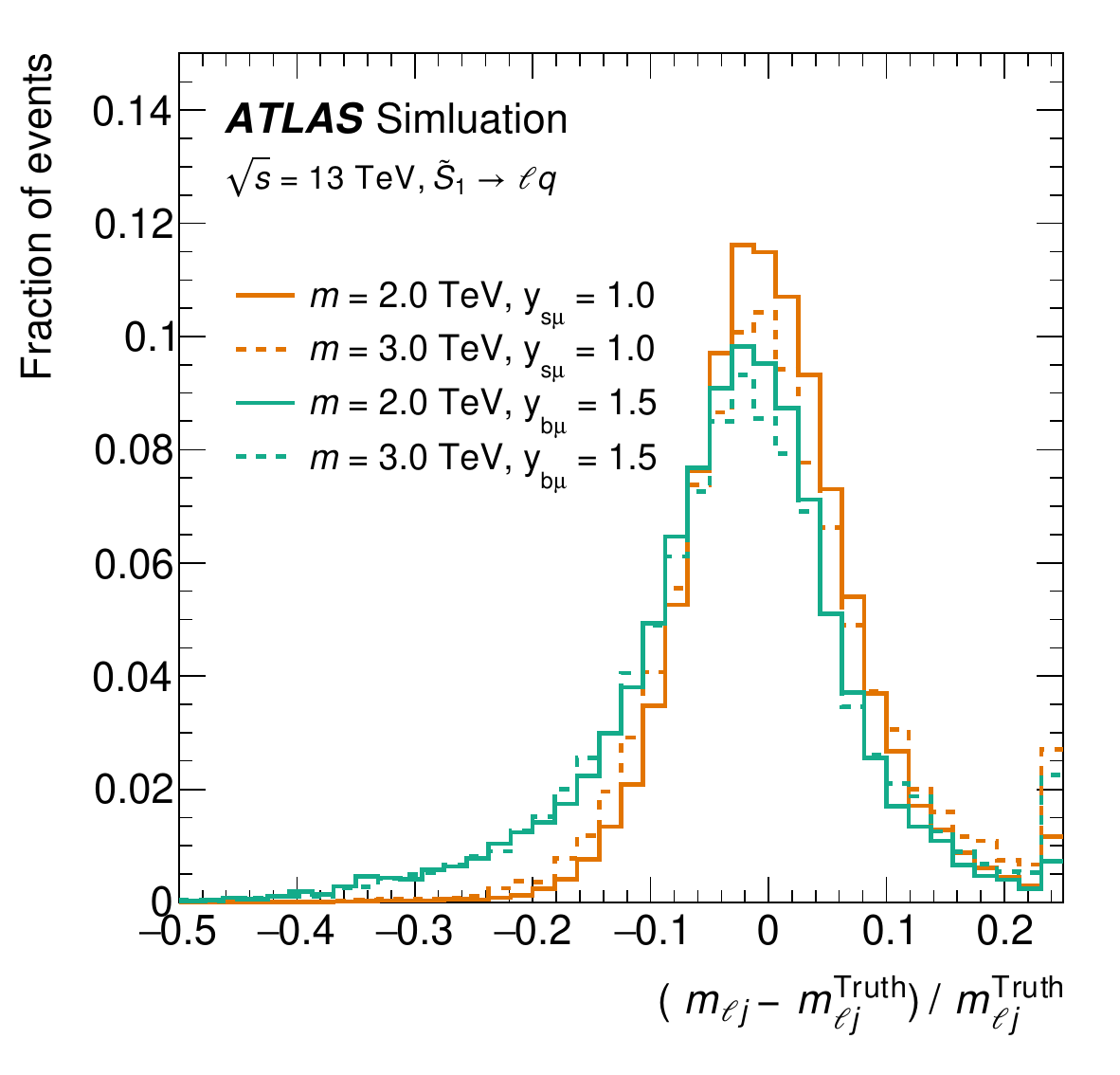}
}
\caption{Relative difference between the reconstructed \mlj and particle-level $\mlj^{\text{Truth}}$ invariant mass of the lepton--jet system for selected example LQ scenarios in the (a) \eljet and \elbjet, and (b) \mujet and \mubjet channels using Run~2 MC simulation.
The dashed lines correspond to an LQ scenario with a larger mass but the same coupling as the one shown by the solid lines.
The particle-level invariant mass is calculated using the particle-level four-momenta of the reconstructed lepton and jet.
The last bin contains the overflow.}
\label{fig:selection:mlj_resolution}
\end{figure}

Table~\ref{tab:selection:binning} provides an overview of how the individual \mlj SR bins are defined for each channel.
In the \eljet channel, the \mlj resolution is found to be approximately 5\%, motivating bin widths of 100~\GeV for LQ masses around 1~\TeV and 200~\GeV for masses near 2~\TeV.
The first bin in the \eljet SRs begins at \mlj = 950~\GeV, while the final bin in \SROneLej (\SRTwoLej) includes events with $\mlj \geq 3100$ (2300)~\GeV.

In the \mujet channel, the \mlj resolution is approximately 10\%, due to the worsened momentum resolution for high-\pt muons compared with electrons.
This suggests a bin width of 200~\GeV for a LQ mass of 1~\TeV and therefore the SRs in this channel begin at $\mlj = 900~\GeV$ with the final bin in both \SROneLmj and \SRTwoLmj includes events with $\mlj \geq 2300~\GeV$.

In the $b$-tagged channels, the \mlj resolution is slightly worse than in the corresponding light-jet channels due to the possible presence of neutrinos in $b$-hadron decays, and is determined to be approximately 8\% in the \elbjet channel and 13\% in the \mubjet channel.
Both \SROneLeb and \SROneLmb begin at \mlj = 900~\GeV, while the final bin in \SROneLeb (\SROneLmb) and \SRTwoLeb (\SRTwoLmb)  includes events with $\mlj \geq 2200$ (2025)~\GeV and $\mlj \geq 1550$ (1550)~\GeV, respectively.

Therefore, the LQ decay width\footnote{The partial decay width $\Gamma_{\text{LQ}}$ of a scalar LQ with mass $m_{\text{LQ}}$ corresponding to coupling $y_{q\ell}$ is given at LO by $\Gamma_{\text{LQ}} = \frac{|y_{q\ell}|^{2}}{16\pi}m_{\text{LQ}}$ in the limit of large LQ masses~\cite{LQ_Pheno_2020_scalar}.} remains well below the experimental resolution for couplings up to about 1.5, corresponding to a width of approximately 80~GeV for a 2~TeV LQ.
For the largest couplings considered, the effects of the intrinsic width and the detector resolution on the reconstructed LQ mass peak become comparable.

\begin{table}[tb]
\centering
\caption{Binning of the \mlj distribution used for \SROneL and \SRTwoL of the \eljet, \mujet, \elbjet and \mubjet channels, respectively.}
\label{tab:selection:binning}
\begingroup
\renewcommand{\arraystretch}{1.25}
\adjustbox{max width=\textwidth}{
\begin{tabular}{cc l}
\toprule
\multicolumn{2}{c}{SR}  & Binning in \mlj[\GeV]  \\
\midrule
\multirow{2}{*}{\eljet} & \SROneLej & [950, 1050, 1150, 1250, 1350, 1450, 1600, 1750, 1900, 2100, 2300, 2550, 2800, 3100, $\infty$) \\
& \SRTwoLej & [950, 1050, 1150, 1250, 1350, 1450, 1600, 1750, 1900, 2100, 2300, $\infty$) \\
\midrule
\multirow{2}{*}{\mujet} & \SROneLmj & [900, 1100, 1300, 1600, 1900, 2300, $\infty$) \\
& \SRTwoLmj & [900, 1100, 1300, 1600, 1900, 2300, $\infty$) \\
\midrule
\multirow{2}{*}{\elbjet} & \SROneLeb & [900, 1100, 1300, 1550, 1850, 2200, $\infty$) \\
& \SRTwoLeb & [900, 1100, 1300, 1550, $\infty$) \\
\midrule
\multirow{2}{*}{\mubjet} & \SROneLmb & [900, 1175, 1550, 2025, $\infty$) \\
& \SRTwoLmb & [900, 1175, 1550, $\infty$) \\
\bottomrule
\end{tabular}
}

\endgroup
\end{table}

The signal composition in the \SRTwoL regions varies across the parameter space.
For coupling values below 0.5, resonant LQ production accounts for approximately 70--80\% of the total signal in these regions.
As the coupling strength increases, the DY+SP production mode becomes increasingly relevant, reaching comparable levels to resonant production at couplings around 3.5.
Within the DY+SP sample, the SP component, which also produces a peak in the \mlj spectrum at the LQ mass, dominates for low couplings and LQ masses up to roughly 2~TeV.
In contrast, the non-resonant DY contribution, which decreases steeply with \mlj and thus has a comparable low selection efficiency in the SRs, gains in relative importance at larger couplings and LQ masses.

The overall signal efficiencies for resonant LQ production, including the detector acceptance, for the \SROneL and \SRTwoL selections in the \eljet channel assuming a LQ with mass of 2~\TeV and coupling of 1.0  are approximately 30\% and 15\%, respectively.
In the \mujet channel these selection efficiencies for a LQ of the same mass and a coupling of 1.5 are approximately 17\% and 10\%, respectively.
The selection efficiencies in the $b$-tagged channels are smaller than in the light-jet channels and are 17\% (12\%) and 7\% (7\%) in the \elbjet (\mubjet) channel, for a LQ signal with mass of 2~\TeV and a coupling of 1.5.


%

%
\section{Background estimation}

\label{sec:bkg_estimation}

The SM processes contributing to the phase space targeted by this analysis can be broadly classified into reducible and irreducible backgrounds.
Reducible backgrounds include events containing at least one FNP lepton, originating from misidentified detector signatures such as jets, or from non-prompt leptons produced in hadron decays involving heavy-flavour quarks.
Irreducible backgrounds arise from processes with prompt, genuine leptons that produce final states resembling the target signal.
The dominant irreducible contributions come from \Wjets and \Zjets production, with top-quark processes providing an additional source in the $b$-tagged channels.

A combination of data-driven and MC-based techniques is used to estimate these SM backgrounds, as detailed in the following sections.
The resulting predictions are validated in dedicated validation regions (VRs), which are typically enriched in specific background processes.

\subsection{Reducible backgrounds}


The particle-level information available in simulation samples enables a study of the origins of reconstructed leptons, allowing a distinction between genuine and misidentified objects.
In the 1L selections of the \eljet and \elbjet channels, a non-negligible fraction of the SM background is predicted to arise from FNP electrons.
The most common origins of these misidentified electrons are prompt photons, arising from e.g. initial- and final-state radiation, and light-hadron decays.
To estimate this contribution, a data-driven technique known as the fake-factor method~\cite{EGAM-2019-01} is employed in the 1L selections of the electron channels.
In contrast, the selections requiring two electrons and all selections in the \mujet and \mubjet channels receive only a small contribution from FNP leptons.
In these cases, the FNP background is instead estimated directly from MC simulation.

In the fake-factor method, the number of FNP leptons entering an analysis region---i.e., those satisfying the \say{tight}, signal selection criteria for leptons, $N_{\text{signal}}^{\text{FNP}}$---is estimated using a separate, independent data sample of \say{loose} leptons in the same region that meet a relaxed set of selection requirements.
The ratio to extrapolate from the loose to the signal lepton sample is referred to as the \say{fake factor} (FF) and is derived from data, as described below.
To isolate the FNP contribution, any contamination from real, genuine leptons in the loose lepton sample, originating mainly from \Wjets events, is estimated by using MC simulation and subtracted from the data.
The resulting estimate of the number of signal FNP leptons $N_{\text{signal}}^{\text{FNP}}$ in a given region is computed as

\begin{equation*}
N_{\text{signal}}^{\text{FNP}} = \text{FF}\cdot (N_{\text{loose}}^{\text{data}} - N_{\text{loose}}^{\text{MC, real}})\,.
\end{equation*}

Here, $N_{\text{loose}}^{\text{data}}$ and $N_{\text{loose}}^{\text{MC, real}}$ are the numbers of loose leptons observed in data and the estimated number of genuine loose leptons from simulation, respectively.
The fake factor is typically parameterised in bins of relevant kinematic variables to account for dependencies in the extrapolation.
These are derived as the ratio of signal to loose electrons in data, using a region enriched in FNP leptons and orthogonal to all other selections used in the analysis.
The contamination from genuine leptons in both the loose and signal samples within the fake factor measurement region is estimated from MC simulation and again subtracted accordingly:

\begin{equation*}
\text{FF} = \frac{N_{\text{signal}}^{\text{data}} - N_{\text{signal}}^{\text{MC, real}}}{N_{\text{loose}}^{\text{data}} - N_{\text{loose}}^{\text{MC, real}}} \,.
\end{equation*}

Loose electrons are defined as preselected electrons that fail to satisfy any of the signal electron selection criteria.
They are further required to satisfy the \wp{Medium} identification and \wp{IsoLoose\_VarRad} isolation working points~\cite{EGAM-2021-01} to match the associated online criteria and avoid any bias originating from the trigger requirements in the measured fake factors.
Moreover, applying these requirements brings the loose electron definition closer to the signal one, reducing the size of the extrapolation between them while maintaining a sufficiently large number of events in the loose electron sample.

Fake factors are measured in events that satisfy the preselection described in Section~\ref{sec:selection}. Such events must contain exactly either one loose or signal electron, no additional preselected leptons, and no $b$-tagged jets.
To ensure orthogonality with the selections in the \eljet channel, a requirement of $\dRlj > 3.7$ is imposed.
An additional $\metsig < 3$ requirement suppresses contributions from \Wjets events.
This selection is referred to as fake factor measurement region, denoted by \MRfake.
Events featuring $3 < \metsig < 5$ instead define the validation region \VRfake, which is used to assess the performance of the measured fake factors.
While the loose electron samples in \MRfake and \VRfake are very pure in FNP leptons, the purity decreases to approximately 40\% and 20\% for signal electrons due to significant contamination from the \Wjets process.
To constrain the \Wjets normalisation when validating the FNP estimate in \VRfake, events with $\metsig > 5$ are used to define a control region, \CRfake.
A summary of the region definitions is provided in Table~\ref{tab:backgrounds:fake_regions_defs}.

\begin{table}[tb]
\centering
\caption{Overview of the definitions for the regions employed to derive and validate the FNP electron estimate.}
\label{tab:backgrounds:fake_regions_defs}
\begingroup
\renewcommand{\arraystretch}{1.25}
\adjustbox{max width=\textwidth}{
\begin{tabular}{c ccc}
\toprule
\multirow{2}{*}{Region}  & \multicolumn{3}{c}{FNP estimation} \\
\cmidrule(lr){2-4}
& \MRfake & \VRfake  & \CRfake  \\
\midrule
\mlj[\GeV] & \multicolumn{3}{c}{$> 700$}  \\
\nel  & \multicolumn{3}{c}{1} \\
\nmu  & \multicolumn{3}{c}{0} \\
\nb   & \multicolumn{3}{c}{0} \\
\dRlj & \multicolumn{3}{c}{$>3.7$} \\
\metsig & $<3$ & 3--5 & $>5.0$\\
\bottomrule
\end{tabular}
}

\endgroup
\end{table}

The fake factors are parameterised as a function of the electron \pt in three bins of $|\eta|$, and are shown separately for Run~2 and Run~3 in Figure~\ref{fig:backgrounds:fake_factors}.
In Run~2, the fake factors range from 0.2 to 0.35 at $\pt \sim 130~\GeV$, depending on $|\eta|$, and typically decrease to values between 0.05 and 0.1 for $\pt > 800~\GeV$.
In Run~3, higher values are observed, with fake factors ranging from approximately 0.35 to 0.5 at low \pt and from 0.05 to 0.25 at high \pt.
This difference between Run~2 and Run~3 is consistently observed in both data and simulation.
The $\dRlj > 3.7$ requirement in the \MRfake region biases the selected electrons towards the more forward direction, limiting the available number of events in the central region and allowing only a coarser binning at high electron \pt.
Although the fake factors are measured in events without $b$-tagged jets, they are applied in both the \eljet and \elbjet channels.
A $b$-tagged counterpart to \MRfake cannot be defined due to low FNP purity and limited number of events.
Since MC simulation shows no strong dependence of the fake factors on the \bjet multiplicity, an additional uncertainty is assigned when applying them to events with $b$-tagged jets instead to cover this dependence, see Section~\ref{sec:systematics}.

\begin{figure}[tb]
\centering
\subfloat[]{
\includegraphics[width=0.48\textwidth]{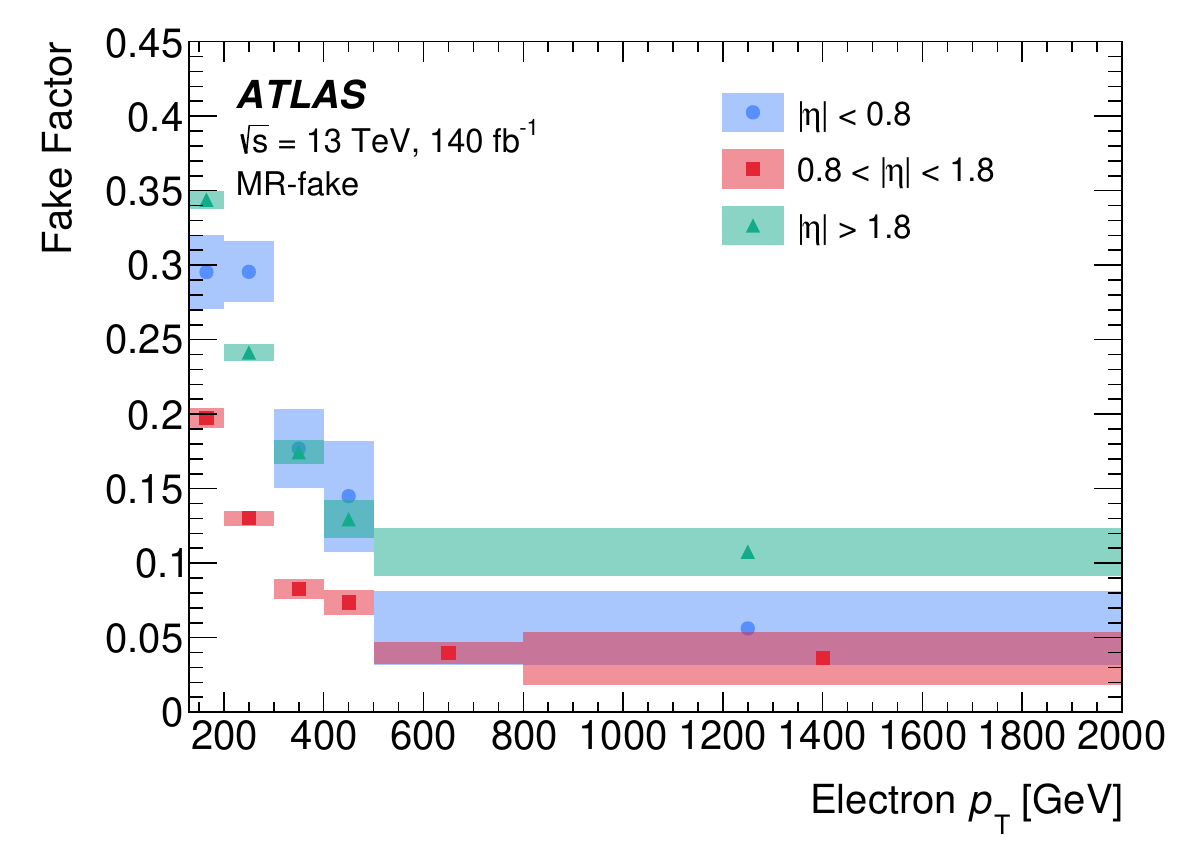}
}
\subfloat[]{
\includegraphics[width=0.48\textwidth]{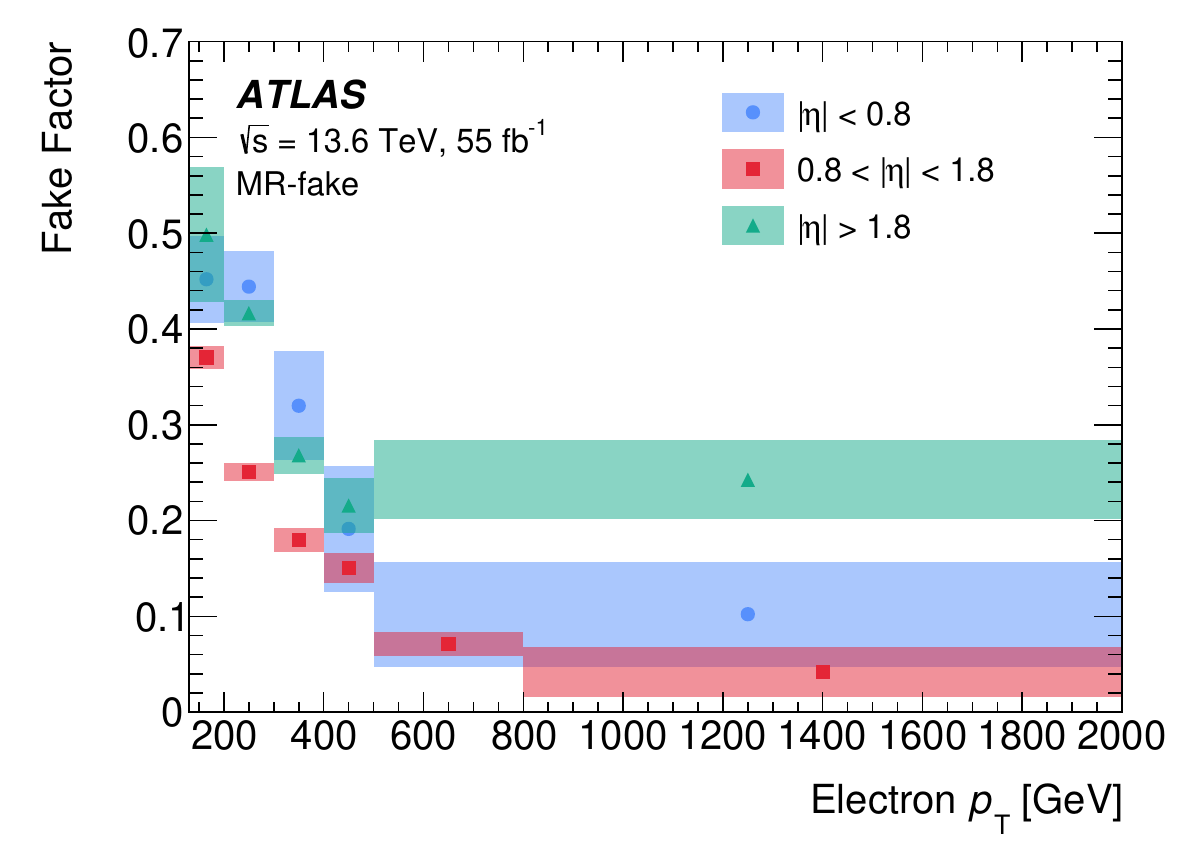}
}
\caption{Measured electron fake factors in \MRfake for (a) Run~2 and (b) Run~3 with respect to \pt in the three $|\eta|$ bins. The uncertainty bands represent the statistical uncertainties in the fake factors. The last bin is inclusive in the electron \pt.}
\label{fig:backgrounds:fake_factors}
\end{figure}

Figure~\ref{fig:results:VRfake} shows the electron \pt distributions in \VRfake separately for Run~2 and Run~3, with the FNP background estimated by using the derived fake factors.
The normalisation of \Wjets in this region is constrained by performing a background-only fit (see Section~\ref{sec:results}) to the corresponding \CRfake region.
The extracted normalisation factors for \Wjets in the \CRfake regions are found to be compatible with unity.
These fit results are used solely to validate the FNP estimate in \VRfake; the following section introduces the dedicated CRs used to constrain the \Wjets normalisation in the SRs.
Good agreement between data and the SM predictions is observed across the \pt spectrum in both data-taking periods within systematic uncertainties (see Section~\ref{sec:systematics}), giving confidence in the procedure used to derive the fake factors.

\begin{figure}[tbh]
\centering
\subfloat[]{
\includegraphics[width=0.48\textwidth]{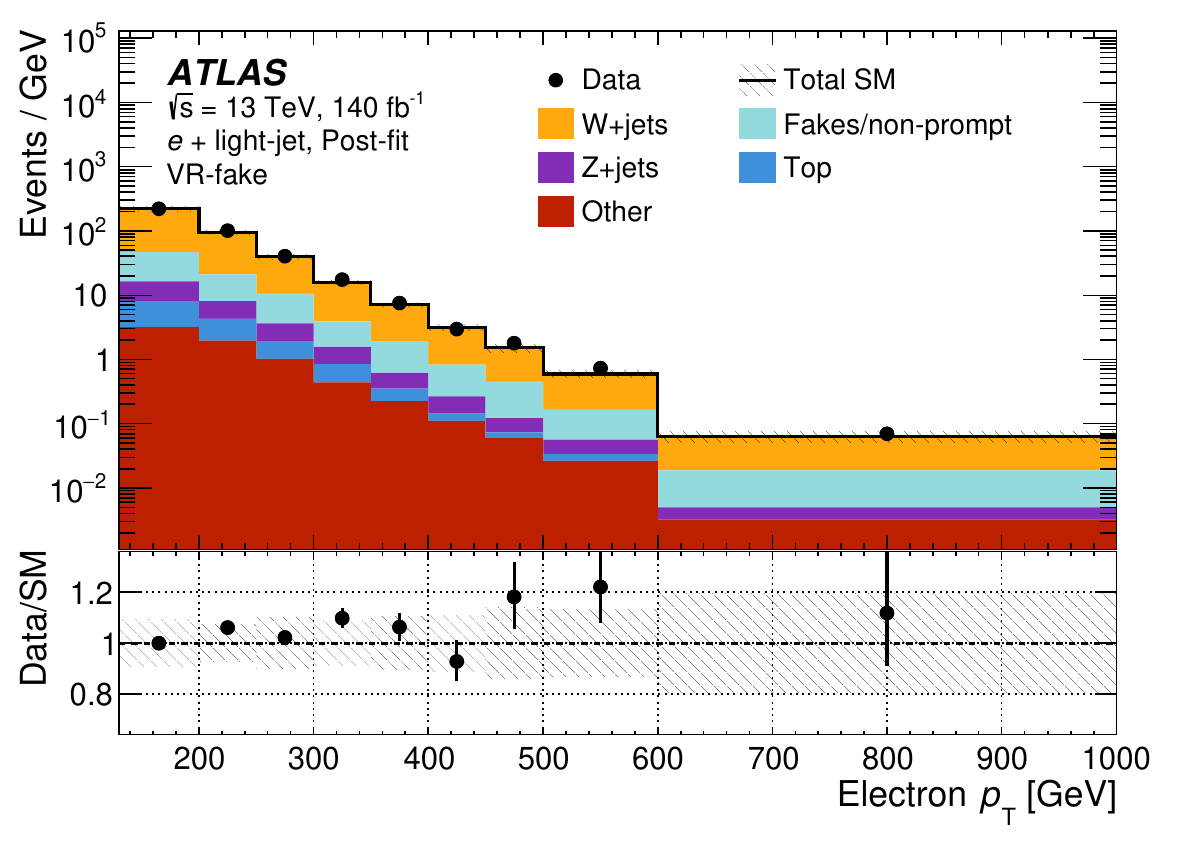}
}
\subfloat[]{
\includegraphics[width=0.48\textwidth]{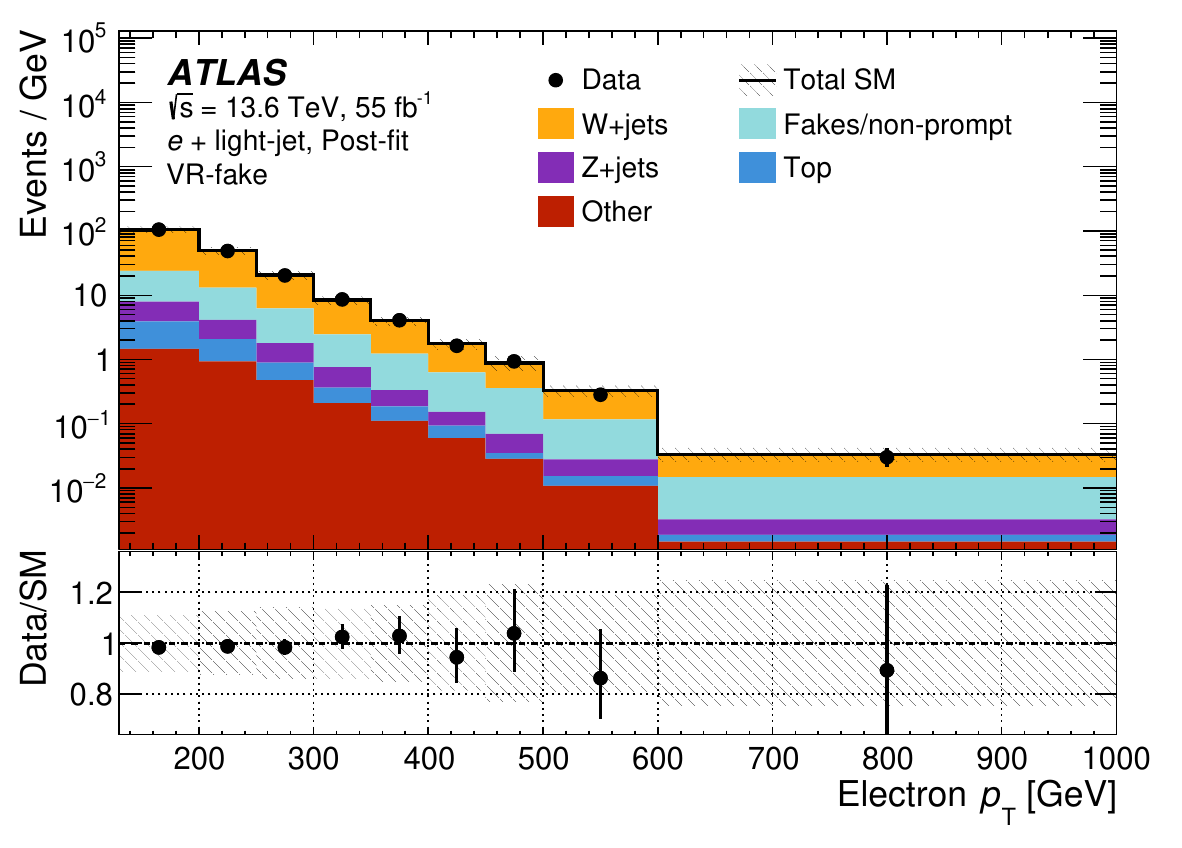}
}
\caption{Post-fit distributions of the electron \pt in \VRfake for (a) Run~2 and (b) Run~3  using a background-only fit of \CRfake.
The error bands include statistical and systematic uncertainties, with correlations between uncertainties taken into account.
The last bin contains the overflow.}
\label{fig:results:VRfake}
\end{figure}


\subsection{Irreducible backgrounds}


Irreducible backgrounds are estimated from MC simulation, with the dominant processes being constrained to data through dedicated control regions.
These CRs are individually optimised for each channel and enriched in a specific background process.
The \Wjets and \Zjets backgrounds are constrained in \CRW and \CRZ, respectively.
The \elbjet and \mubjet channels define additional \CRT regions to normalise backgrounds containing top quarks.
Events from \ttbar and single-top production are therefore grouped into a common category labelled \say{Top} and constrained using a common normalisation factor.
Other, rarer processes involving genuine leptons—such as diboson production, $\ttV$, and \ljSM—are estimated from MC simulation and grouped into a category labelled \say{Others}.
MC events used to estimate the irreducible backgrounds are required to contain only genuine leptons to avoid double-counting with the FNP estimate described in the previous subsection.
The extracted background normalisations are validated in the corresponding VRs which also adopt the \mlj binning of the SRs.
Summaries of the CR and VR definitions are provided in Table~\ref{tab:backgrounds:CRs_lightjet} for the light-jet and Tables~\ref{tab:backgrounds:CRs_elbjet} and~\ref{tab:backgrounds:CRs_mubjet} for the \elbjet and \mubjet channels, respectively.

\begin{table}[tb]
\centering
\caption{Definitions of the CRs and VRs for the \eljet and \mujet channels.
The requirements listed are placed on top of the preselection introduced in the main text.
The lepton requirements indicate an additional veto of preselected leptons, e.g. $\nel = 1$ requires the presence of exactly one signal electron but no other preselected lepton in the event.
The \VRW and \VRZ regions follow the \mlj binning of the associated SRs as described in the main text. The suffixes $\text{-}ej$ (e.g. in \CRWej) and ${-}\mu j$  of each region name are dropped in the table for brevity, respectively.}
\label{tab:backgrounds:CRs_lightjet}
\begingroup
\renewcommand{\arraystretch}{1.25}
\setlength{\tabcolsep}{3pt}
\footnotesize
\begin{tabular}{c cccccc cccc}
\toprule
\multirow[c]{2}{*}[-2mm]{Region}  & \multicolumn{6}{c}{\eljet} & \multicolumn{4}{c}{\mujet}\\
\cmidrule(lr){2-7} \cmidrule(lr){8-11}
& \CRW & \VRW & \VRMETSig & \CRZ & \VRZ & \VRMll & \CRW & \VRW & \CRZ & \VRZ\\
\cmidrule(lr){2-3} \cmidrule(lr){4-4} \cmidrule(lr){5-6} \cmidrule(lr){7-7} \cmidrule(lr){8-9} \cmidrule(lr){10-11}
\midrule
\mlj[\GeV] & [700, 950] & $> 950$ & [700, 950] & [700, 950] & $> 950$ & [700,950] & [700, 900] & $> 900$ & [700, 900] & $> 900$ \\
\nel & \multicolumn{2}{c}{1} & 1 & \multicolumn{2}{c}{2} & 2 & \multicolumn{2}{c}{0} & \multicolumn{2}{c}{0} \\
\nmu & \multicolumn{2}{c}{0} & 0 & \multicolumn{2}{c}{0} & 0 & \multicolumn{2}{c}{1} & \multicolumn{2}{c}{2} \\
\nb & \multicolumn{2}{c}{0} & 0 & \multicolumn{2}{c}{0}  & 0 & \multicolumn{2}{c}{0} & \multicolumn{2}{c}{0} \\
\dRlj & \multicolumn{2}{c}{$<3.7$} & $<3.7$ & \multicolumn{2}{c}{--} & -- & \multicolumn{2}{c}{[2.9, 3.6]} & \multicolumn{2}{c}{--}\\
\mll[\GeV] & \multicolumn{2}{c}{--} & -- & \multicolumn{2}{c}{[70, 160]} & [160, 250] & \multicolumn{2}{c}{--} & \multicolumn{2}{c}{[70, 160]} \\
\metsig & \multicolumn{2}{c}{$>3.5$} & $<3.5$ & \multicolumn{2}{c}{$<3.5$} & $<3.5$ & \multicolumn{2}{c}{$>3.5$} & \multicolumn{2}{c}{$<3.5$}\\
\LepOverMlj & \multicolumn{2}{c}{$>0.4$} & $>0.4$ & \multicolumn{2}{c}{$>0.3$} & $>0.3$ & $>0.4$ & $>0.2$ & \multicolumn{2}{c}{$>0.3$} \\
\bottomrule
\end{tabular}

\endgroup
\end{table}

\begin{table}[tb]
\centering
\caption{Definitions of the CRs and VRs for the \elbjet channel.
The requirements listed are placed on top of the preselection introduced in the main text.
The lepton requirements indicate an additional veto of preselected leptons, e.g. $\nel = 1$ requires the presence of exactly one signal electron but no other preselected lepton in the event. The \VRW, \VRZ and \VRThigh regions follow the \mlj binning of the associated SRs as described in the main text. The suffix $\text{-}eb$ (e.g. in \CRWeb) of each region name is dropped in the table for brevity.}
\label{tab:backgrounds:CRs_elbjet}
\begingroup
\renewcommand{\arraystretch}{1.25}
\setlength{\tabcolsep}{3pt}
\footnotesize
\begin{tabular}{c cccccccc}
\toprule
\multirow{2}{*}[-2mm]{Region}  & \multicolumn{6}{c}{\elbjet} \\
\cmidrule(lr){2-9}

& \CRW & \VRW & \CRZ & \VRZ & \CRThigh & \VRThigh & \CRTlow & \VRMETSig \\
\cmidrule(lr){2-3} \cmidrule(lr){4-5} \cmidrule(lr){6-7} \cmidrule(lr){8-8} \cmidrule(lr){9-9}
\midrule
\mlj[\GeV] & [700, 900] & $> 900$ & [700, 900] & $> 900$ & [900, 1300] & $> 1300$ & [700, 900] & [700, 900] \\
\nel & \multicolumn{2}{c}{1} & \multicolumn{2}{c}{2} & \multicolumn{2}{c}{1} & 1 & 1 \\
\nmu & \multicolumn{2}{c}{0} & \multicolumn{2}{c}{0} & \multicolumn{2}{c}{0} & 0 & 0 \\
\nb  & \multicolumn{2}{c}{1} & \multicolumn{2}{c}{1} & \multicolumn{2}{c}{$\geq 2$} & $\geq 2$ & 1 \\
\mll[\GeV] & \multicolumn{2}{c}{--} & \multicolumn{2}{c}{[70, 110]} & \multicolumn{2}{c}{--} & -- & -- \\
\metsig & \multicolumn{2}{c}{$>3.0$} & \multicolumn{2}{c}{$<5.0$} & \multicolumn{2}{c}{$<4.0$} & $>3.0$ & $<3.0$ \\
\LepOverMlj & \multicolumn{2}{c}{$>0.3$} & \multicolumn{2}{c}{$>0.25$} & \multicolumn{2}{c}{$>0.3$} & $>0.3$ & $>0.3$  \\
\dPhilv & \multicolumn{2}{c}{$<1.0$} & \multicolumn{2}{c}{--} & \multicolumn{2}{c}{$<1.0$} & $<1.0$ & $<1.0$ \\
\bottomrule
\end{tabular}
\endgroup
\end{table}

\begin{table}[bt]
\centering
\caption{Definitions of the CRs and VRs for the and \mubjet channel.
The requirements listed are placed on top of the preselection introduced in the main text.
The lepton requirements indicate an additional veto of preselected leptons, e.g. $\nel = 1$ requires the presence of exactly one signal electron but no other preselected lepton in the event. The \VRW, \VRZ and \VRThigh regions follow the \mlj binning of the associated SRs as described in the main text. The suffix $\text{-}\mu b$ (e.g. in \CRWmb) of each region name is dropped in the table for brevity.}
\label{tab:backgrounds:CRs_mubjet}
\begingroup
\renewcommand{\arraystretch}{1.25}
\setlength{\tabcolsep}{3pt}
\adjustbox{max width=\textwidth}{
\footnotesize
\begin{tabular}{c ccccccccc}
\toprule
\multirow{2}{*}[-2mm]{Region}  & \multicolumn{9}{c}{\mubjet} \\
\cmidrule(lr){2-10}
& \CRW & \VRW & \CRZ & \VRZ & \CRThigh & \VRThigh & \CRTlow & \VRTlow & \VRMETSig \\
\cmidrule(lr){2-3} \cmidrule(lr){4-5} \cmidrule(lr){6-7} \cmidrule(lr){8-9} \cmidrule(lr){10-10}
\midrule
\mlj[\GeV] & [700, 900] & $> 900$ & [700, 900] & $> 900$ & $> 900$ & $> 900$ & \multicolumn{2}{c}{[700, 900]} & [700, 900] \\
\nel & \multicolumn{2}{c}{0} & \multicolumn{2}{c}{0} & \multicolumn{2}{c}{0} & \multicolumn{2}{c}{0} & 0 \\
\nmu & \multicolumn{2}{c}{1} & \multicolumn{2}{c}{2} & \multicolumn{2}{c}{1} & \multicolumn{2}{c}{1} & 1 \\
\nb  & \multicolumn{2}{c}{1} & \multicolumn{2}{c}{1} & \multicolumn{2}{c}{$\geq 2$} & \multicolumn{2}{c}{$\geq 2$} & 1 \\
\dRlj & \multicolumn{2}{c}{[2.4, 4.2]} & \multicolumn{2}{c}{[2.4, 4.2]} & \multicolumn{2}{c}{[2.4, 4.2]} & \multicolumn{2}{c}{[2.4, 4.2]} & [2.4, 4.2] \\
\mll[\GeV] & \multicolumn{2}{c}{--} & \multicolumn{2}{c}{[70, 120]} & \multicolumn{2}{c}{--} & \multicolumn{2}{c}{--} & -- \\
\metsig & \multicolumn{2}{c}{$>3.0$} & \multicolumn{2}{c}{$<5.0$} & $>4.0$ & [3.0, 4.0] & $>4.0$ & [3.0, 4.0] & $<3.0$ \\
\LepOverMlj & \multicolumn{2}{c}{$>0.2$} & \multicolumn{2}{c}{$>0.2$} & \multicolumn{2}{c}{$>0.3$} & \multicolumn{2}{c}{$>0.3$} & $>0.3$ \\
\bottomrule
\end{tabular}
}
\endgroup
\end{table}

\begin{figure}[tb]
\centering
\subfloat[]{
\includegraphics[width=0.48\textwidth]{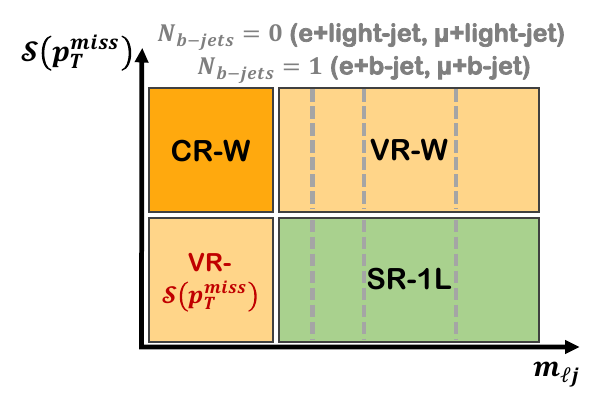}
}
\subfloat[]{
\includegraphics[width=0.45\textwidth]{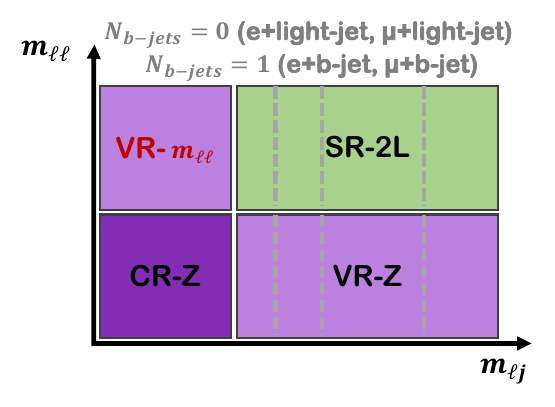}
}
\caption{Illustrations of the CRs and VRs definitions for the (a) \Wjets and (b) \Zjets backgrounds.
The phase space of \SROneL and \SRTwoL is also indicated, respectively.
Grey vertical dashed lines indicate that a region is binned in \mlj.}
\label{fig:bkg:CRsVjets}
\end{figure}

All four channels follow a similar strategy to derive and validate the normalisations for \Wjets and \Zjets backgrounds.
Both the \CRW and \CRZ are defined at lower \mlj values than the SRs to ensure sufficiently large event counts in these regions.
Illustrations showing the region layout of the \CRW and \CRZ are presented in Figure~\ref{fig:bkg:CRsVjets}.
These CRs span the \mlj range from 700~\GeV up to the start of the SRs, which begins at either 900 or 950~\GeV depending on the channel.
The \VRW and \VRZ regions apply the same selection as their respective CRs but cover the \mlj range used in the SRs.
To maintain kinematic similarity to the SRs, the CRs and VRs mirror the SR requirements on \dRlj, \mll, \metsig, and \LepOverMlj as closely as possible, with one selection inverted to ensure orthogonality.
In some of these CRs and VRs, additional or relaxed requirements relative to the SRs are applied to ensure that the contamination from signals not already excluded by previous ATLAS searches remains below 10\%.
\CRW and \VRW replicate the \SROneL selection but invert the \metsig requirement to enrich \Wjets events.
Similarly, \CRZ and \VRZ mimic the \SRTwoL selection but require \mll to lie within a window around the $Z$ boson mass.
To facilitate a validation of the background normalisations across \mlj, \VRW and \VRZ adopt the same \mlj binning as \SROneL and \SRTwoL, respectively.
If the expected number of events in a VR bin falls below approximately 10, the corresponding bin is made inclusive in \mlj to ensure adequate event counts for validation.

\begin{figure}[tb]
\centering
\subfloat[]{
\includegraphics[width=0.48\textwidth]{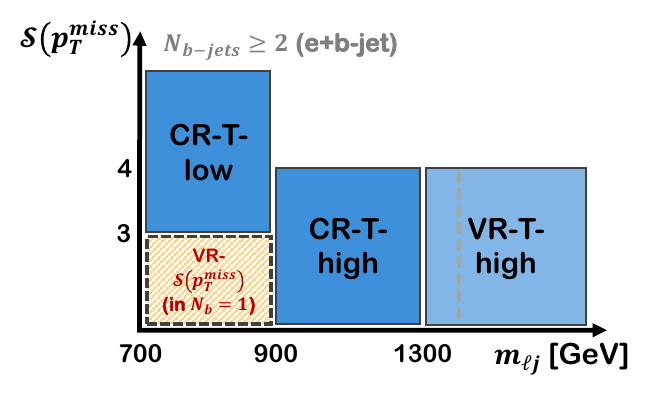}
}
\subfloat[]{
\includegraphics[width=0.48\textwidth]{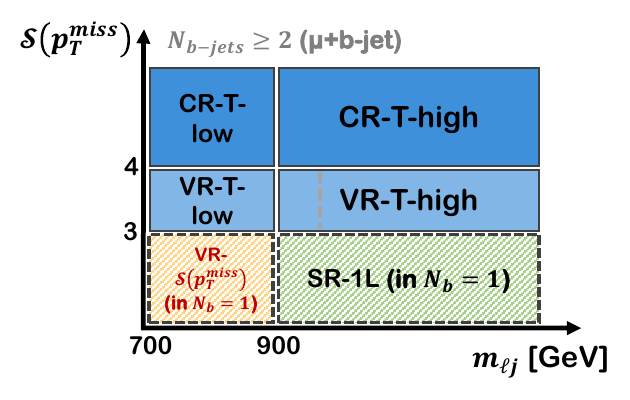}
}
\caption{Illustrations of the CRs and VRs definitions for the top background in the (a) \elbjet and (b) \mubjet channel, respectively.
The phase spaces of \VRMETSig and \SROneL which select event with exactly one $b$-tagged jet are indicated as hatched boxes.
The phase space \SROneL is only indicated in (b) as it overlaps with the phase space of CR/VR-T-high in (a) in the \mlj--\metsig plane.
Grey vertical dashed lines indicate that a region is binned in \mlj.}
\label{fig:bkg:CRsTop}
\end{figure}

The estimation strategy for the top background is adapted relative to that for the \Vjets processes, as its normalisation is found to depend on \mlj.
The associated \CRT and \VRT regions select 1L events with two or more $b$-tagged jets, with the $b$-jet requirement ensuring orthogonality with the SRs.
Both the \elbjet and \mubjet channels define a \CRThigh region at $\mlj > 900~\GeV$ to normalise top events in the high-\mlj regime.

Since top-quark decays involving leptons yield genuine \met, \CRThighmb applies the same kinematic selections as \SROneLmb but requires $\metsig > 4.0$ to enhance the top fraction while suppressing signal contamination.
Events with $3.0 < \metsig < 4.0$ define \VRThighmb, which otherwise matches the CR selection and is binned in \mlj for validation.

In contrast, \CRThigheb applies $\metsig < 4.0$, i.e. an upper requirement on this variables similar as in the SRs, since the top normalisation also shows some dependence on \metsig in the \eljet channel.
\CRThigheb extends up to $\mlj = 1300~\GeV$, while higher-\mlj events are covered by \VRThigheb, which otherwise mirrors the CR selection and is binned in \mlj.

The \CRW regions in both the \elbjet and \mubjet channels, which cover \mlj values below 900~\GeV, receive substantial contributions from top-quark backgrounds.
To ensure proper normalisation of these events, dedicated \CRTlow regions are defined to normalise top events in the low-\mlj regime.
\CRTloweb adopts the same requirements as \CRWeb, differing only in the $b$-jet multiplicity.
\CRTlowmb reflects the kinematic selections of \SROneLmb but enforces $\metsig > 4.0$.
An illustration of the CRs and VRs definitions for the top background is shown in Figure~\ref{fig:bkg:CRsTop}.

A set of additional VRs is defined to test the robustness of the background normalisations when extrapolated over variables other than \mlj.
The \VRMETSig regions invert the \metsig requirement of the corresponding \CRW selections to test the modelling of \Wjets at low \metsig.
The extrapolation of the \Zjets normalisation across \mll is validated with \VRMll.
This validation region is defined only for the \eljet channel, as the other three channels show significant signal contamination that precludes its use.
The \mubjet channel defines \VRTlowmb, to validate the low-\mlj top background normalisation across \metsig as the associated \VRMETSig receives a substantial contribution from this background.



%

%
\section{Systematic uncertainties}

\label{sec:systematics}

While this search is predominantly limited by statistical constraints, experimental, theory, and modelling uncertainties have non-negligible contributions to the total uncertainty. These systematic effects are quantified and incorporated into the statistical model (see Section~\ref{sec:results}), with their key components detailed below.
The overall effect of systematic uncertainties in the LQ mass sensitivity reach is found to be below 5\%.

Figure~\ref{fig:systematics:breakdown} shows a decomposition into individual categories of the systematic uncertainties in the total SM predictions for the SRs in Run~2 of each channel, with a similar breakdown observed for the corresponding Run~3 regions.
This decomposition is obtained by performing a series of fits in which the parameters associated with a given category are fixed to their best-fit values and held constant, effectively removing their contribution from the systematic model.
The systematic uncertainty attributed to each category is then computed as the quadratic difference between the total background uncertainty in the nominal fit and that in the fit with the category fixed~\cite{Baak:2014wma}.

\begin{figure}[tbh]
\centering
\subfloat[]{
\includegraphics[width=0.45\textwidth]{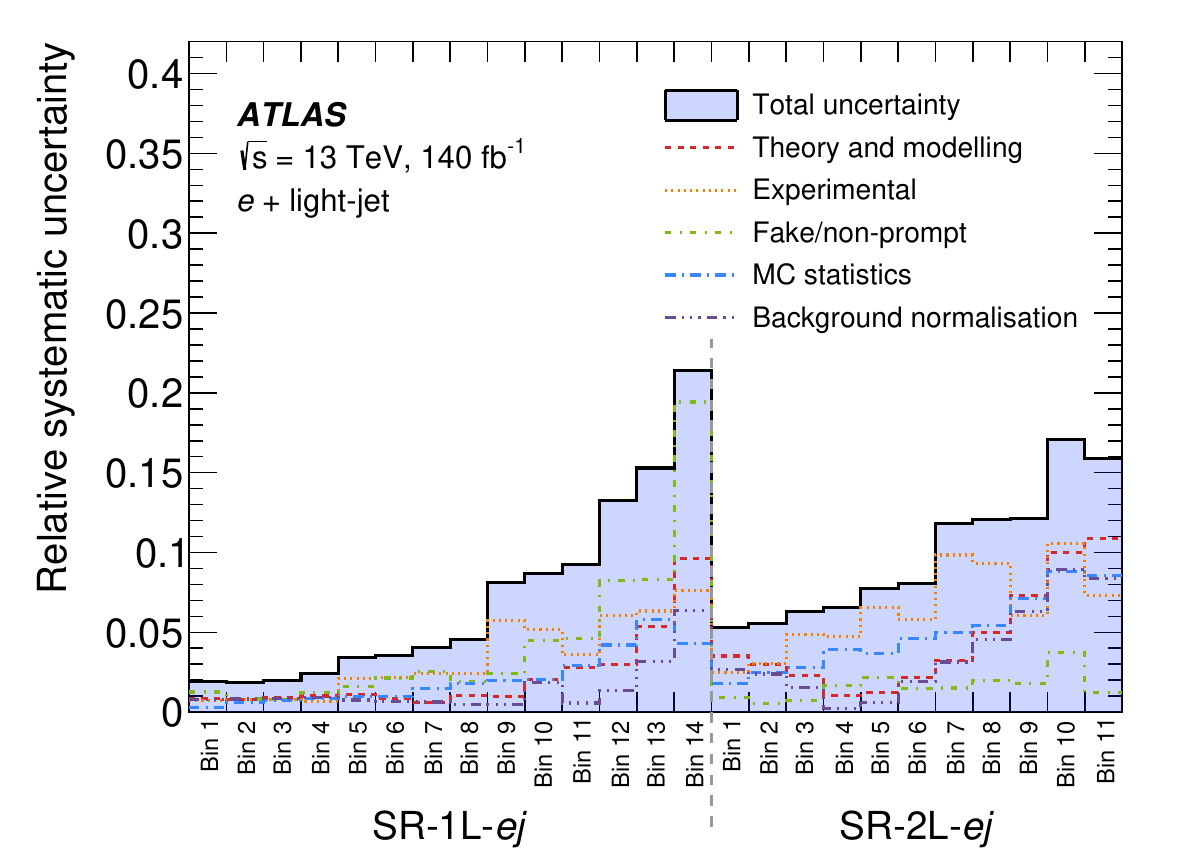}
}
\subfloat[]{
\includegraphics[width=0.45\textwidth]{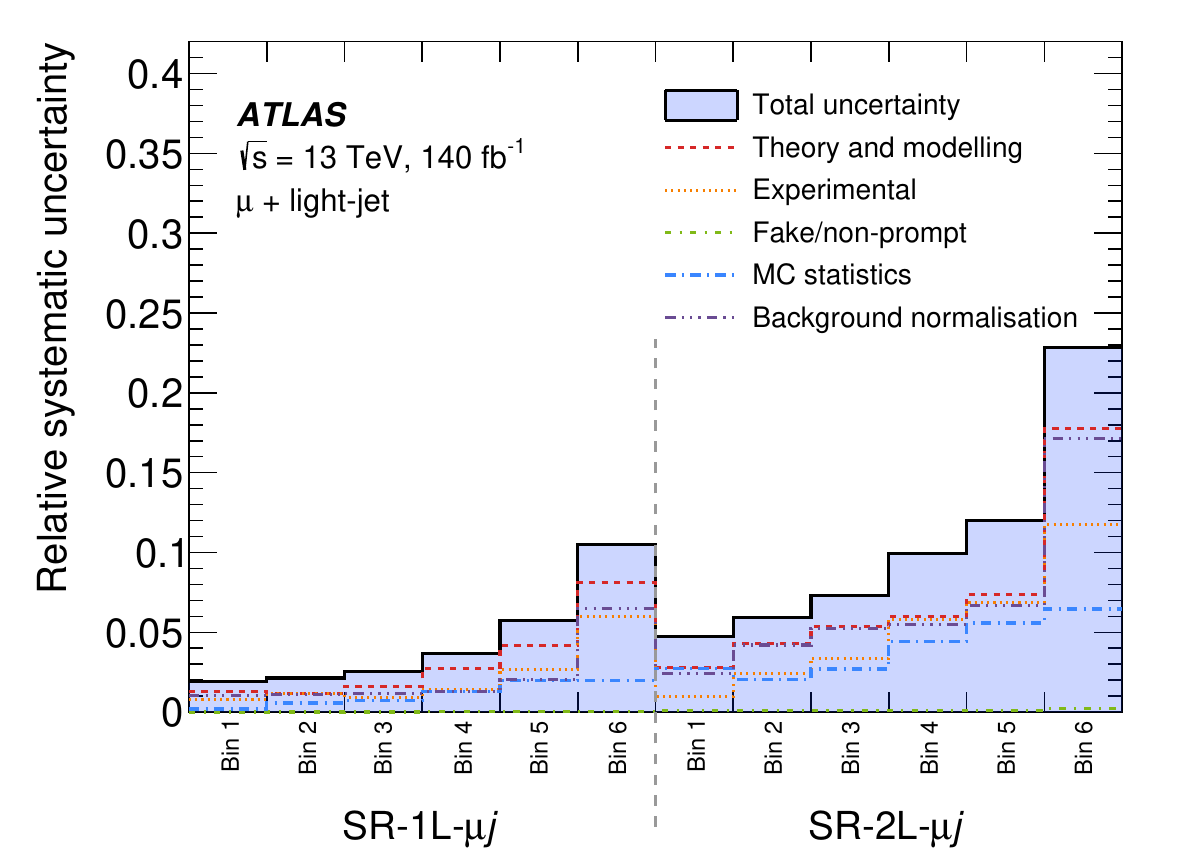}
}\\
\subfloat[]{
\includegraphics[width=0.45\textwidth]{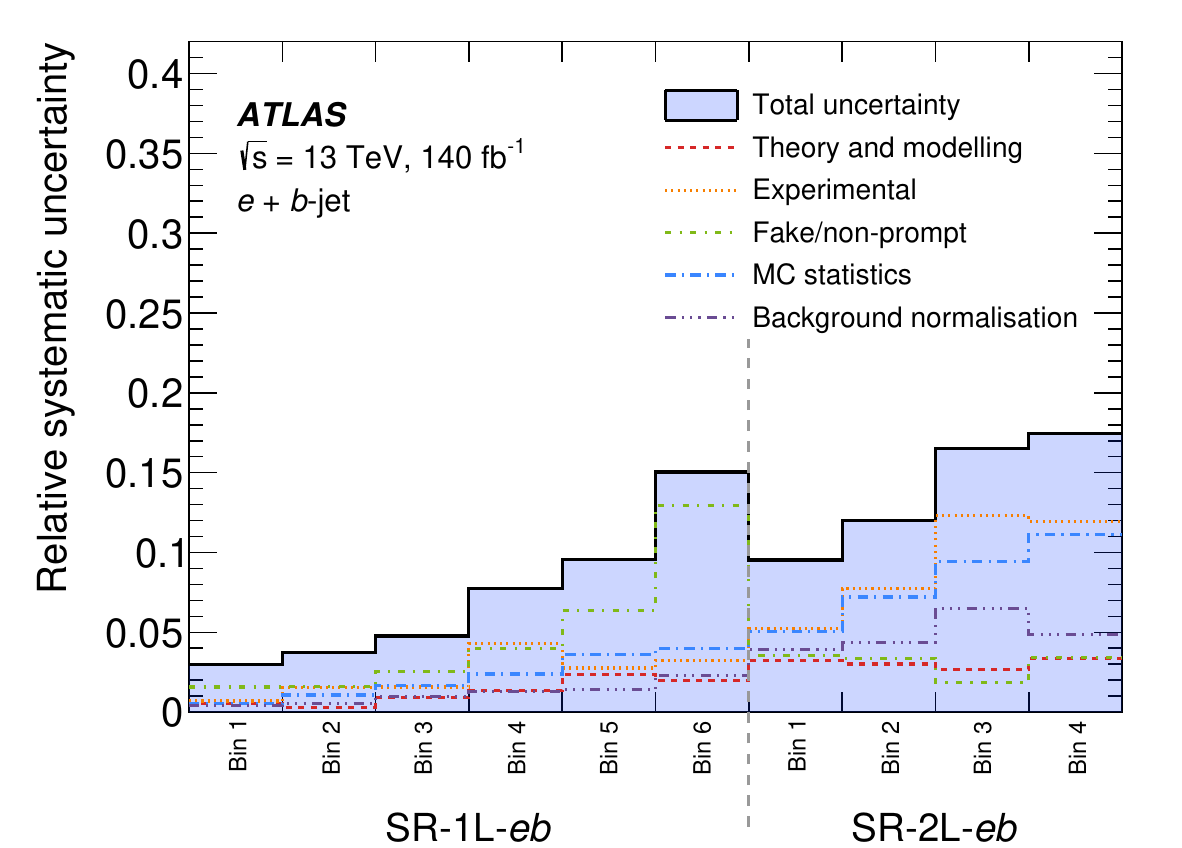}
}
\subfloat[]{
\includegraphics[width=0.45\textwidth]{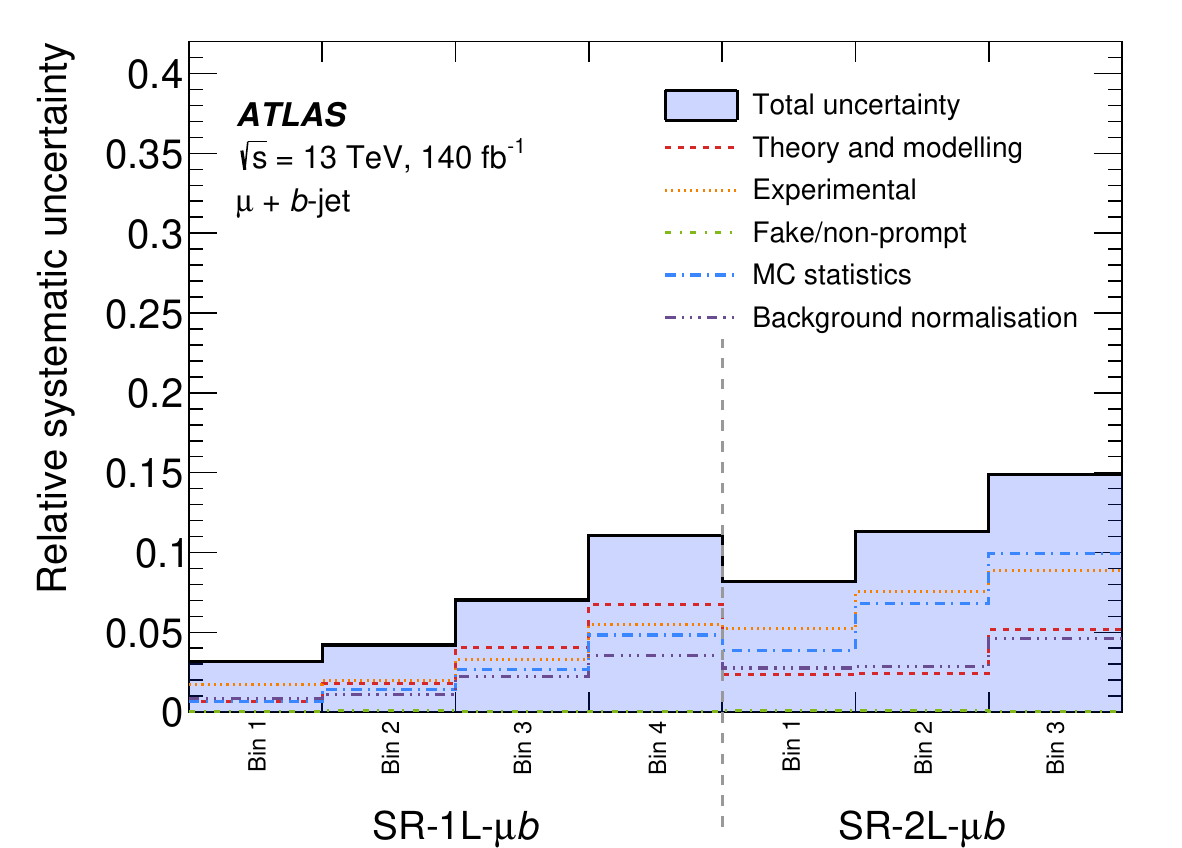}
}
\caption{Relative systematic uncertainties in the post-fit SM background estimates in the (a) \eljet, (b) \mujet, (c) \elbjet and (d) \mubjet channels obtained from a background-only fit to the respective CRs and SRs.
The ``Fake/non-prompt'' category reflects uncertainties impacting the FNP background estimate.
Uncertainties originating from the limited size of the MC samples used to model the irreducible background contributions are contained in the ``MC statistics'' category.
The ``Background normalisation'' category reflects uncertainty in normalisation factors for the \Wjets, \Zjets and top backgrounds extracted from the respective \CRW, \CRZ and \CRT regions.
The ``Theory and modelling'' category includes the different sources of theory modelling uncertainties for the \Wjets, \Zjets and top backgrounds.
The ``Experimental'' category covers detector related uncertainties from the reconstruction and selection of objects in the analysis.
The individual uncertainties are correlated and do not necessarily add up in quadrature to the total uncertainty.}
\label{fig:systematics:breakdown}
\end{figure}

A detailed evaluation of detector-related systematic uncertainties is performed.
These include uncertainties in lepton performance, covering trigger, reconstruction, identification, and isolation efficiencies for electrons~\cite{EGAM-2018-01} and muons~\cite{MUON-2018-03}, along with momentum calibration uncertainties for both lepton species.
Jet energy calibration uncertainties are also considered, including those in the jet energy scale (JES) and jet energy resolution (JER)~\cite{JETM-2018-05}.
Additional jet-related uncertainties arise from efficiency corrections applied to pile-up jet tagging~\cite{PERF-2014-03} and $b$-jet identification~\cite{FTAG-2018-01,FTAG-2020-08,FTAG-2019-02}.
Missing transverse momentum uncertainties originate from the propagation of JES and JER uncertainties to the \met calculation, supplemented by uncertainties related to tracks associated with the primary vertex but unmatched to reconstructed objects~\cite{JETM-2020-03}.
The uncertainties in the combined 2015--2018 and 2022--2023 integrated luminosities are~0.83\%~\cite{DAPR-2021-01} and 2.0\%~\cite{ATL-DAPR-PUB-2023-001,ATL-DAPR-PUB-2024-001}, respectively, obtained using the LUCID-2 detector~\cite{LUCID2} for the primary luminosity measurements, complemented by measurements using the inner detector and calorimeters.
Additionally, a dedicated uncertainty accounts for discrepancies between data and simulation in pile-up profile modelling.

The treatment of FNP background employs distinct strategies across analysis channels.
Simulated samples directly model these backgrounds in the $\mu+\text{jet}$ and 2$e$ electron channels, with a conservative 80\%
normalisation uncertainty applied to account for potential mismodelling. In the 1$e$ selections of the \eljet and \elbjet
channels, systematic uncertainties associated with the data-driven FNP estimate (Section~\ref{sec:bkg_estimation})
are characterised through multiple dedicated studies.
The dominant uncertainty originates from the subtraction of the prompt-lepton contamination in the loose and tight electron samples, originating in particular from \Wjets events.
Therefore, $\pm$15\% variations in the MC-derived real lepton fractions are propagated through the fake-factor calculation, with the variation magnitude chosen to cover the size of the \Wjets normalisation factors observed in the \eljet channel, see Section~\ref{sec:results}.
To account for the reduced binning granularity in \pt for central electrons with $\pt > 800~\GeV$ in the fake-factor measurements, an additional 70\% uncertainty, based on the \pt dependence of the fake factors observed in MC simulation, is assigned on the FNP estimate from such electrons.
A non-negligible dependence of the fake factors on \dRlj is observed.
Comparing fake factors from MC simulation from events with \dRlj smaller and larger than 3.7, results in uncertainties of 50\% ($\pt < 200~\GeV$) and 25\% ($\pt > 200~\GeV$), respectively.
Differences between the FNP composition between the fake-factor measurement and application domains are evaluated using MC simulation and translated into an additional 50\% uncertainty in the FNP estimate.
Furthermore, potential $b$-jet induced modifications to the fake factors were investigated through dedicated simulation studies comparing $b$-jet enriched and depleted phase spaces in the \elbjet channel.
The observed weak correlation between fake factor magnitudes and $b$-jet presence was conservatively accommodated through an additional 25\% uncertainty component in the FNP estimates derived for the \elbjet channel.

The background composition exhibits a strong channel dependence, with \Wjets~production
constituting the dominant contribution in SR-1L and \Zjets~prevailing in SR-2L.
Systematic uncertainties in these \Vjets~processes incorporate three principal components: variations of the QCD renormalisation $\mu_R$ and factorisation $\mu_F$ scales by a factor two evaluated through the envelope of seven $\mu_R$--$\mu_F$ combinations, PDF eigenvector variations
using the \NNPDF[3.1] NNLO set, and $\alpha_S$ variations.
Additional uncertainties are applied to take into account electroweak corrections on the \Vjets samples~\cite{Kallweit:2014xda}.

Top-quark backgrounds, while small in $b$-vetoed regions, are subleading contributions
in the \elbjet and \mubjet channels.
Their theory uncertainties mirror the \Vjets~framework in scale and PDF treatment,
augmented by process-specific considerations including parton shower modelling differences quantified
via \PYTHIA[8.230] versus \HERWIG[7.2.1] (for \ttbar) or \HERWIG[7.1.6] (for single top) generator comparisons.
Uncertainties associated with the level of initial- (ISR) and final-state radiation (FSR) are estimated by variations of $\alpha_S$ in the A14 tune~\cite{ATL-PHYS-PUB-2014-021} and variations of the renormalization scale for FSR branchings, respectively.
The matrix element-parton shower interface uncertainty is assessed through a variation of the $\pt^{\mathrm{hard}}$ parameter that regulates how the radiation phase space of the parton shower is determined, following the prescription in Ref.~\cite{Hoche:2021mkv}.
Interference effects between $\ttbar$ and $Wt$ processes are evaluated through diagram subtraction/removal scheme comparisons~\cite{Frixione:2008yi}.

Similarly, uncertainties in the signal predictions from scale and PDF variations were evaluated using generator-level reweighting, focusing on \mlj~spectrum distortions in both the SR-1L and SR-2L regions.
These uncertainties were found to have a negligible impact on the results and were therefore neglected.

As seen in Figure~\ref{fig:systematics:breakdown}, the dominant sources of systematic uncertainties vary depending on the channel and signal region.
In the \SROneL selections of the \eljet and \elbjet channels, the leading uncertainty arises from the estimate of backgrounds with FNP electrons.
Other major experimental uncertainties are associated with the jet energy scale and resolution, as well as the $b$-jet identification efficiency in the \elbjet and \mubjet channels.
Among the modelling uncertainties, the most significant contributions come from QCD scale variations and electroweak corrections to the \Vjets backgrounds.

\FloatBarrier


%

%
\section{Results}

\label{sec:results}

Observed data in the CRs, VRs, and SRs are compared with the SM predictions using a profile likelihood method~\cite{Cowan:2010js}.
The statistical model is constructed using the \texttt{cabinetry}~\cite{cabinetry} package, which interfaces with \texttt{pyhf}~\cite{my_pyhf_joss}, a Python-based implementation of the \textsc{HistFactory}~\cite{histfactory} template.
Parameters of interest, such as the signal strength, along with other floating parameters like normalisation factors, are determined via a maximum-likelihood fit to data.
In each region, the expected event yield is modelled as the sum of contributions from the individual physics processes (samples).
The predicted rate for each sample may depend on a set of free parameters $\vb*{\psi}$ such as normalisation factors and the signal strength, and a set of constrained nuisance parameters $\vb*{\theta}$, which encode the effect of systematic uncertainties.

The probability density function for bin $b$ in region $r$ is modelled as a Poisson distribution, $\text{Pois}(n_{rb} \mid \nu_{rb}(\vb*{\psi}, \vb*{\theta}))$, where $n_{rb}$ is the observed number of events and $\nu_{rb}(\vb*{\psi}, \vb*{\theta})$ is the predicted event yield.
To account for systematic uncertainties, additional constraint terms are introduced, which control the allowed deviations of the nuisance parameters from their nominal values.
These constraints are interpreted as auxiliary measurements, with associated global observables $\vb*{a}$, such that the full set of observations is denoted by $\vb*{x} = (\vb*{n}, \vb*{a})$, where $\vb*{n} = {n_{rb}}$ represents the set of observed yields across all bins and regions.
The full likelihood function is then constructed as the product of the Poisson likelihoods for each bin and region and the constraint terms for each nuisance parameter:

\begin{equation}
L(\vb*{x} \mid \vb*{\psi}, \vb*{\theta}) = \prod_{r \, \in \, \text{regions}} \ \prod_{b \, \in \, \text{bins}} \text{Pois}(n_{rb} \mid \nu_{rb}(\vb*{\psi}, \vb*{\theta})) \prod_{\theta \, \in \, \vb*{\theta}} f_{\theta}(a_\theta \mid \theta)\,,
\end{equation}

where $f_\theta(a_\theta \mid \theta)$ is the constraint term associated with nuisance parameter $\theta$, typically modelled as a Gaussian distribution centred on the nominal value with a width reflecting the corresponding uncertainty.

Different fit configurations are employed to derive the results presented below. All fits are performed independently for each channel and for the validation of the FNP background estimate, and generally also individually for the Run~2 and Run~3 datasets.

The first configuration, referred to as \emph{CR-only fit}, is a background-only fit to the \CRW and \CRZ regions to validate the background estimates by extracting normalisation factors \muW and \muZ for the \Wjets and \Zjets backgrounds, respectively.
In the \elbjet and \mubjet channels, the \CRTlow and \CRThigh regions are additionally included to extract normalisation factors \muTlow and \muThigh for top backgrounds in the low- and high-\mlj regimes, respectively.
The results of the CR-only fit are extrapolated to the corresponding VRs to compare the post-fit SM predictions with the observed data.
For the validation of the FNP estimate, a separate CR-only fit is performed using only the \CRfake region to derive predictions in \VRfake, see Section~\ref{sec:bkg_estimation}.
This CR is not included in any other fit.

The second configuration, referred to as \emph{CR+SR fit}, simultaneously fits the \mlj distributions in \SROneL and \SRTwoL along with the associated CRs.
This configuration probes for potential BSM contributions while constraining the background components, and is also used to evaluate the compatibility of the observed data with specific signal hypotheses.
The normalisation factors derived in this configuration are found to be compatible with the ones from a CR-only fit.

A third configuration, the \emph{Run~2+3 combination}, maximizes sensitivity to a given signal model by fitting the \mlj spectra in the SRs of both Run 2 and Run 3 simultaneously.
In this configuration, all CRs and SRs from both data-taking periods are included, with independent normalisation factors for the backgrounds in Run 2 and Run 3.
All nuisance parameters are treated as uncorrelated between the two data-taking periods, except for most of the JES and JER variations, which are derived from a shared set of in situ calibrations.
Treating all nuisance parameters as fully correlated between Run~2 and Run~3 was not found to have a notable impact on the final sensitivity reach of the search.

\subsection{Results for control and validation regions}

\begin{figure}[phtb]
\centering
\subfloat[]{
\label{fig:results:VRs:lightjet:eljet}
\includegraphics[width=0.7\textwidth]{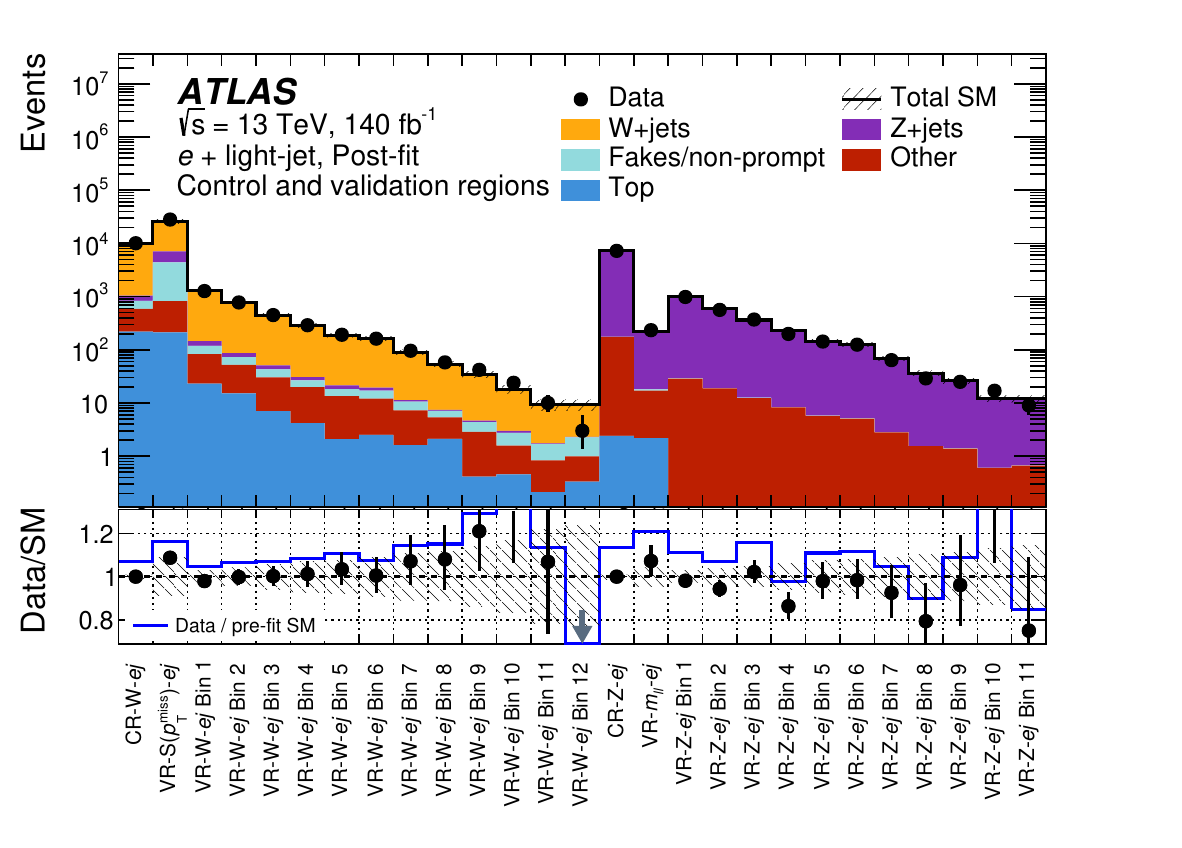}
}\\
\subfloat[]{
\label{fig:results:VRs:lightjet:mujet}
\includegraphics[width=0.7\textwidth]{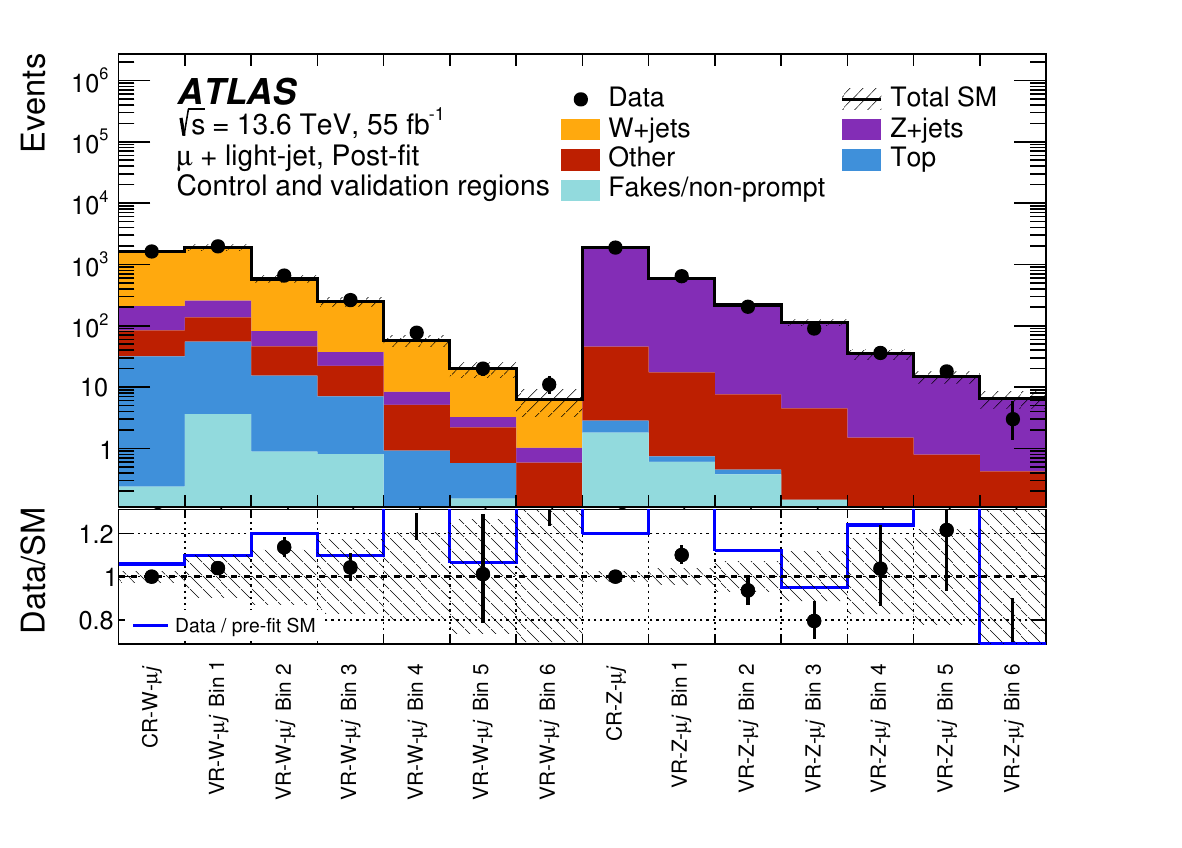}
}
\caption{Data (dots) and post-fit SM predictions (histograms) in the VRs for (a) Run~2 of the \eljet and (b) Run~3 of the \mujet channels obtained by a CR-only fit.
The lower panel shows the ratio of observed data to the total post- and pre-fit SM prediction.
Uncertainties in the background estimates include both the statistical and systematic uncertainties, with correlations between uncertainties taken into account.
A grey arrow in the lower panel indicates a data point outside the vertical range shown.
}
\end{figure}

\begin{figure}[phtb]
\centering
\subfloat[]{
\includegraphics[width=0.7\textwidth]{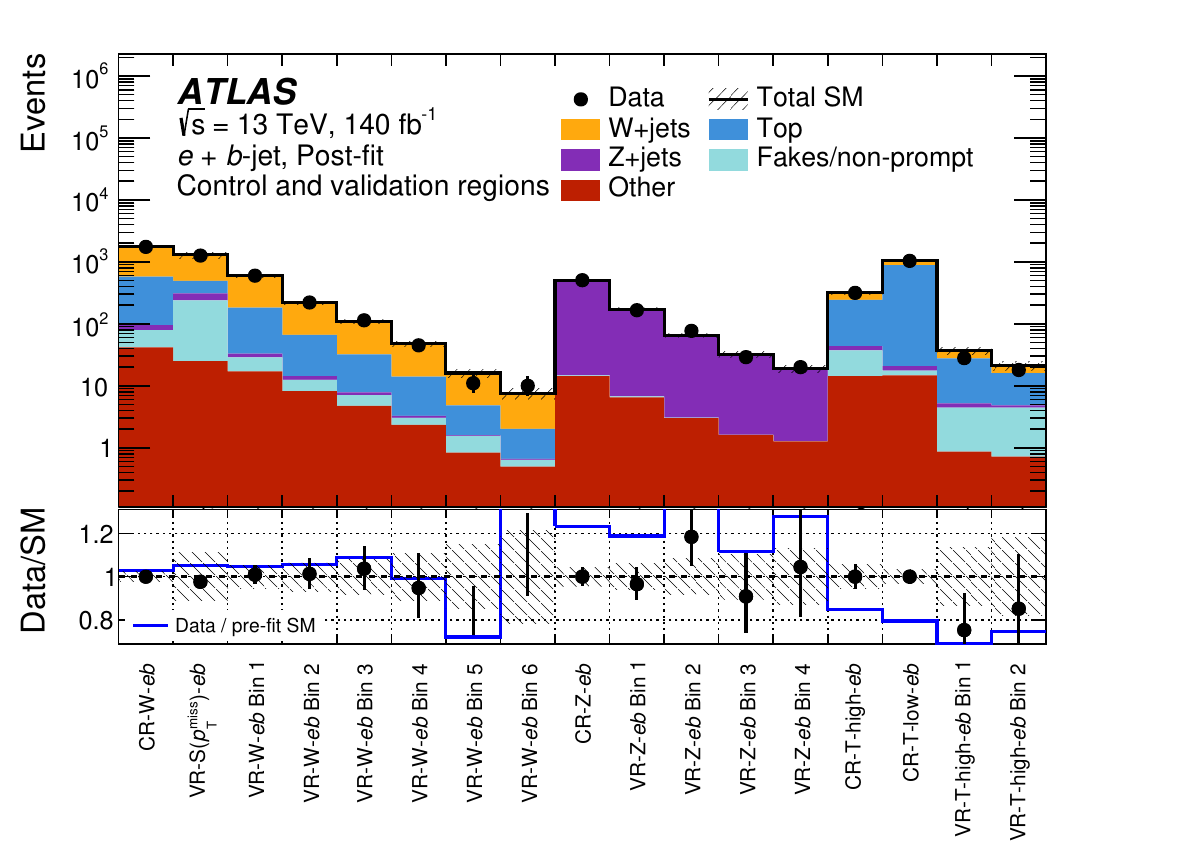}
}\\
\subfloat[]{
\includegraphics[width=0.7\textwidth]{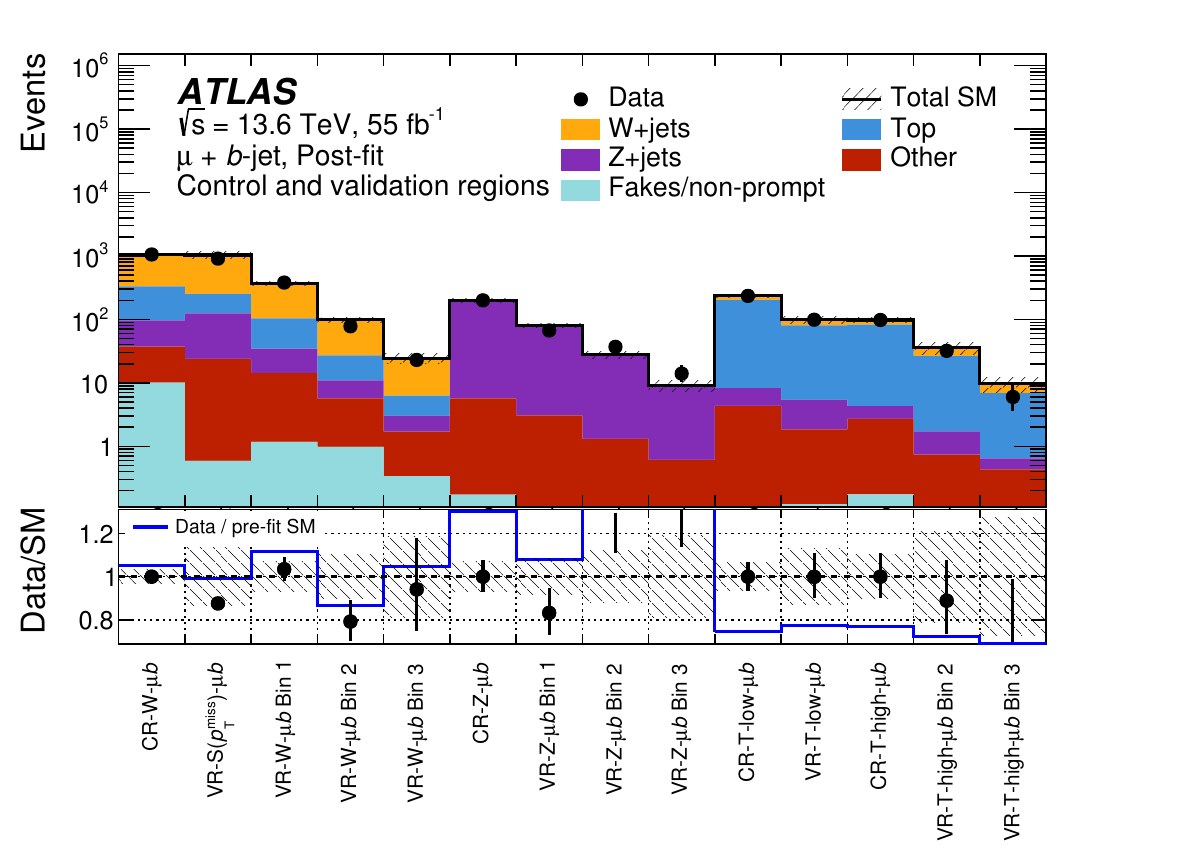}
}
\caption{Data (dots) and post-fit SM predictions (histograms) in the VRs for (a) Run~2 of the \elbjet and (b) Run~3 of the \mubjet channels obtained by a CR-only fit.
The lower panel shows the ratio of observed data to the total post- and pre-fit SM prediction.
Uncertainties in the background estimates include both the statistical and systematic uncertainties, with correlations between uncertainties taken into account.
}
\label{fig:results:VRs:bjet}
\end{figure}

The agreement between observed data and SM predictions in the \eljet VRs after a fit to \CRWej and \CRZej for Run~2 is shown in Figure~\ref{fig:results:VRs:lightjet:eljet}.
Overall, the data are consistent with the SM predictions within the associated uncertainties.
A notable deficit is observed in the last bin of \VRWej in Run~2, which is not reproduced in the corresponding Run~3 region and is therefore attributed to a statistical fluctuation.
Overall, the observations support the robustness of the \Wjets and \Zjets normalisation extrapolation over \mlj.

\begin{table}[tb]
\centering
\caption{Normalisation factors \muW, \muZ, \muThigh and \muTlow for the \Wjets, \Zjets and top backgrounds, respectively, extracted from CR-only fits using the respective CRs of each channel. The associated uncertainties include all statistical and systematic contributions.}
\label{tab:results:normfactors}
\begingroup
\renewcommand{\arraystretch}{1.25}
\adjustbox{max width=\textwidth}{
\begin{tabular}{c cc cc cc cc}
\toprule
\multirow{2}{*}{Norm. factor}  & \multicolumn{2}{c}{\eljet} & \multicolumn{2}{c}{\mujet} & \multicolumn{2}{c}{\elbjet} & \multicolumn{2}{c}{\mubjet}\\
\cmidrule(lr){2-3} \cmidrule(lr){4-5} \cmidrule(lr){6-7}  \cmidrule(lr){8-9}
& Run~2 & Run~3  & Run~2 & Run~3 & Run~2 & Run~3 & Run~2 & Run~3 \\
\midrule
\muW  & $1.08 \pm 0.21$ & $1.13 \pm 0.25$ & $1.04 \pm 0.18$ & $1.05 \pm 0.21$ & $1.21 \pm 0.28$ & $1.22 \pm 0.33$ & $1.24 \pm 0.27$ & $1.24 \pm 0.31$ \\
\muZ  & $1.14 \pm 0.25$ & $1.20 \pm 0.23$ & $1.08 \pm 0.20$ & $1.20 \pm 0.22$ & $1.24 \pm 0.28$ & $1.19 \pm 0.26$ & $1.11 \pm 0.23$ & $1.31 \pm 0.29$ \\
\muThigh & \multicolumn{2}{c}{--} & \multicolumn{2}{c}{--} & $0.74 \pm 0.17$ & $0.79 \pm 0.26$ & $0.55 \pm 0.13$ & $0.71 \pm 0.27$ \\
\muTlow  & \multicolumn{2}{c}{--} & \multicolumn{2}{c}{--} & $0.75 \pm 0.12$ & $0.77 \pm 0.23$ & $0.71 \pm 0.12$ & $0.69 \pm 0.21$ \\
\bottomrule
\end{tabular}
}
\endgroup
\end{table}

Similar conclusions are drawn from the CR-only fit results in the VRs of the \mujet, \elbjet, and \mubjet channels.
As an example, the Run~3 VR results for the \mujet channel are shown in Figure~\ref{fig:results:VRs:lightjet:mujet}, while Figure~\ref{fig:results:VRs:bjet} presents the corresponding VRs for the \elbjet and \mubjet channels.
The data generally agree with the SM predictions within uncertainties, and no consistent mis-modelling is observed between Run~2 and Run~3.
Table~\ref{tab:results:normfactors} reports the extracted normalisation factors for all channels.

\subsection{Results for signal regions}

The observed and predicted \mlj distributions in \SROneLej and \SRTwoLej of the \eljet channel for Run~2 and Run~3 are shown in Figure~\ref{fig:results:eljet:SRs}.
The SM predictions agree well with the data within their uncertainties, and no significant excess of events is observed.
Among the selections, \SRTwoL provides the strongest sensitivity to the LQ signals of interest due to its much higher signal to background ratio, as illustrated by the overlaid nominal predictions of two example signal models in Figure~\ref{fig:results:eljet:SRs}.

Figure~\ref{fig:results:mujet:SRs} presents the results of \SROneLmj and \SRTwoLmj of the \mujet channel for both Run~2 and Run~3.
The data agree well with the SM predictions within uncertainties in the \SROneLmj regions.
Deficits of approximately 50\% are observed in the last two bins of \SRTwoLmj in Run~2, although compatible with the predictions within uncertainties.

The results of \SROneLeb and \SRTwoLeb of the \elbjet channel for both data-taking periods are shown in Figure~\ref{fig:results:elbjet:SRs}, while Figure~\ref{fig:results:mubjet:SRs} presents the corresponding SRs, \SROneLeb and \SRTwoLeb, of the \mubjet channel.
In both channels no significant excesses are observed and the SM predictions generally agree with the data within uncertainties.

To quantitatively probe the observed \mlj spectra in each channel for a new-physics signal, the $p_{0}$ value of the background-only hypothesis is evaluated relative to each signal model and translated into a discovery significance.
No deviations above the $1\sigma$ level are observed in any channel.

\begin{figure}[tbh]
\centering
\subfloat[]{
\includegraphics[width=0.48\textwidth]{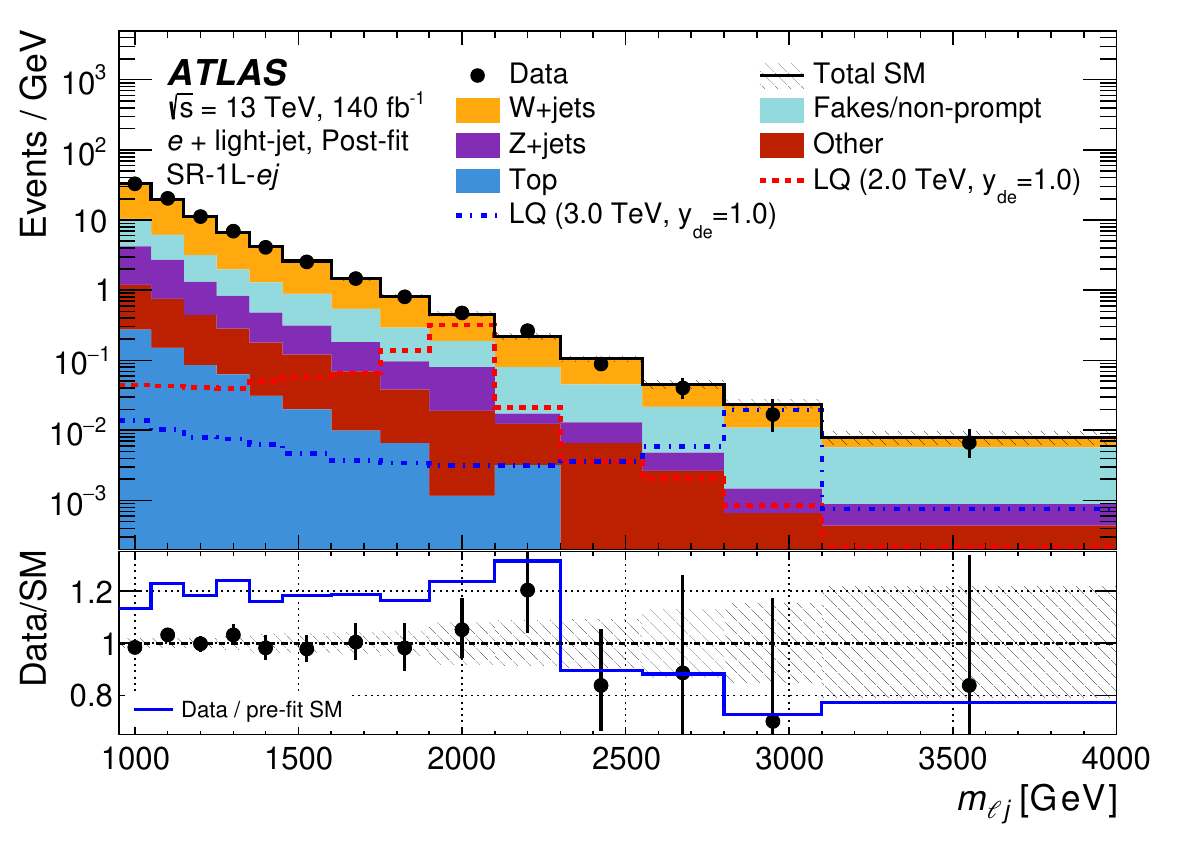}
}
\subfloat[]{
\includegraphics[width=0.48\textwidth]{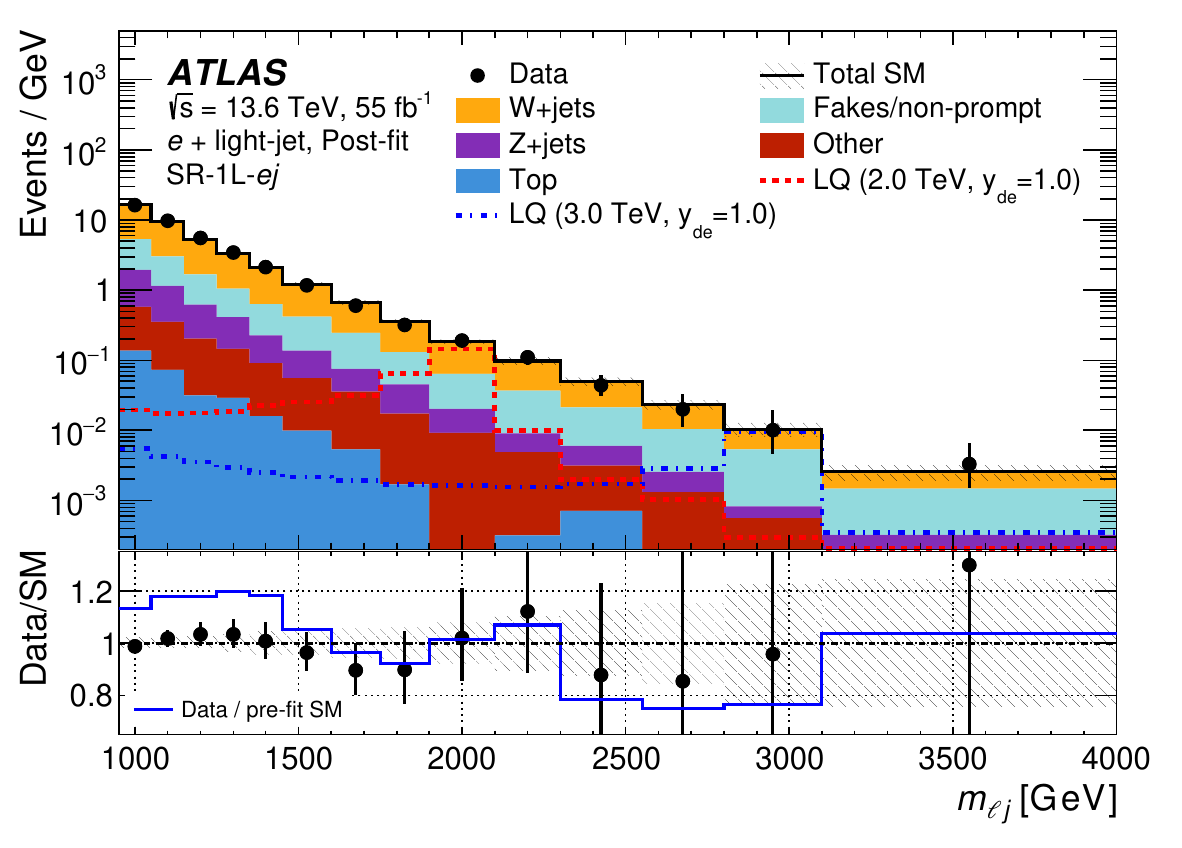}
}\\
\subfloat[]{
\includegraphics[width=0.48\textwidth]{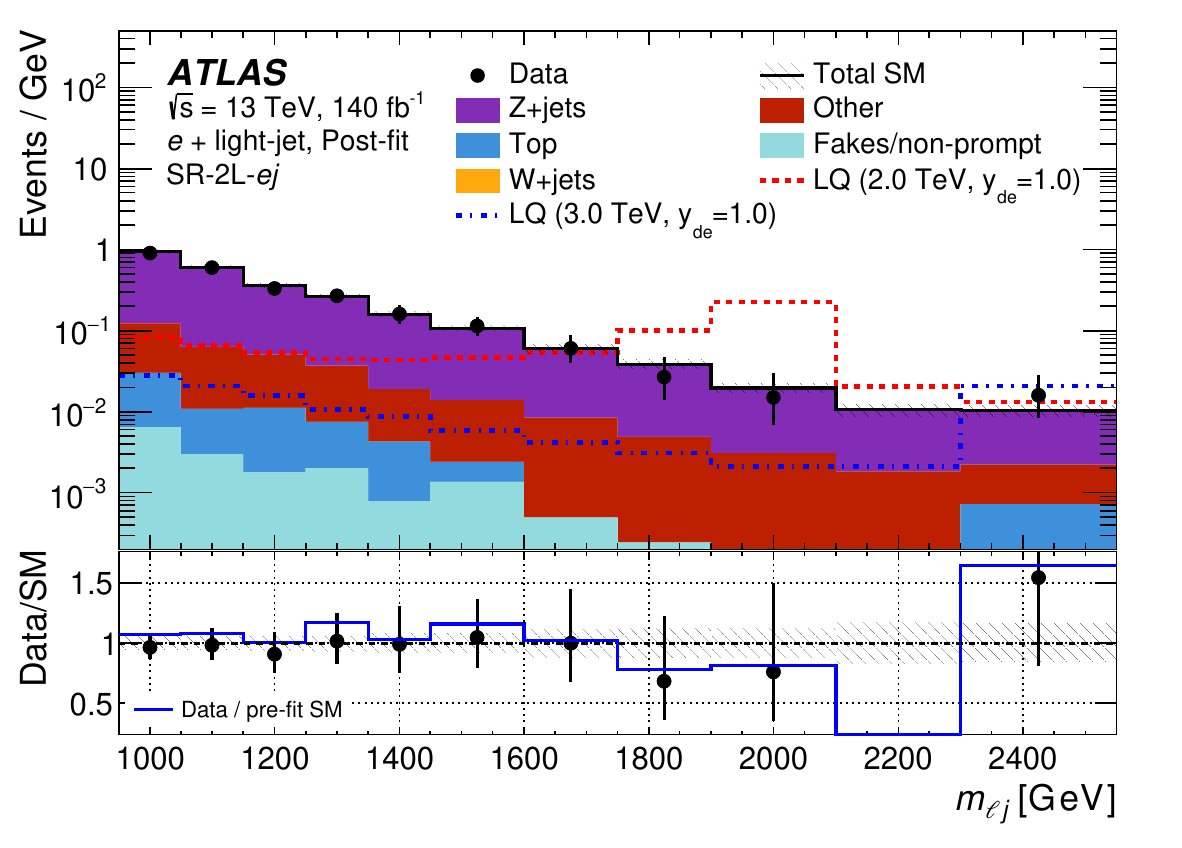}
}
\subfloat[]{
\includegraphics[width=0.48\textwidth]{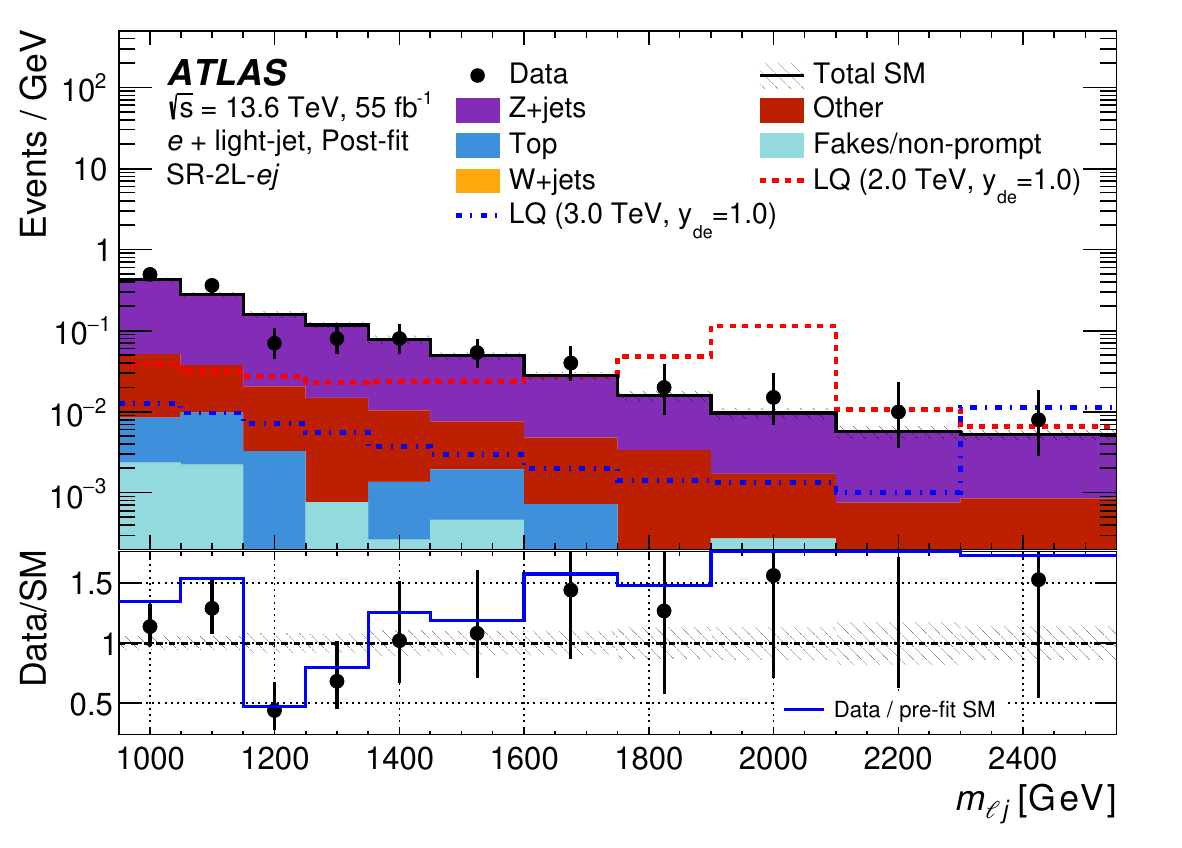}
}
\caption{Data (dots) and post-fit SM distribution (histograms) of \mlj in (a, b) \SROneLej and (c, d) \SRTwoLej of the \eljet channel obtained by a CR+SR background-only fit for Run~2 and Run~3, respectively.
The lower panel shows the ratio of observed data to the total post- and pre-fit SM prediction.
The last bin includes the overflow.
Uncertainties in the background estimates include both the statistical and systematic uncertainties, with correlations between uncertainties taken into account.
The dashed lines show the predicted yields for two benchmark signal models corresponding to $\Stilde(m, y_{de}) = (2.0~\TeV, 1.0)$ and $\Stilde(m, y_{de}) = (3.0~\TeV, 1.0)$, respectively.}
\label{fig:results:eljet:SRs}
\end{figure}

\begin{figure}[tbh]
\centering
\subfloat[]{
\includegraphics[width=0.48\textwidth]{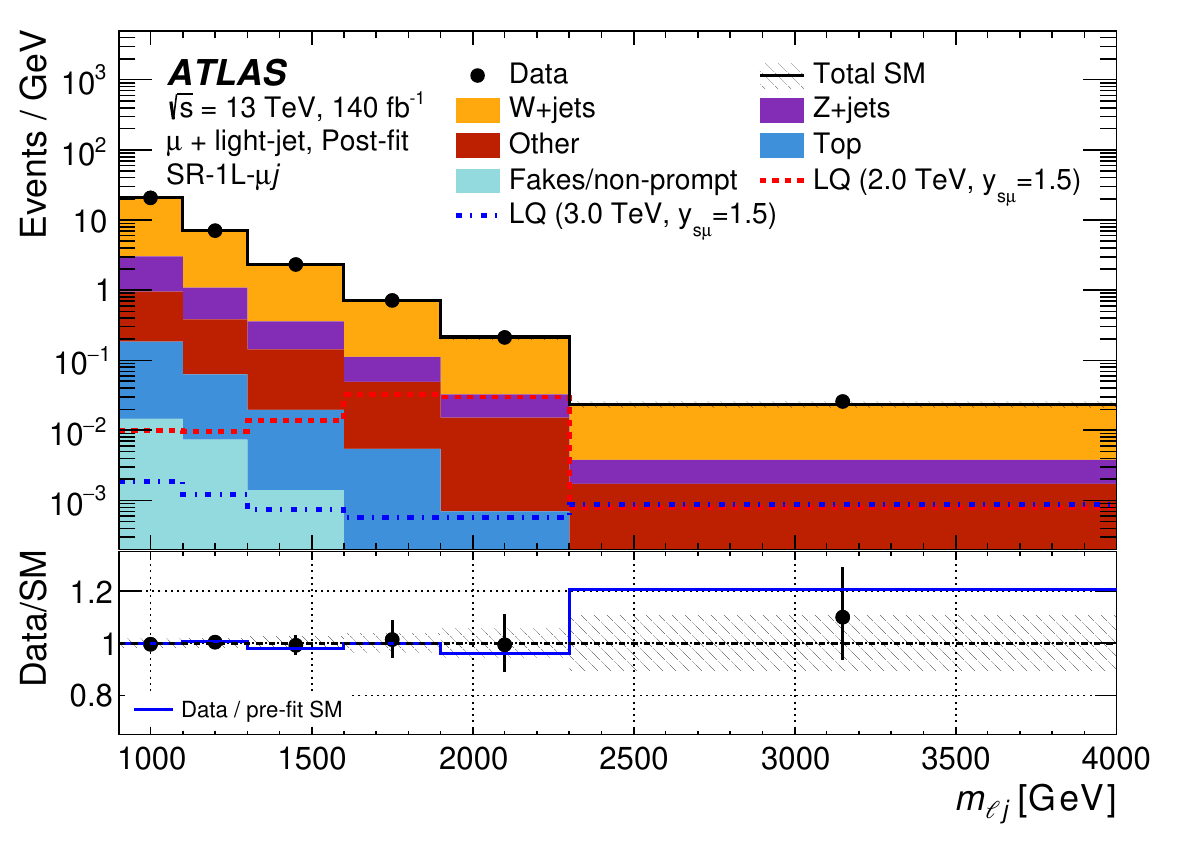}
}
\subfloat[]{
\includegraphics[width=0.48\textwidth]{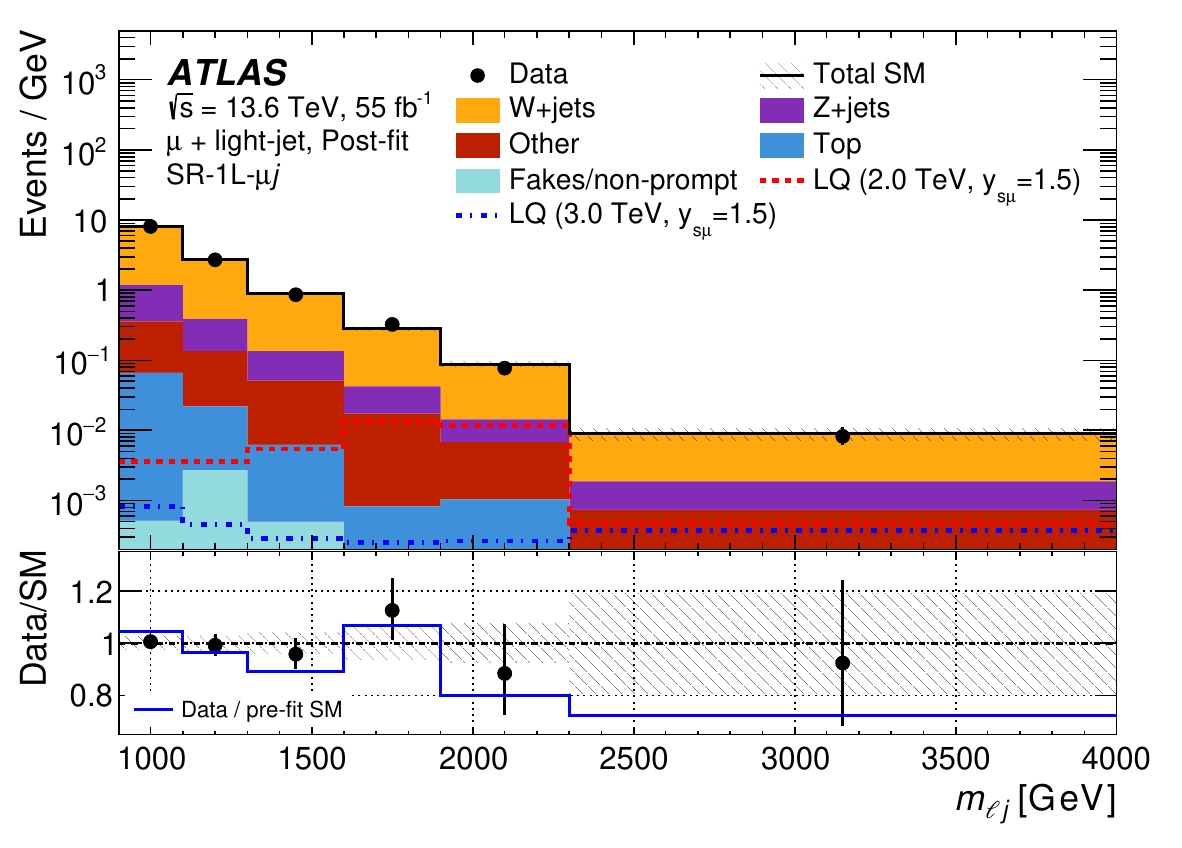}
}\\
\subfloat[]{
\includegraphics[width=0.48\textwidth]{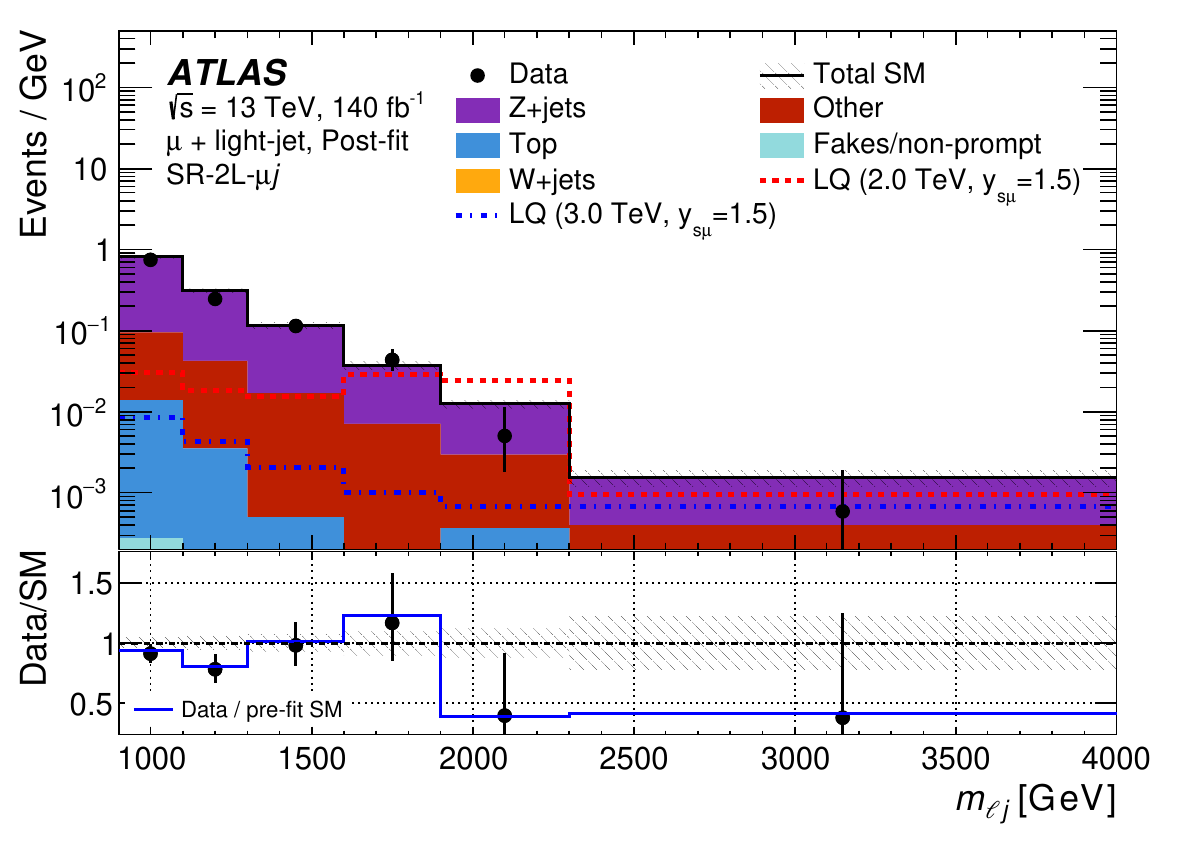}
}
\subfloat[]{
\includegraphics[width=0.48\textwidth]{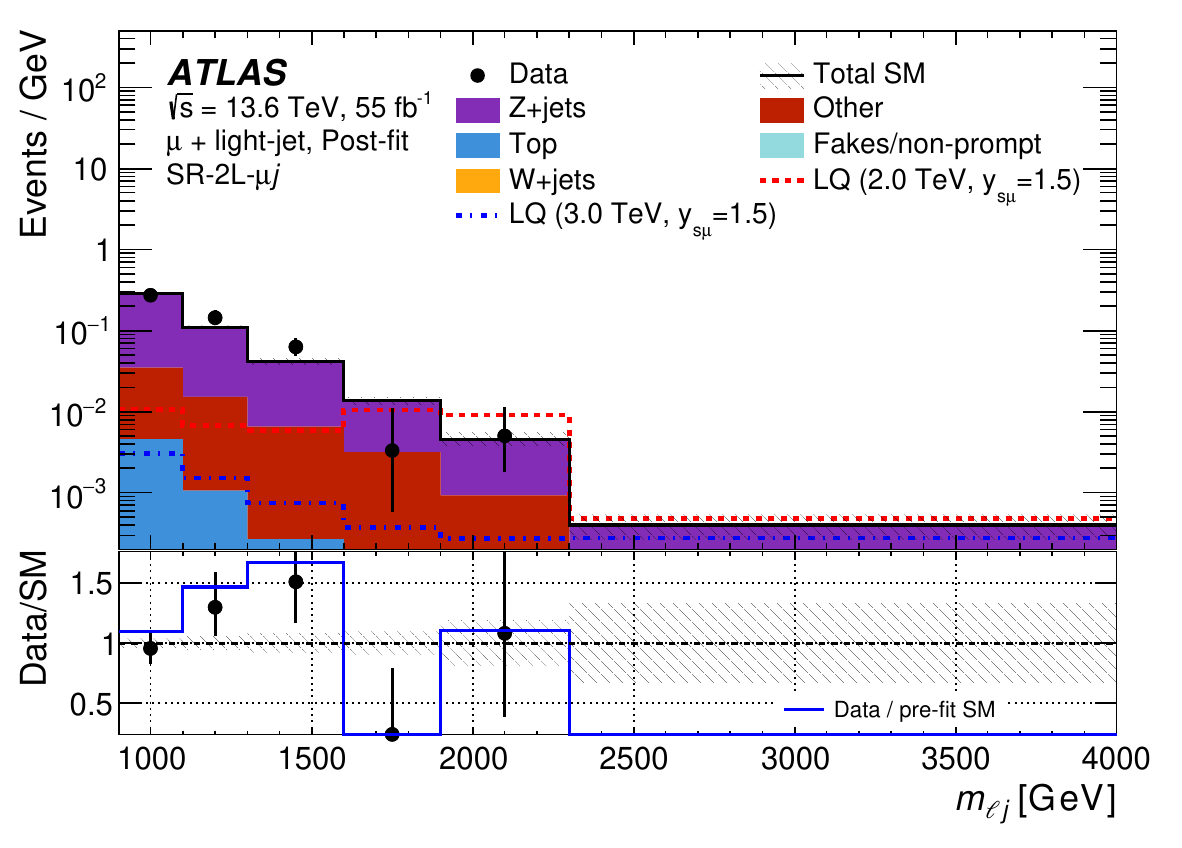}
}
\caption{Data (dots) and post-fit SM distribution (histograms) of \mlj in (a, b) \SROneLmj and (c, d) \SRTwoLmj of the \mujet channel obtained by a CR+SR background-only fit for Run~2 and Run~3, respectively.
The lower panel shows the ratio of observed data to the total post- and pre-fit SM prediction.
The last bin includes the overflow.
Uncertainties in the background estimates include both the statistical and systematic uncertainties, with correlations between uncertainties taken into account.
The dashed lines show the predicted yields for two benchmark signal models corresponding to $\Stilde(m, y_{s\mu}) = (2.0~\TeV, 1.5)$ and $\Stilde(m, y_{s\mu}) = (3.0~\TeV, 1.5)$, respectively.}
\label{fig:results:mujet:SRs}
\end{figure}

\begin{figure}[tbh]
\centering
\subfloat[]{
\includegraphics[width=0.48\textwidth]{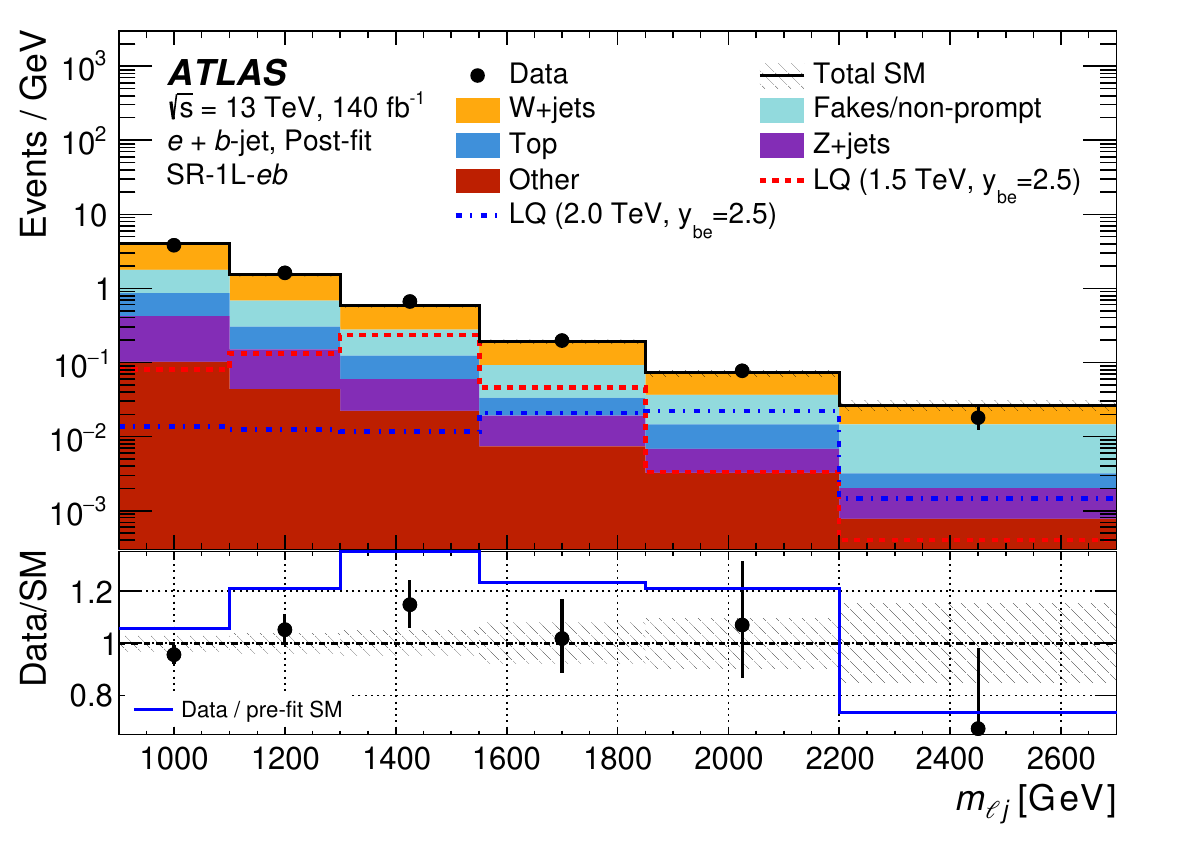}
}
\subfloat[]{
\includegraphics[width=0.48\textwidth]{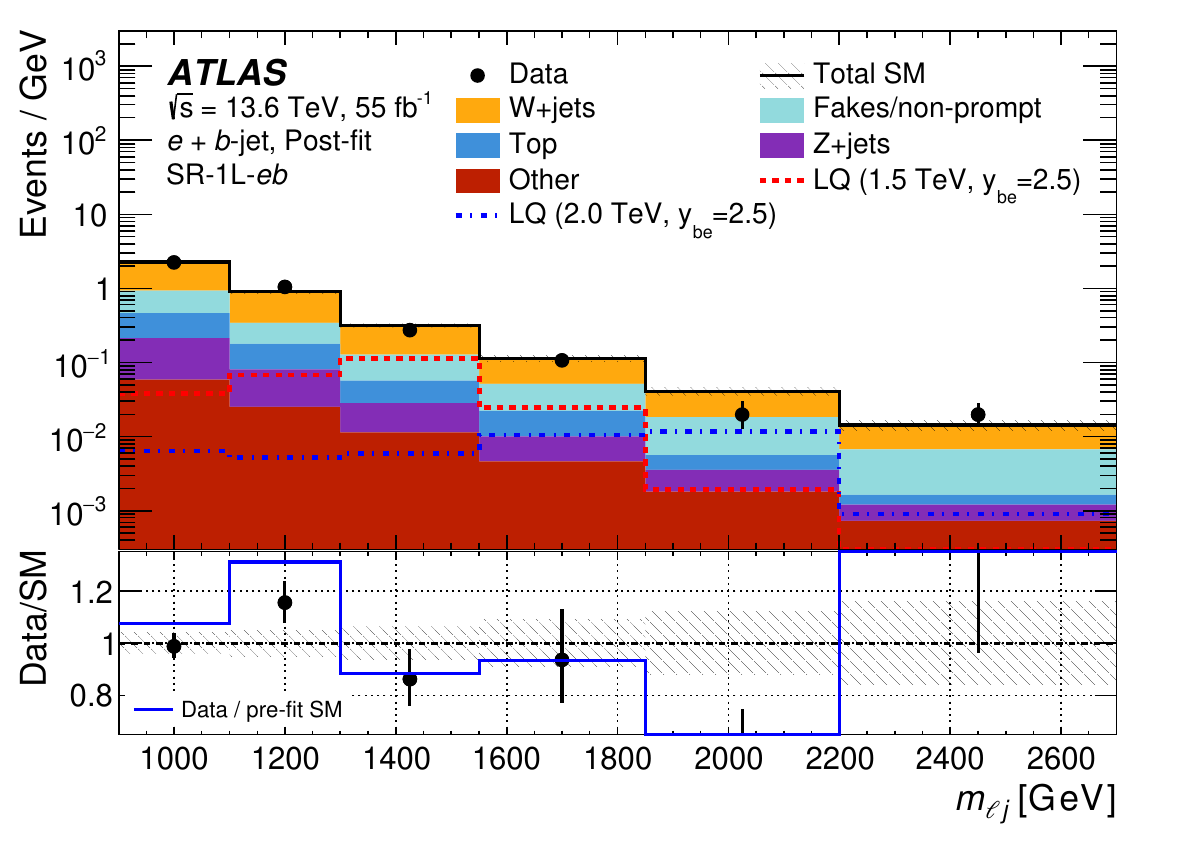}
}\\
\subfloat[]{
\includegraphics[width=0.48\textwidth]{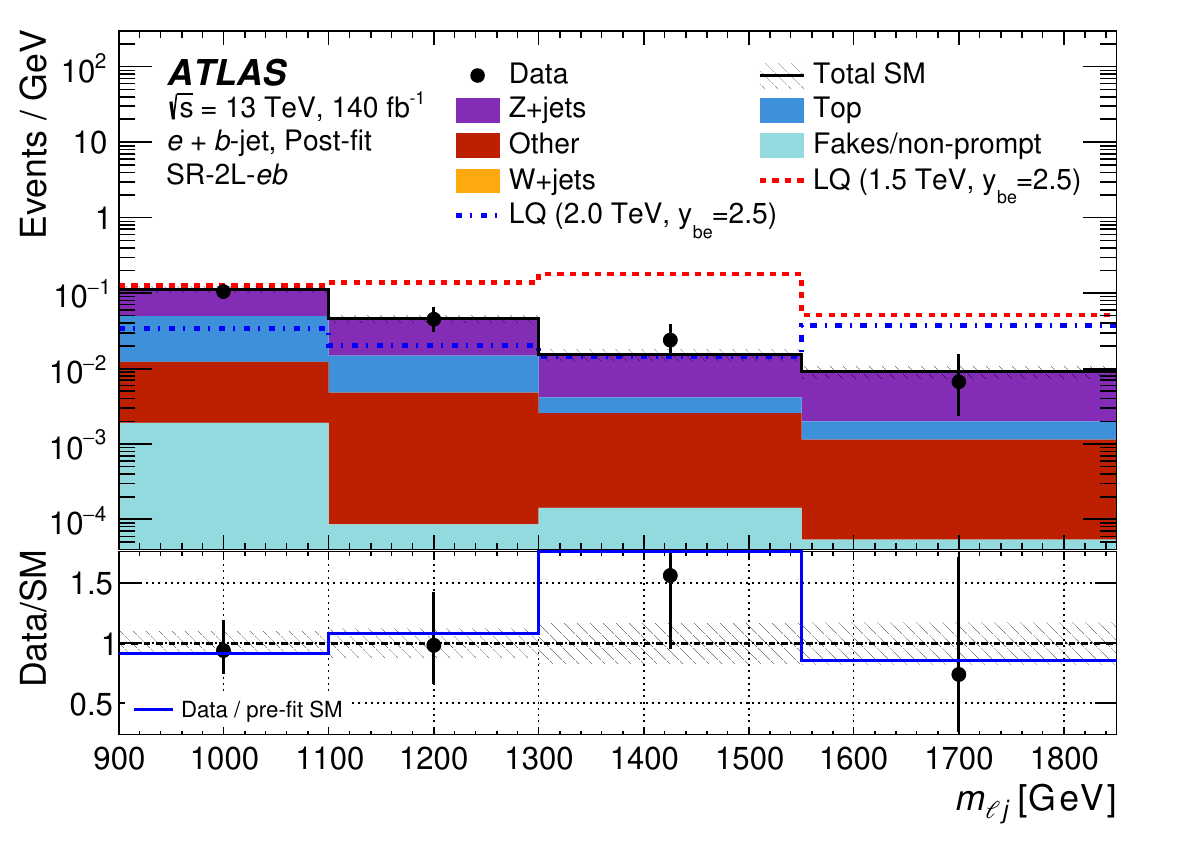}
}
\subfloat[]{
\includegraphics[width=0.48\textwidth]{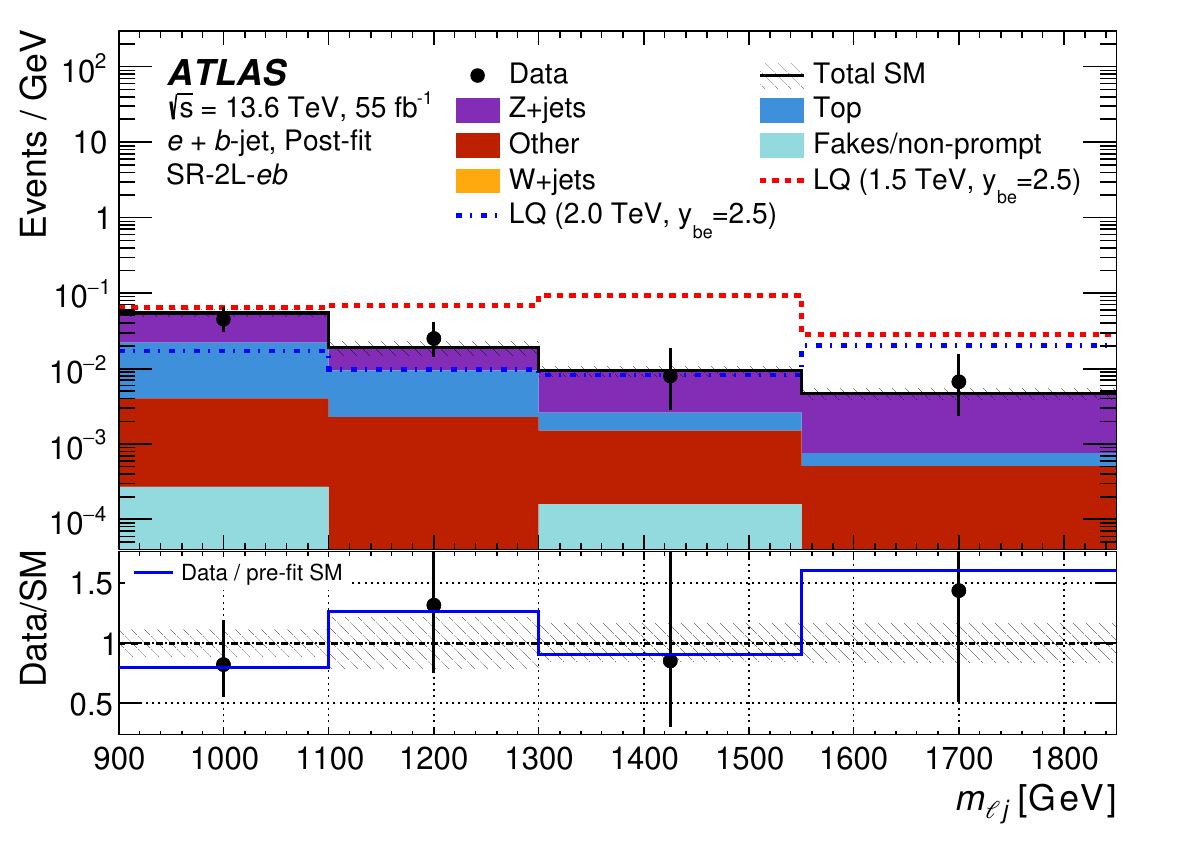}
}
\caption{Data (dots) and post-fit SM distribution (histograms) of \mlj in (a, b) \SROneLeb and (c, d) \SRTwoLeb of the \elbjet channel obtained by a CR+SR background-only fit for Run~2 and Run~3, respectively.
The lower panel shows the ratio of observed data to the total post- and pre-fit SM prediction.
The last bin includes the overflow.
Uncertainties in the background estimates include both the statistical and systematic uncertainties, with correlations between uncertainties taken into account.
The dashed lines show the predicted yields for two benchmark signal models corresponding to $\Stilde(m, y_{be}) = (1.5~\TeV, 2.5)$ and $\Stilde(m, y_{be}) = (2.0~\TeV, 2.5)$, respectively.}
\label{fig:results:elbjet:SRs}
\end{figure}

\begin{figure}[tbh]
\centering
\subfloat[]{
\includegraphics[width=0.48\textwidth]{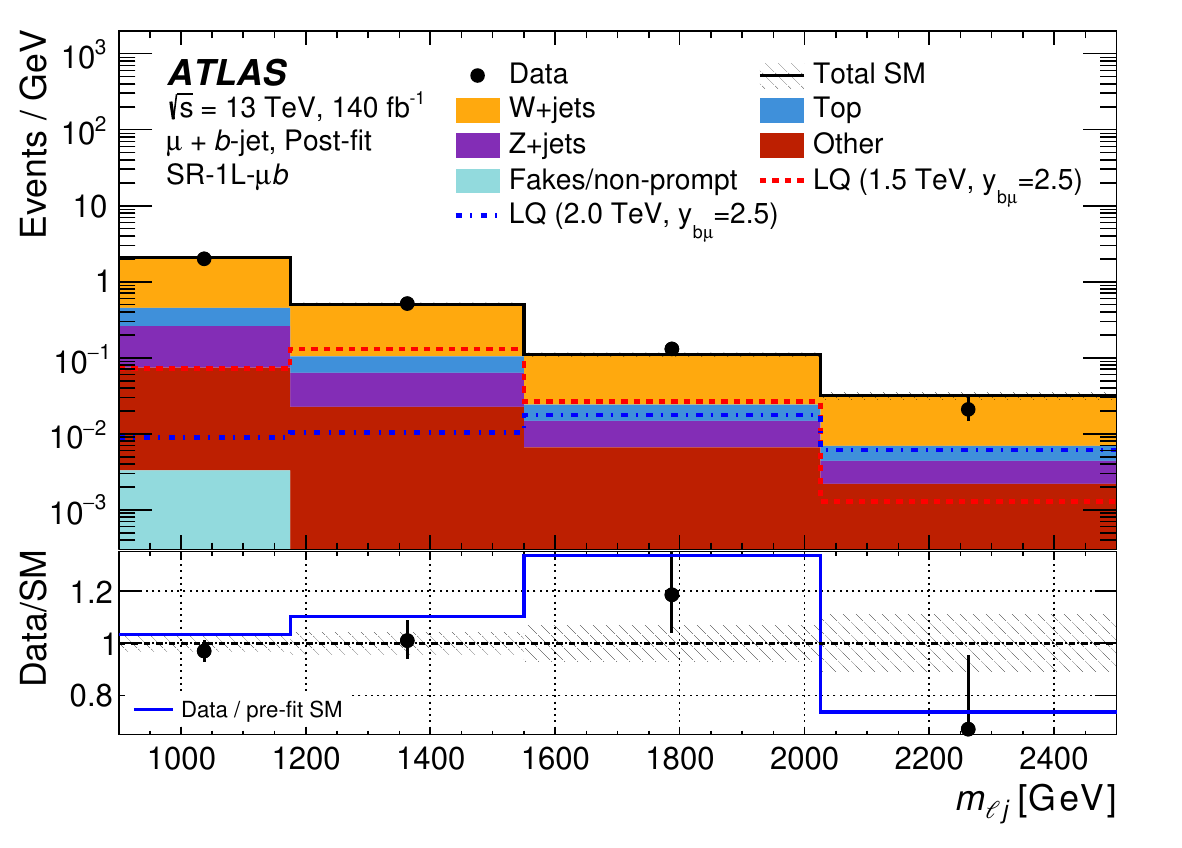}
}
\subfloat[]{
\includegraphics[width=0.48\textwidth]{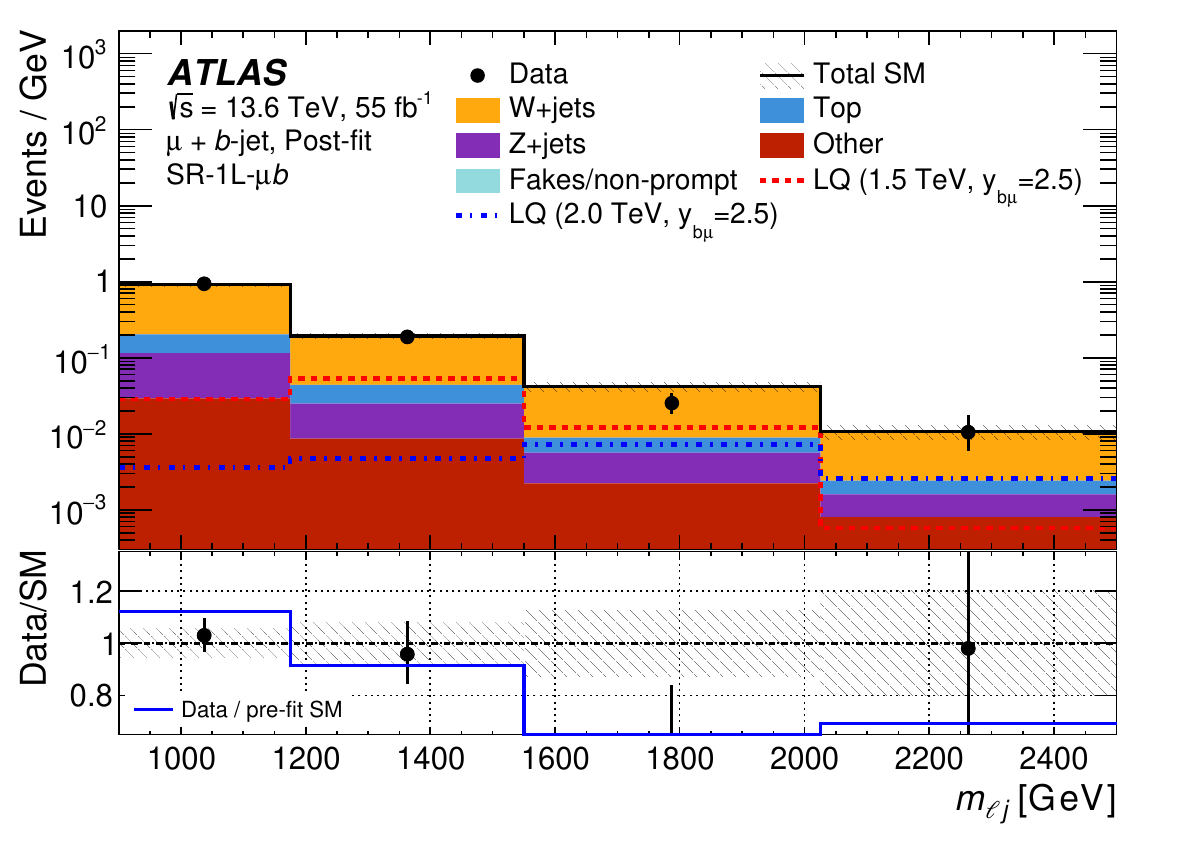}
}\\
\subfloat[]{
\includegraphics[width=0.48\textwidth]{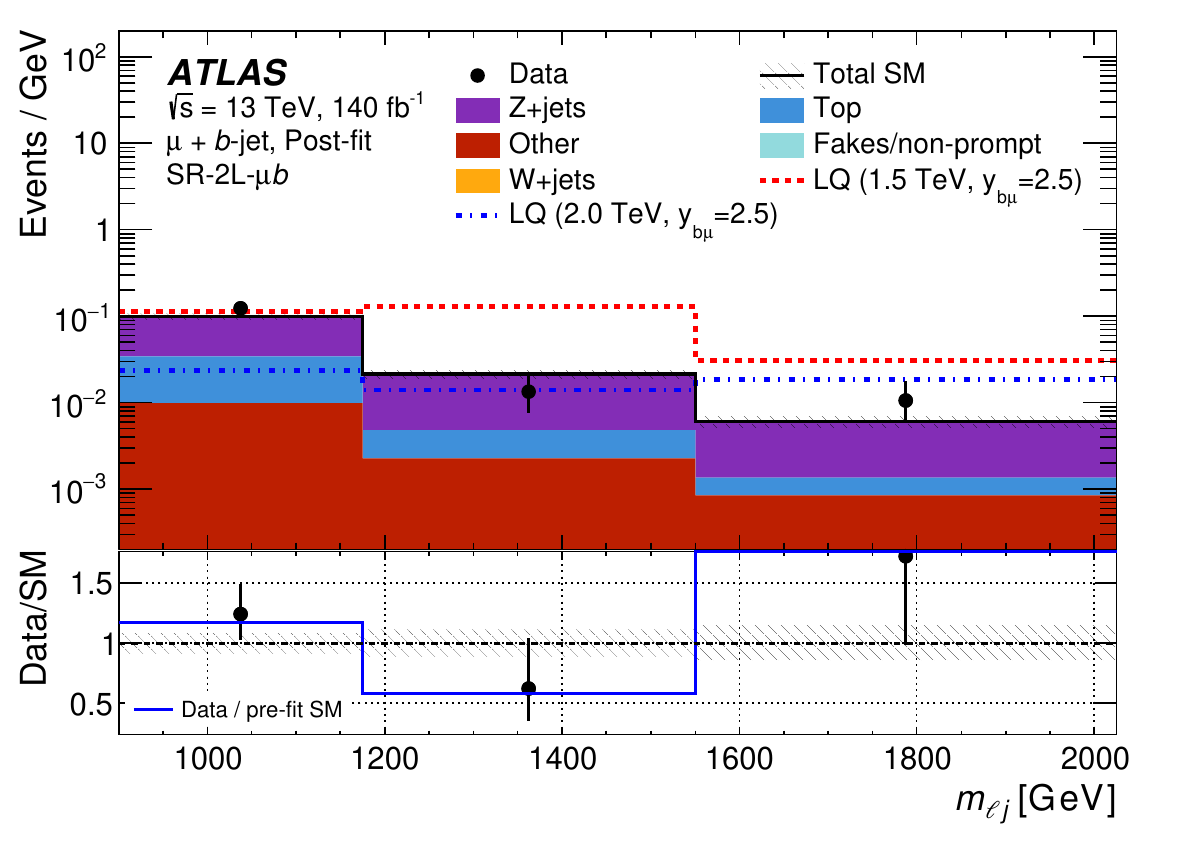}
}
\subfloat[]{
\includegraphics[width=0.48\textwidth]{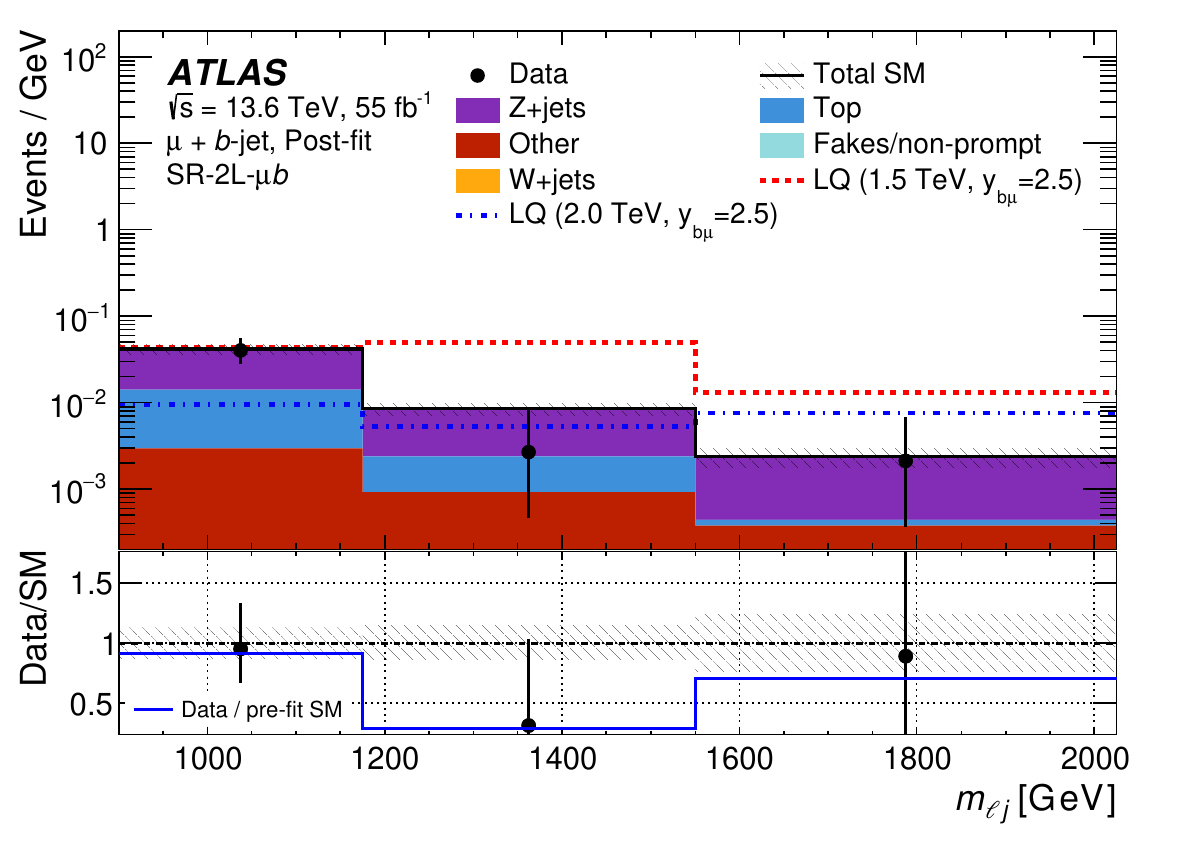}
}
\caption{Data (dots) and post-fit SM distribution (histograms) of \mlj in (a, b) \SROneLmb and (c, d) \SRTwoLmb of the \mubjet channel obtained by a CR+SR background-only fit for Run~2 and Run~3, respectively.
The lower panel shows the ratio of observed data to the total post- and pre-fit SM prediction.
The last bin includes the overflow.
Uncertainties in the background estimates include both the statistical and systematic uncertainties, with correlations between uncertainties taken into account.
The dashed lines show the predicted yields for two benchmark signal models corresponding to $\Stilde(m, y_{b\mu}) = (1.5~\TeV, 1.5)$ and $\Stilde(m, y_{b\mu}) = (2.0~\TeV, 1.5)$, respectively.}
\label{fig:results:mubjet:SRs}
\end{figure}

\subsection{Interpretations}

In the absence of any significant indications for new-physics contributions in the SRs, the observations are interpreted as constraints on minimal LQ production models featuring \Stilde with either $y_{de}$, $y_{s\mu}$, $y_{be}$, or $y_{b\mu}$ couplings as benchmark scenarios.
To evaluate the compatibility of the observed data with a given \Stilde model, a combined fit to the Run-2 and Run-3 datasets is performed, effectively fitting the \mlj spectra from both data-taking periods simultaneously.
In these fits, the signal strength is a free parameter and coherently scales the nominal signal predictions across all regions, accounting for the difference of the \Stilde production cross-sections between Run~2 and Run~3.
The \cls prescription~\cite{Read:2002hq} is used to perform hypothesis tests and set exclusion limits at 95\% confidence level (CL), employing the asymptotic approximation~\cite{Cowan:2010js} for the calculation of the \cls values.
The results were cross-checked using pseudo experiments and found to agree within 10\%.
In each channel the constraints are dominated by the Run-2 results due to the larger dataset size and the \SRTwoL regions which are significantly more sensitive to the signal of interest than their respective \SROneL counterparts.

Figure~\ref{fig:results:exlusion_contour_eljet} shows the observed and expected exclusion limits for the \eljet channel.
The contours are presented in the $m(\Stilde)$--$y_{de}$ plane, with couplings ranging from 0.1 to 1.0.
In this minimal LQ model, \Stilde masses up to approximately 3.4~\TeV are excluded for a coupling value of $y_{de} = 1.0$.
For couplings larger than roughly 0.25, this search surpasses the limits set by a previous ATLAS search for LQ pair production~\cite{EXOT-2019-13}, which are independent of the coupling strength.
Strong indirect constraints on $y_{de}$ arise from weak charge measurements of protons and nuclei~\cite{Schmaltz_2019}, which impose $y_{de} \leq  0.17 \frac{m(S)}{[\TeV]}$, where $m(S)$ denotes the scalar LQ mass.
The results from this search exceed those limits in a small region of parameter space below approximately 2.3~\TeV.
Although the weak charge constraints dominate across most of the parameter space, they may be relaxed or even vanish in scenarios with more than one LQ present in the mass range of interest~\cite{Dorsner:2022ibm}, rendering the constraints from this search complementary.

The observed and expected exclusion contours for the \mujet channel are shown in Figure~\ref{fig:results:exlusion_contour_mujet}.
The deficits observed in the highest \mlj bins of \SRTwoLmj for Run~2 shift the observed limits beyond the $1\sigma$ uncertainty band, though they are found to remain well within the $2\sigma$ band.
\Stilde masses up to approximately 4.3~\TeV are excluded assuming coupling values of 3.5.

Figure~\ref{fig:results:exlusion_contour_elbjet} shows the observed and expected exclusion limits for the \elbjet channel.
The derived constraints on the \Stilde mass depend again on the coupling and range up to 3.1~\TeV for $y_{be} = 3.5$.
The mild excess observed in the third \mlj bin of \SRTwoLeb in Run~2 slightly weakens the observed limit relative to the expected one for masses below approximately 2~\TeV.

Exclusion contours for the \mubjet channel are presented in Figure~\ref{fig:results:exlusion_contour_mubjet} where constraints on the \Stilde mass extend up to 2.8~\TeV at couplings of 3.5.

In summary, all channels improve upon existing ATLAS constraints for scalar LQ models at large coupling values—approximately above 0.25 in the \eljet channel, above 0.7 in the \mujet channel and above 1.0 in the other channels.
At lower couplings, the LQ pair production searches remain more sensitive due to their independence from the LQ coupling strength.

\begin{figure}[th]
\centering
\includegraphics[width=0.85\textwidth]{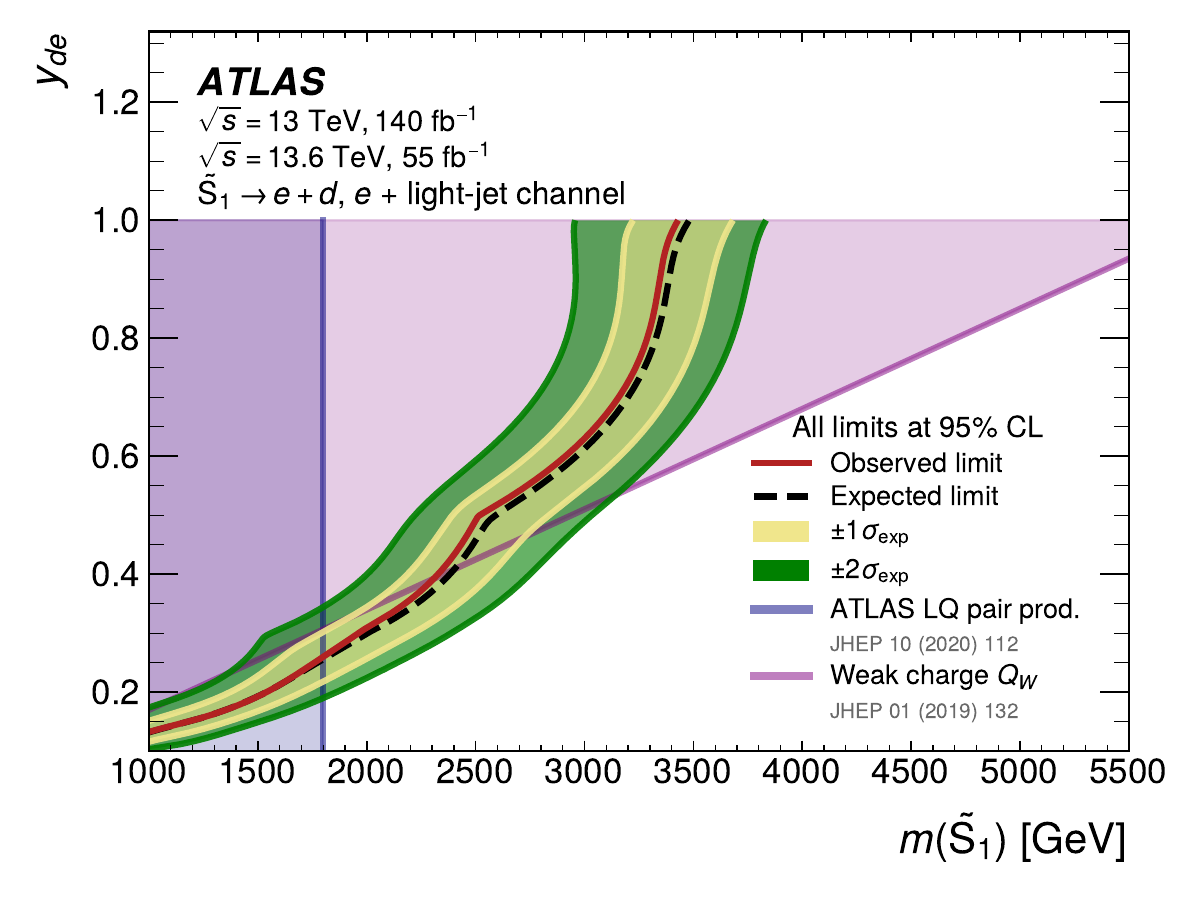}
\caption{Exclusion limits for minimal models of $\Stilde$ production with only $y_{de}$ being non-zero obtained from a simultaneous fit to \SROneL and \SRTwoL of the \eljet channel combining Run-2 and Run-3 data.
All limits are computed at 95\% CL and the observed (red solid lines) and expected (black dashed lines) exclusion limits are shown in the $m(\Stilde)$ -- $y_{de}$ plane.
The observed exclusion should be interpreted as the region above the red line.
The yellow inner (green outer) shaded band around the expected limits corresponds to the $\pm 1\,\sigma$ ($\pm 2\,\sigma$) variations of the expected limit, accounting for all uncertainties.
The observed limit obtained from ATLAS searches for LQ pair production is also shown as dark blue line~\cite{EXOT-2019-13}.
Constraints from weak charge measurements of protons and nuclei on $y_{de}$ couplings derived by Ref.~\cite{Schmaltz_2019} are shown as light magenta line.\\
}
\label{fig:results:exlusion_contour_eljet}
\end{figure}

\begin{figure}[th]
\centering
\includegraphics[width=0.85\textwidth]{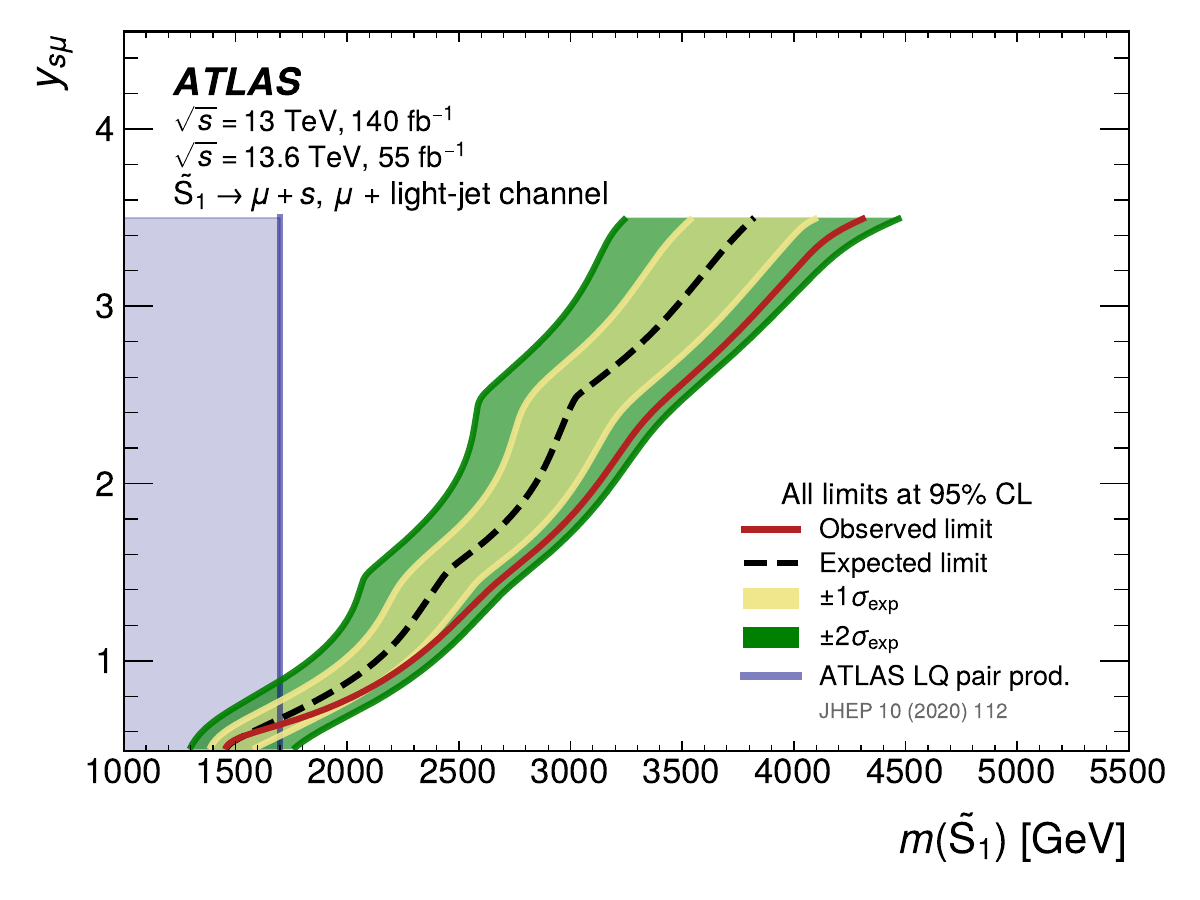}
\caption{Exclusion limits for minimal models of $\Stilde$ production with only $y_{s\mu}$ being non-zero obtained from a simultaneous fit to \SROneL and \SRTwoL of the \mujet channel combining Run-2 and Run-3 data.
All limits are computed at 95\% CL and the observed (red solid lines) and expected (black dashed lines) exclusion limits are shown in the $m(\Stilde)$ -- $y_{s\mu}$ plane.
The observed exclusion should be interpreted as the region above the red line.
The yellow inner (green outer) shaded band around the expected limits corresponds to the $\pm 1\,\sigma$ ($\pm 2\,\sigma$) variations of the expected limit, accounting for all uncertainties.
The observed limit obtained from ATLAS searches for LQ pair production is also shown as dark blue line~\cite{EXOT-2019-13}.\\
}
\label{fig:results:exlusion_contour_mujet}
\end{figure}

\begin{figure}[th]
\centering
\includegraphics[width=0.85\textwidth]{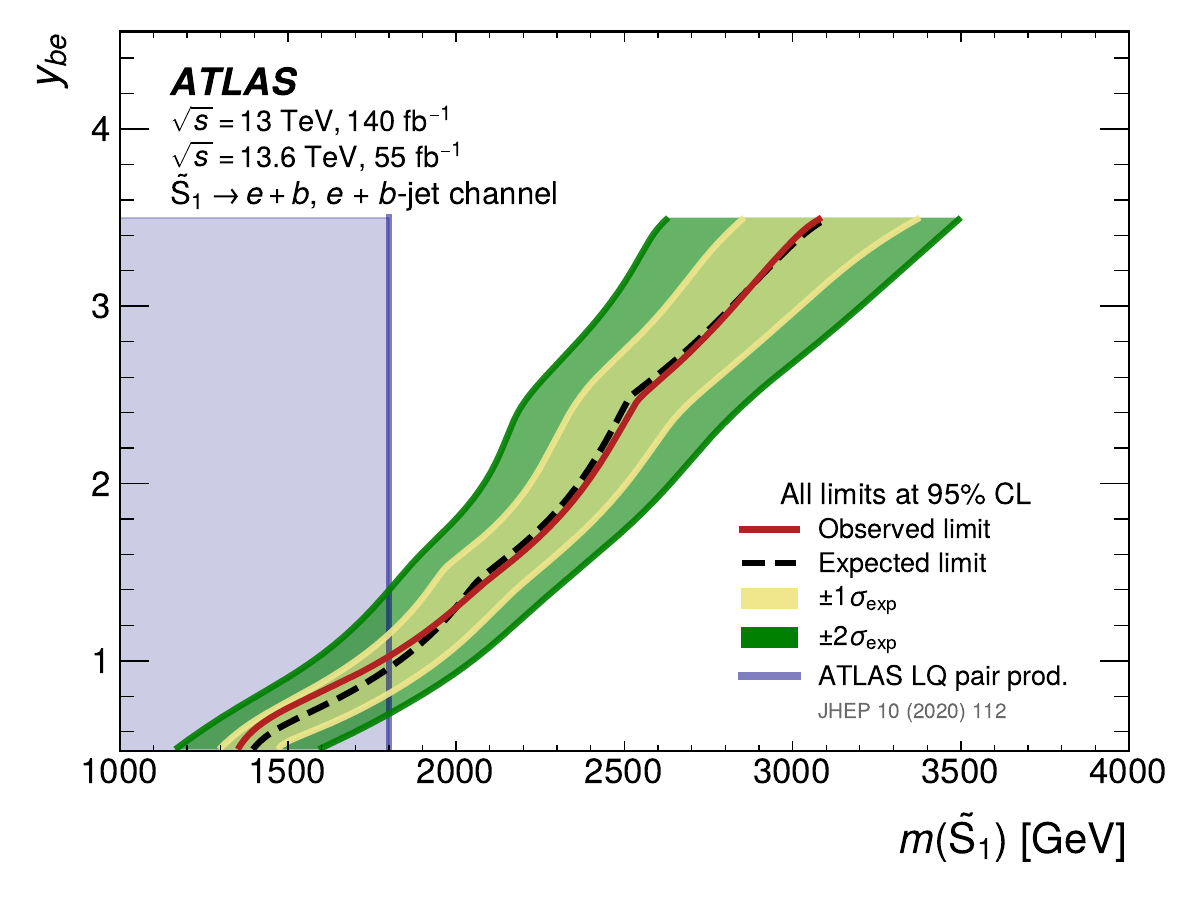}
\caption{Exclusion limits for minimal models of $\Stilde$ production with only $y_{be}$ being non-zero obtained from a simultaneous fit to \SROneL and \SRTwoL of the \elbjet channel combining Run-2 and Run-3 data.
All limits are computed at 95\% CL and the observed (red solid lines) and expected (black dashed lines) exclusion limits are shown in the $m(\Stilde)$ -- $y_{be}$ plane.
The observed exclusion should be interpreted as the region above the red line.
The yellow inner (green outer) shaded band around the expected limits corresponds to the $\pm 1\,\sigma$ ($\pm 2\,\sigma$) variations of the expected limit, accounting for all uncertainties.
The observed limit obtained from ATLAS searches for LQ pair production is also shown as dark blue line~\cite{EXOT-2019-13}.\\
}
\label{fig:results:exlusion_contour_elbjet}
\end{figure}

\begin{figure}[th]
\centering
\includegraphics[width=0.85\textwidth]{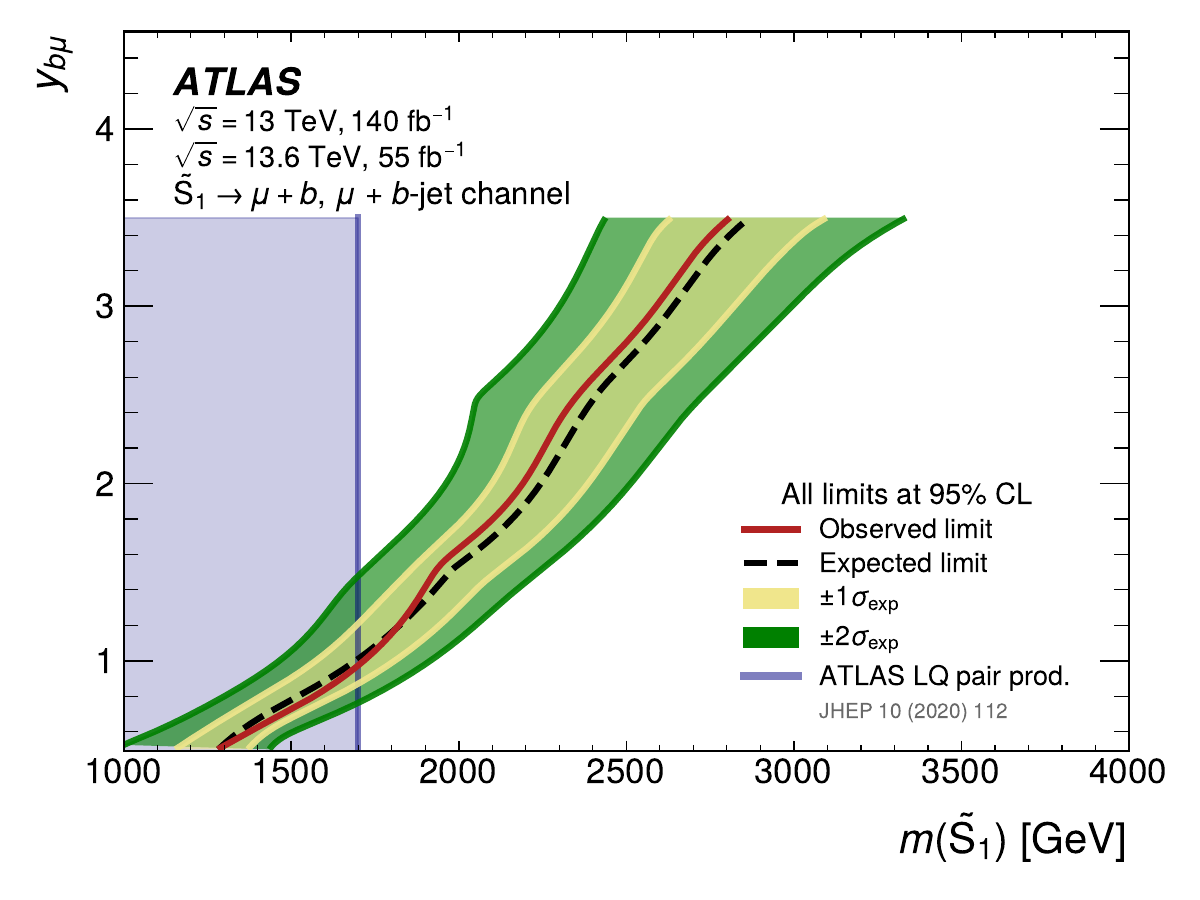}
\caption{Exclusion limits for minimal models of $\Stilde$ production with only $y_{b\mu}$ being non-zero obtained from a simultaneous fit to \SROneL and \SRTwoL of the \mubjet channel combining Run-2 and Run-3 data.
All limits are computed at 95\% CL and the observed (red solid lines) and expected (black dashed lines) exclusion limits are shown in the $m(\Stilde)$ -- $y_{b\mu}$ plane.
The observed exclusion should be interpreted as the region above the red line.
The yellow inner (green outer) shaded band around the expected limits corresponds to the $\pm 1\,\sigma$ ($\pm 2\,\sigma$) variations of the expected limit, accounting for all uncertainties.
The observed limit obtained from ATLAS searches for LQ pair production is also shown as dark blue line~\cite{EXOT-2019-13}.\\
}
\label{fig:results:exlusion_contour_mubjet}
\end{figure}

\FloatBarrier


%

%
%
%
\FloatBarrier
\section{Conclusion}

\label{sec:conclusion}

This paper reports a search for resonantly produced LQs using proton--proton collision data
from the full Run-2 (2015--2018) and partial Run-3 (2022--2023) datasets at the LHC\@.
The analysis probes the $s$-channel production of LQs, which exploits the lepton content of the proton
to yield a distinctive lepton+jet final state.
The signal topology features a narrow \mlj peak near the LQ mass, and the inclusion of 2-lepton+jet final states---motivated by NLO contributions to resonant LQ production---is crucial for maximizing the analysis sensitivity as it also provides acceptance for additional LQ production modes beyond the lepton- and photon-induced processes.

Four orthogonal channels---\eljet, \mujet, \elbjet, and \mubjet
---are analysed, each with individually optimised SRs binned in $m_{\ell j}$
to maximize coverage across a broad range of LQ masses.
Dominant SM backgrounds, including \Wjets and \Zjets, are constrained through dedicated CRs at low $m_{\ell j}$, with extrapolations validated in VRs.
For $b$-jet channels, the top background is further controlled using two CRs spanning distinct
kinematic regimes of $m_{\ell j}$, addressing observed dependencies in top processes modelling.

No significant excesses beyond SM predictions are observed in any of the SRs. The results are interpreted within
a minimal $\tilde{S}_1$-model framework, where only a single LQ coupling is non-zero. Combining Run-2 and partial Run-3 data,
the analysis achieves stringent exclusion limits on LQ masses: for
electron+jet channels (\eljet and \elbjet), LQ masses below 3.4~\TeV
(coupling $y_{de} = 1.0$) and 3.1~\TeV ($y_{be} = 3.5$) are excluded at 95\% CL\@.
For muon+jet channels (\mujet and \mubjet), exclusion reaches extend to
4.3~\TeV ($y_{s\mu} = 3.5$) and 2.8~\TeV ($y_{b\mu} = 3.5$), respectively.
These limits surpass those from previous ATLAS searches for LQ pair production for sufficiently large couplings ($y_{de} \gtrsim 0.25, y_{s\mu} \gtrsim 0.7, y_{be, b\mu} \gtrsim 1.0$).
By establishing robust exclusion limits for LQ masses and couplings beyond the reach of pair-production searches, this work underscores the critical role of resonant production as a complementary probe of \TeV-scale LQ scenarios.
The results also highlight the potential of early Run-3 data to constrain exotic physics, paving the way for
future high-luminosity LHC analyses to explore uncharted parameter space in LQ models.


%

%
%
%
%
%
\section*{Acknowledgements}
%

%

%
%

%
%

We thank CERN for the very successful operation of the LHC and its injectors, as well as the support staff at
CERN and at our institutions worldwide without whom ATLAS could not be operated efficiently.

The crucial computing support from all WLCG partners is acknowledged gratefully, in particular from CERN, the ATLAS Tier-1 facilities at TRIUMF/SFU (Canada), NDGF (Denmark, Norway, Sweden), CC-IN2P3 (France), KIT/GridKA (Germany), INFN-CNAF (Italy), NL-T1 (Netherlands), PIC (Spain), RAL (UK) and BNL (USA), the Tier-2 facilities worldwide and large non-WLCG resource providers. Major contributors of computing resources are listed in Ref.~\cite{ATL-SOFT-PUB-2025-001}.

We gratefully acknowledge the support of ANPCyT, Argentina; YerPhI, Armenia; ARC, Australia; BMWFW and FWF, Austria; ANAS, Azerbaijan; CNPq and FAPESP, Brazil; NSERC, NRC and CFI, Canada; CERN; ANID, Chile; CAS, MOST and NSFC, China; Minciencias, Colombia; MEYS CR, Czech Republic; DNRF and DNSRC, Denmark; IN2P3-CNRS and CEA-DRF/IRFU, France; SRNSFG, Georgia; BMFTR, HGF and MPG, Germany; GSRI, Greece; RGC and Hong Kong SAR, China; ICHEP and Academy of Sciences and Humanities, Israel; INFN, Italy; MEXT and JSPS, Japan; CNRST, Morocco; NWO, Netherlands; RCN, Norway; MNiSW, Poland; FCT, Portugal; MNE/IFA, Romania; MSTDI, Serbia; MSSR, Slovakia; ARIS and MVZI, Slovenia; DSI/NRF, South Africa; MICIU/AEI, Spain; SRC and Wallenberg Foundation, Sweden; SERI, SNSF and Cantons of Bern and Geneva, Switzerland; NSTC, Taipei; TENMAK, T\"urkiye; STFC/UKRI, United Kingdom; DOE and NSF, United States of America.

Individual groups and members have received support from BCKDF, CANARIE, CRC and DRAC, Canada; CERN-CZ, FORTE and PRIMUS, Czech Republic; COST, ERC, ERDF, Horizon 2020, ICSC-NextGenerationEU and Marie Sk{\l}odowska-Curie Actions, European Union; Investissements d'Avenir Labex, Investissements d'Avenir Idex and ANR, France; DFG and AvH Foundation, Germany; Herakleitos, Thales and Aristeia programmes co-financed by EU-ESF and the Greek NSRF, Greece; BSF-NSF and MINERVA, Israel; NCN and NAWA, Poland; La Caixa Banking Foundation, CERCA Programme Generalitat de Catalunya and PROMETEO and GenT Programmes Generalitat Valenciana, Spain; G\"{o}ran Gustafssons Stiftelse, Sweden; The Royal Society and Leverhulme Trust, United Kingdom.

In addition, individual members wish to acknowledge support from CERN: European Organization for Nuclear Research (CERN DOCT); Chile: Agencia Nacional de Investigaci\'on y Desarrollo (FONDECYT 1230812, FONDECYT 1240864, Fondecyt 3240661); China: Chinese Ministry of Science and Technology (MOST-2023YFA1605700, MOST-2023YFA1609300), National Natural Science Foundation of China (NSFC - 12175119, NSFC 12275265); Czech Republic: Czech Science Foundation (GACR - 24-11373S), Ministry of Education Youth and Sports (ERC-CZ-LL2327, FORTE CZ.02.01.01/00/22\_008/0004632), PRIMUS Research Programme (PRIMUS/21/SCI/017); EU: H2020 European Research Council (ERC - 101002463); European Union: European Research Council (BARD No. 101116429, ERC - 948254, ERC 101089007), European Regional Development Fund (SMASH COFUND 101081355, SLO ERDF), Horizon 2020 Framework Programme (MUCCA - CHIST-ERA-19-XAI-00), European Union, Future Artificial Intelligence Research (FAIR-NextGenerationEU PE00000013), Italian Center for High Performance Computing, Big Data and Quantum Computing (ICSC, NextGenerationEU); France: Agence Nationale de la Recherche (ANR-21-CE31-0022, ANR-22-EDIR-0002, ANR-24-CE31-0504-01); Germany: Baden-Württemberg Stiftung (BW Stiftung-Postdoc Eliteprogramme), Deutsche Forschungsgemeinschaft (DFG - 469666862, DFG - CR 312/5-2); China: Research Grants Council (GRF); Italy: Istituto Nazionale di Fisica Nucleare (ICSC, NextGenerationEU), Ministero dell'Università e della Ricerca (NextGenEU 153D23001490006 M4C2.1.1, NextGenEU I53D23000820006 M4C2.1.1, NextGenEU I53D23001490006 M4C2.1.1, SOE2024\_0000023); Japan: Japan Society for the Promotion of Science (JSPS KAKENHI JP22H01227, JSPS KAKENHI JP22H04944, JSPS KAKENHI JP22KK0227, JSPS KAKENHI JP24K23939, JSPS KAKENHI JP24KK0251, JSPS KAKENHI JP25H00650, JSPS KAKENHI JP25H01291, JSPS KAKENHI JP25K01023); Norway: Research Council of Norway (RCN-314472); Poland: Ministry of Science and Higher Education (IDUB AGH, POB8, D4 no 9722), Polish National Science Centre (NCN 2021/42/E/ST2/00350, NCN OPUS 2023/51/B/ST2/02507, NCN OPUS nr 2022/47/B/ST2/03059, NCN UMO-2019/34/E/ST2/00393, UMO-2022/47/O/ST2/00148, UMO-2023/49/B/ST2/04085, UMO-2023/51/B/ST2/00920, UMO-2024/53/N/ST2/00869); Portugal: Foundation for Science and Technology (FCT); Spain: Ministry of Science and Innovation (MCIN \& NextGenEU PCI2022-135018-2, MICIN \& FEDER PID2021-125273NB, RYC2019-028510-I, RYC2020-030254-I, RYC2021-031273-I, RYC2022-038164-I); Sweden: Carl Trygger Foundation (Carl Trygger Foundation CTS 22:2312), Swedish Research Council (Swedish Research Council 2023-04654, VR 2021-03651, VR 2022-03845, VR 2022-04683, VR 2023-03403, VR 2024-05451), Knut and Alice Wallenberg Foundation (KAW 2018.0458, KAW 2022.0358, KAW 2023.0366); Switzerland: Swiss National Science Foundation (SNSF - PCEFP2\_194658); United Kingdom: Royal Society (NIF-R1-231091); United States of America: U.S. Department of Energy (ECA DE-AC02-76SF00515), Neubauer Family Foundation.

%
%


%
\clearpage
\appendix
\printbibliography
\clearpage
 
\begin{flushleft}
\hypersetup{urlcolor=black}
{\Large The ATLAS Collaboration}

\bigskip

\AtlasOrcid[0000-0002-6665-4934]{G.~Aad}$^\textrm{\scriptsize 104}$,
\AtlasOrcid[0000-0001-7616-1554]{E.~Aakvaag}$^\textrm{\scriptsize 17}$,
\AtlasOrcid[0000-0002-5888-2734]{B.~Abbott}$^\textrm{\scriptsize 123}$,
\AtlasOrcid[0000-0002-0287-5869]{S.~Abdelhameed}$^\textrm{\scriptsize 119a}$,
\AtlasOrcid[0000-0002-1002-1652]{K.~Abeling}$^\textrm{\scriptsize 55}$,
\AtlasOrcid[0000-0001-5763-2760]{N.J.~Abicht}$^\textrm{\scriptsize 49}$,
\AtlasOrcid[0000-0002-8496-9294]{S.H.~Abidi}$^\textrm{\scriptsize 30}$,
\AtlasOrcid[0009-0003-6578-220X]{M.~Aboelela}$^\textrm{\scriptsize 45}$,
\AtlasOrcid[0000-0002-9987-2292]{A.~Aboulhorma}$^\textrm{\scriptsize 36e}$,
\AtlasOrcid[0000-0001-5329-6640]{H.~Abramowicz}$^\textrm{\scriptsize 157}$
% ...
\bigskip
\\

% ...
$^{17}$Department for Physics and Technology, University of Bergen, Bergen; Norway.\\
% ...
$^{30}$Physics Department, Brookhaven National Laboratory, Upton NY; United States of America.\\
% ...
$^{36}$$^{(a)}$Facult\'e des Sciences Ain Chock, Universit\'e Hassan II de Casablanca;$^{(b)}$Facult\'{e} des Sciences, Universit\'{e} Ibn-Tofail, K\'{e}nitra;$^{(c)}$Facult\'e des Sciences Semlalia, Universit\'e Cadi Ayyad, LPHEA-Marrakech;$^{(d)}$LPMR, Facult\'e des Sciences, Universit\'e Mohamed Premier, Oujda;$^{(e)}$Facult\'e des sciences, Universit\'e Mohammed V, Rabat;$^{(f)}$Institute of Applied Physics, Mohammed VI Polytechnic University, Ben Guerir; Morocco.\\
% ...
$^{45}$Physics Department, Southern Methodist University, Dallas TX; United States of America.\\
% ...
$^{49}$Fakult\"{a}t Physik , Technische Universit{\"a}t Dortmund, Dortmund; Germany.\\
% ...
$^{55}$II. Physikalisches Institut, Georg-August-Universit\"{a}t G\"ottingen, G\"ottingen; Germany.\\
% ...
$^{104}$CPPM, Aix-Marseille Universit\'e, CNRS/IN2P3, Marseille; France.\\
% ...
$^{119}$$^{(a)}$New York University Abu Dhabi, Abu Dhabi;$^{(b)}$United Arab Emirates University, Al Ain; United Arab Emirates.\\
% ...
$^{123}$Homer L. Dodge Department of Physics and Astronomy, University of Oklahoma, Norman OK; United States of America.\\
% ...
$^{157}$Raymond and Beverly Sackler School of Physics and Astronomy, Tel Aviv University, Tel Aviv; Israel.\\
% ...
$^{a}$ Also at Affiliated with an institute formerly covered by a cooperation agreement with CERN.\\
$^{b}$ Also at An-Najah National University, Nablus; Palestine.\\
$^{c}$ Also at Borough of Manhattan Community College, City University of New York, New York NY; United States of America.\\
$^{d}$ Also at Center for Interdisciplinary Research and Innovation (CIRI-AUTH), Thessaloniki; Greece.\\
$^{e}$ Also at Centre of Physics of the Universities of Minho and Porto (CF-UM-UP); Portugal.\\
$^{f}$ Also at CERN, Geneva; Switzerland.\\
% ...

\end{flushleft}


%

%
%

%

%
%
%
%

\end{document}